\documentclass[usenatbib]{mn2e}

\usepackage{graphicx}	
\usepackage{amssymb}
\usepackage[fleqn]{amsmath}

\def\jnl@style{}
\def\aaref@jnl#1{{\jnl@style#1}}

\def\aaref@jnl#1{{\jnl@style#1}}

\def\aj{\aaref@jnl{AJ}}                   
\def\araa{\aaref@jnl{ARA\&A}}             
\def\apj{\aaref@jnl{ApJ}}                 
\def\apjl{\aaref@jnl{ApJ}}                
\def\apjs{\aaref@jnl{ApJS}}               
\def\ao{\aaref@jnl{Appl.~Opt.}}           
\def\apss{\aaref@jnl{Ap\&SS}}             
\def\aap{\aaref@jnl{A\&A}}                
\def\aapr{\aaref@jnl{A\&A~Rev.}}          
\def\aaps{\aaref@jnl{A\&AS}}              
\def\azh{\aaref@jnl{AZh}}                 
\def\baas{\aaref@jnl{BAAS}}               
\def\jrasc{\aaref@jnl{JRASC}}             
\def\memras{\aaref@jnl{MmRAS}}            
\def\mnras{\aaref@jnl{MNRAS}}             
\def\pra{\aaref@jnl{Phys.~Rev.~A}}        
\def\prb{\aaref@jnl{Phys.~Rev.~B}}        
\def\prc{\aaref@jnl{Phys.~Rev.~C}}        
\def\prd{\aaref@jnl{Phys.~Rev.~D}}        
\def\pre{\aaref@jnl{Phys.~Rev.~E}}        
\def\prl{\aaref@jnl{Phys.~Rev.~Lett.}}    
\def\pasp{\aaref@jnl{PASP}}               
\def\pasj{\aaref@jnl{PASJ}}               
\def\qjras{\aaref@jnl{QJRAS}}             
\def\skytel{\aaref@jnl{S\&T}}             
\def\solphys{\aaref@jnl{Sol.~Phys.}}      
\def\sovast{\aaref@jnl{Soviet~Ast.}}      
\def\ssr{\aaref@jnl{Space~Sci.~Rev.}}     
\def\zap{\aaref@jnl{ZAp}}                 
\def\nat{\aaref@jnl{Nature}}              
\def\iaucirc{\aaref@jnl{IAU~Circ.}}       
\def\aplett{\aaref@jnl{Astrophys.~Lett.}} 
\def\apspr{\aaref@jnl{Astrophys.~Space~Phys.~Res.}}
\def\bain{\aaref@jnl{Bull.~Astron.~Inst.~Netherlands}} 
\def\fcp{\aaref@jnl{Fund.~Cosmic~Phys.}}  
\def\gca{\aaref@jnl{Geochim.~Cosmochim.~Acta}}   
\def\grl{\aaref@jnl{Geophys.~Res.~Lett.}} 
\def\jcp{\aaref@jnl{J.~Chem.~Phys.}}      
\def\jgr{\aaref@jnl{J.~Geophys.~Res.}}    
\def\jqsrt{\aaref@jnl{J.~Quant.~Spec.~Radiat.~Transf.}}
\def\memsai{\aaref@jnl{Mem.~Soc.~Astron.~Italiana}}
\def\nphysa{\aaref@jnl{Nucl.~Phys.~A}}   
\def\physrep{\aaref@jnl{Phys.~Rep.}}   
\def\physscr{\aaref@jnl{Phys.~Scr}}   
\def\planss{\aaref@jnl{Planet.~Space~Sci.}}   
\def\procspie{\aaref@jnl{Proc.~SPIE}}   

\begin{document}

\title[Resolved DEBRIS Discs]{Resolved Debris Discs Around A Stars in the Herschel DEBRIS Survey\thanks{Herschel is an ESA space observatory with science instruments provided by European-led Principal Investigator consortia and with important participation from NASA.}}
\author[M. Booth et al.]{Mark Booth$^{1,2}$\thanks{E-mail: markbooth@cantab.net}, Grant Kennedy$^{3}$, Bruce Sibthorpe$^4$, Brenda C. Matthews$^{2,1}$,\newauthor Mark C. Wyatt$^3$, Gaspard Duch{\^e}ne$^{5,6}$, J. J. Kavelaars$^{2,1}$, David Rodriguez$^7$,\newauthor Jane S. Greaves$^8$, Alice Koning$^{1,2}$, Laura Vican$^9$, George H. Rieke$^{10}$, \newauthor Kate Y. L. Su$^{10}$, Amaya Moro-Mart{\'i}n$^{11}$ and Paul Kalas$^{5,12}$ \\
$^{1}$ Dept. of Physics \& Astronomy, University of Victoria, Elliott Building, 3800 Finnerty Rd, Victoria, BC, V8P 5C2 Canada \\
$^{2}$ National Research Council of Canada, 5071 West Saanich Road, Victoria,
BC, V9E 2E7 Canada\\
$^{3}$ Institute of Astronomy, Madingley Rd, Cambridge CB3 0HA, UK \\
$^{4}$ UK Astronomy Technology Center, Royal Observatory, Blackford
              Hill, Edinburgh EH9 3HJ, UK\\
$^{5}$ Astronomy Department, UC Berkeley, Hearst Field Annex B-20, Berkeley, CA 94720-3411, USA\\
$^{6}$ UJF-Grenoble 1 / CNRS-INSU, Institut de Plan\'etologie et d'Astrophysique de Grenoble (IPAG) UMR 5274, Grenoble, F-38041,\\ France\\
$^{7}$ Universidad de Chile,
Camino el Observatorio 1515,
Las Condes, Santiago,
Chile\\
$^{8}$ School of Physics and Astronomy, University of St Andrews,
  North Haugh, St Andrews, Fife KY16 9SS, UK \\
$^{9}$ Dept. of Physics \& Astronomy, University of California, Los
Angeles, 475 Portola Plaza, Los Angeles, CA 90095-1547, USA \\
$^{10}$ Steward Observatory, University of Arizona 933 N Cherry Avenue Tucson, AZ 85721 \\
$^{11}$ Centro de Astrobiolog{\'i}a (CSIC-INTA), 28850 Torrej{\'o}n de Ardoz,
Madrid, Spain \\
$^{12}$ SETI Institute, 515 North Whisman Rd., Mountain View, CA 94043, USA 
}

\date{Accepted 2012 September 28.  Received 2012 September 28; in original form 2012 June 18}

\maketitle

\begin{abstract}
The majority of debris discs discovered so far have only been detected through infrared excess emission above stellar photospheres. While disc properties can be inferred from unresolved photometry alone
under various assumptions for the physical properties of dust grains,
there is a degeneracy between disc radius and dust temperature that
depends on the grain size distribution and optical properties. By resolving the disc we can measure the actual location of the dust. The launch of Herschel, with an angular resolution superior to previous
far-infrared telescopes, allows us to spatially resolve more discs and
locate the dust directly. Here we present the nine resolved discs around A stars between 20 and 40~pc observed by the DEBRIS survey. We use these data to investigate the disc radii by fitting narrow
ring models to images at 70, 100 and 160~$\mu$m and by fitting
blackbodies to full spectral energy distributions. We do this with the aim of finding an improved way of estimating disc radii for unresolved systems. The ratio between the resolved and blackbody radii varies between 1 and 2.5. This ratio is inversely correlated with luminosity and any remaining discrepancies are most likely explained by differences to the minimum size of grain in the size distribution or differences in composition. We find that three of the systems are well fit by a narrow ring, two systems are borderline cases and the other four likely require wider or multiple rings to fully explain the observations, reflecting the diversity of planetary systems.
\end{abstract}


\begin{table*}
\begin{minipage}{166mm}
	\caption{Star information. The first column gives the Unbiased Nearby Stars (UNS) identifier from the DEBRIS survey target list \citep{phillips10}. The first letter gives the spectral type of the primary and the number gives the rank order of distance from the Sun.}
\begin{tabular}{llllllll}
	\hline
  UNS ID & HD & Name & Spectral Type & Distance (pc) & Age (Gyr) & RA & DEC \\
	\hline
  A018 & 139006 & $\alpha$ CrB & A1IV & 23.0 & 0.27$\pm$0.1 & 15$^{\rm{h}}$34$^{\rm{m}}$41$\fs$268 & +26$^\circ$42\arcmin52\farcs89  \\
  A024 & 95418 & $\beta$ Uma & A1IVps (Sr II) & 24.5 & 0.31$\pm$0.1 & 11$^{\rm{h}}$01$^{\rm{m}}$50$\fs$477 & +56$^\circ$22\arcmin56\farcs73 \\
  A053 & 125162 & $\lambda$ Boo & A3Va & 30.4 & 0.29$\pm$0.1 & 14$^{\rm{h}}$16$^{\rm{m}}$23$\fs$019 & +46$^\circ$05\arcmin17\farcs90 \\
  A061 & 188228 & $\epsilon$ Pav & A0Va & 32.2 & 0.25$\pm$0.1 & 20$^{\rm{h}}$00$^{\rm{m}}$35$\fs$556 & $-$72$^\circ$54\arcmin37\farcs82 \\
  A064 & 20320 & $\zeta$ Eri & kA4hA9mA9V & 33.6 & 0.80$\pm$0.1 & 03$^{\rm{h}}$15$^{\rm{m}}$50$\fs$027 & $-$08$^\circ$49\arcmin11\farcs02 \\
  A067 & 14055 & $\gamma$ Tri & A1Vnn & 34.4 & 0.16$\pm$0.3 & 02$^{\rm{h}}$17$^{\rm{m}}$18$\fs$867 & +33$^\circ$50\arcmin49\farcs90 \\
  A076 & 110411 & $\rho$ Vir & A3Va & 36.3 & 0.50$\pm$0.1 & 12$^{\rm{h}}$41$^{\rm{m}}$53$\fs$057 & +10$^\circ$14\arcmin08\farcs25 \\
  A082 & 71155 & 30 Mon & A0Va & 37.5 & 0.17$\pm$0.3 & 08$^{\rm{h}}$25$^{\rm{m}}$39$\fs$632 & $-$03$^\circ$54\arcmin23\farcs12 \\
  A086 & 13161 & $\beta$ Tri & A5IV & 38.9 & 0.73$\pm$0.3 & 02$^{\rm{h}}$09$^{\rm{m}}$32$\fs$627 & +34$^\circ$59\arcmin14\farcs27 \\
	\hline
\end{tabular}
	\medskip \\
	Spectral types and distances are from \protect\citet{phillips10}. Ages are from \protect\citet{vican12}. Right ascensions and declinations  are the J2000 epoch \citep{leeuwen07}.
	\label{tsparam}
\end{minipage}
\end{table*}

\section{Introduction}
Debris discs have now been discovered around hundreds of stars. In most cases we only know of their existence due to observations in the infrared showing a greater flux than is expected for the star's photosphere alone, which is attributed to dust grains produced by collisions
of planetesimals. Knowing the radial structure of the parent planetesimals is key to probing
many aspects of debris discs and, by extension, the planetary system as a whole.
When no spatial is information available, fitting a blackbody to the spectral
energy distribution (SED) provides an estimate of the dust temperature that
can be converted into a distance from the star. However, a blackbody is a
poor approximation for small dust grains, which are inefficient emitters
at long wavelengths and are generally hotter than blackbodies \citep[e.g.][]{backman93}.

So far, only a small fraction of these disc candidates have
been spatially resolved.  For a limited number of systems, scattered light
from the dusty debris can be seen when the light from the star is carefully
removed and/or blocked by a coronagraph,
though the interpretation of these data are complicated by the fact that
the distribution of the submicron scattering grains is strongly affected
by radiation forces. At longer infrared wavelengths, thermal emission from
the dusty debris can be detected and contrast with the star becomes less
problematic. Moreover, far-infrared observations probe grains that are above the size at which grains are blown out of the system by radiation pressure and are thus expected to trace the
locations of the parent planetesimals. 

Together with the
SED, resolved images allows us to directly measure the radial location of
emitting dust, and provide constraints on the grain size and composition. Studies based on a combination of observations at different
wavelengths and using different methods have suggested that resolved debris disc radii
are 1-5 times larger than inferred using blackbody fits (e.g.  Rodriguez and
Zuckerman 2012).
In addition, resolved images can show us radial and vertical asymmetries \citep[e.g.][]{kalas95}, azimuthal structure \citep[e.g.][]{greaves05} and inner edges \citep[e.g.][]{boley12}, which in turn can be used to infer the existence of planets in the system, as was successfully done with $\beta$ Pic \citep{mouillet97,lagrange10} and Fomalhaut \citep{kalas05,quillen06}.

Space based observatories have been at the forefront of debris disc observations ever since the first, Vega, was discovered with the Infrared Astronomy Satellite \citep{aumann84}. The Herschel Space Observatory \citep[hereafter Herschel]{pilbratt10} is the latest infrared satellite and has two key programmes dedicated to searching for debris discs: Disc Emission via a Bias-free Reconnaissance in the Infrared/Submillimetre \citep[DEBRIS][]{matthews10} and DUst around NEarby Stars \citep[DUNES][]{eiroa10}. Herschel provides higher
sensitivity and resolution than previously available at far-IR wavelengths, and therefore can
spatially resolve many debris discs for the first time.

In this paper we present images (section \ref{sobs}) of nine debris discs around A stars from the DEBRIS survey ($\alpha$ CrB, $\beta$ Uma, $\lambda$ Boo, $\epsilon$ Pav, $\zeta$ Eri, $\gamma$ Tri, $\rho$ Vir, 30 Mon, $\beta$ Tri), the cold components of which have never been previously resolved. These 9 stars constitute the full sample of resolved A star discs between 20 and 40~pc. In section \ref{ssedfit} we present the spectral energy distributions of these systems. In section \ref{smod} we model the resolved images as narrow rings. In section \ref{sddr} we use our results from a consistent fitting procedure to compare the blackbody radii and resolved radii. In section \ref{sindiv} we summarise the results for each system individually. In section \ref{sext} we discuss differences between systems that are well described
by a narrow ring and those that are radially broad. The conclusions are presented in section \ref{sconc}.

\begin{figure*}

  \begin{center}

    \vspace{-0.2in}

    \begin{tabular}{ccc}

			\vspace{-0.2in}
      \hspace{-0.5in} \includegraphics[width=0.31\textwidth]{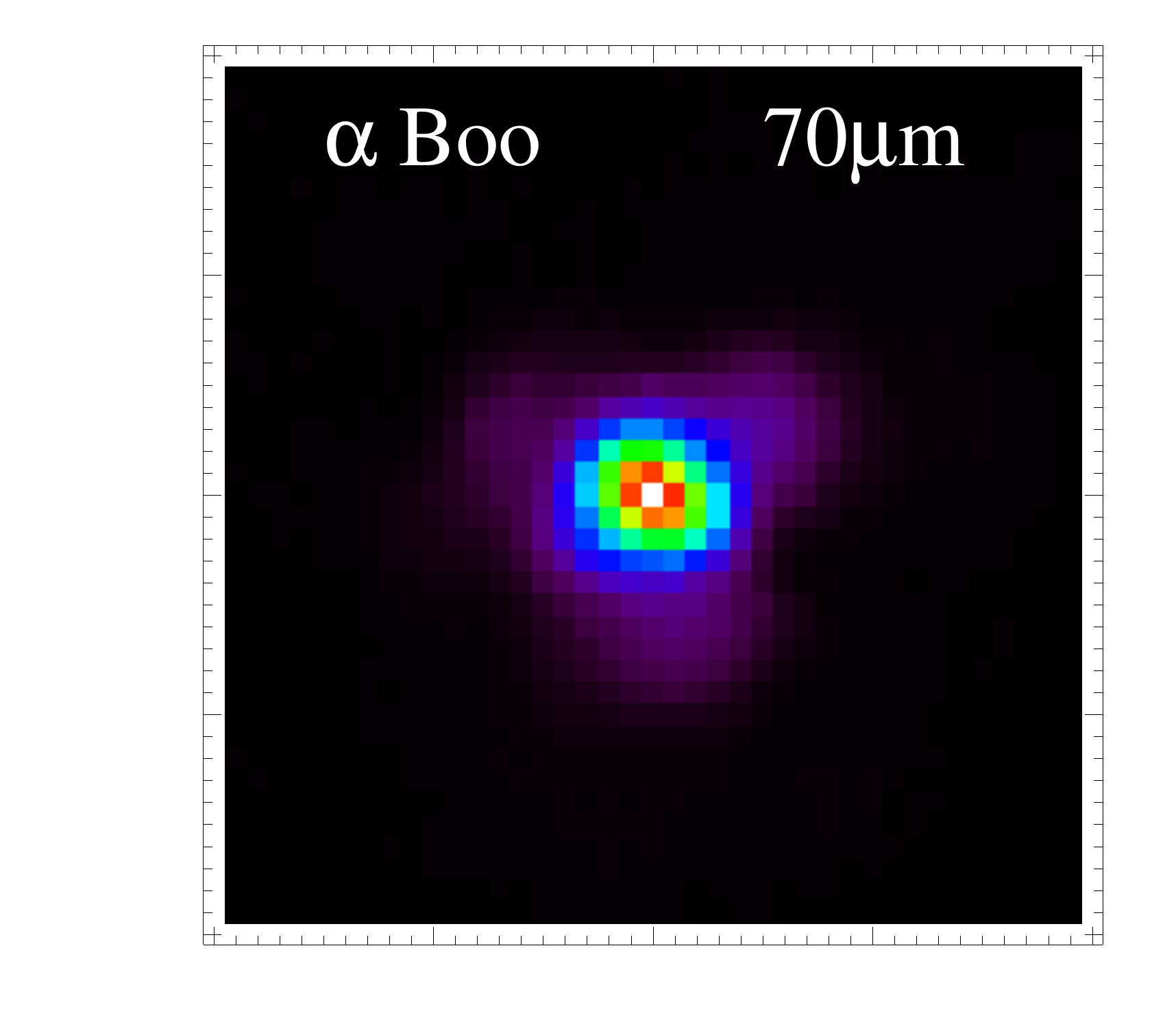} &

      \hspace{-0.5in} \includegraphics[width=0.31\textwidth]{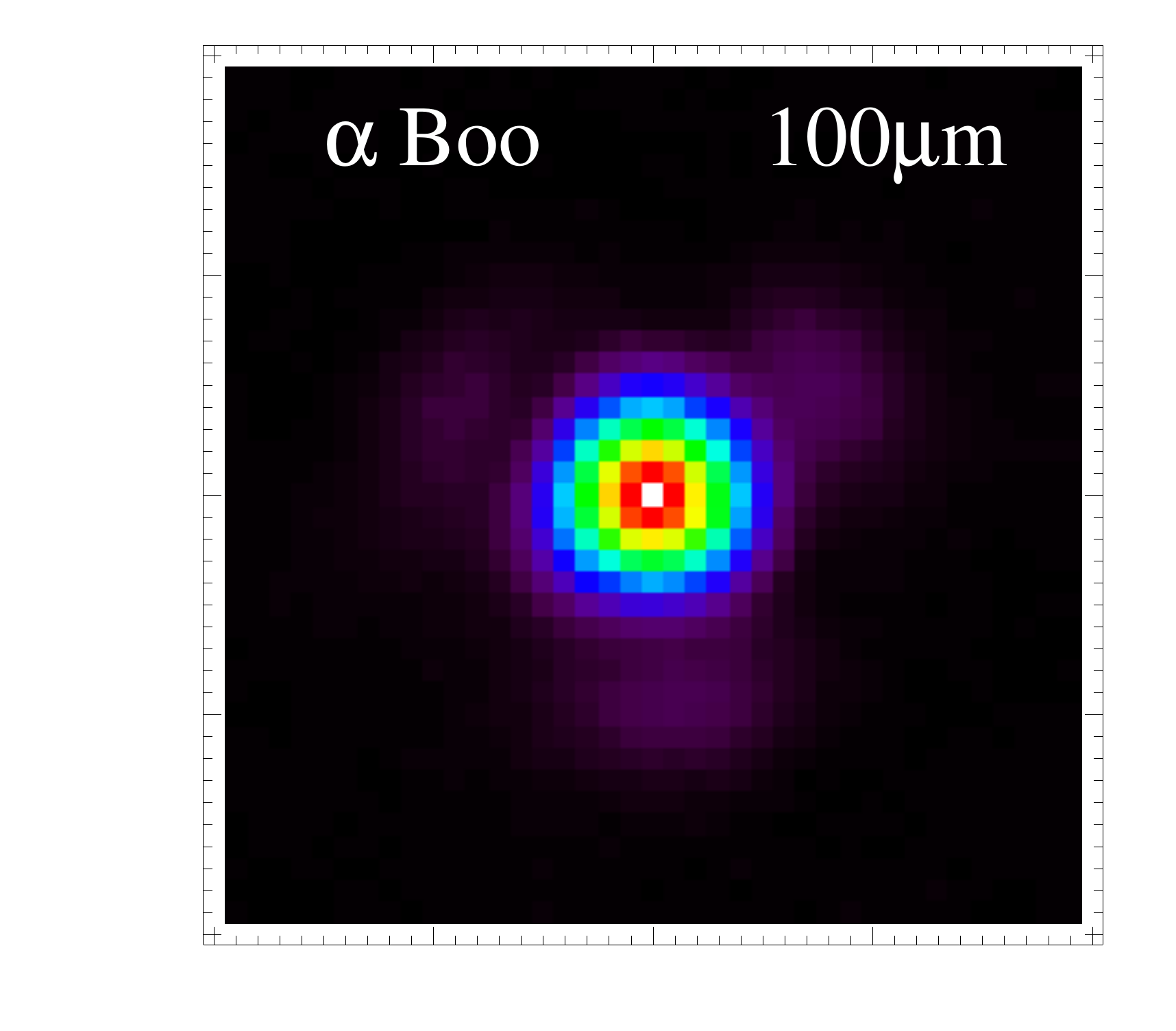} &

      \hspace{-0.5in} \includegraphics[width=0.31\textwidth]{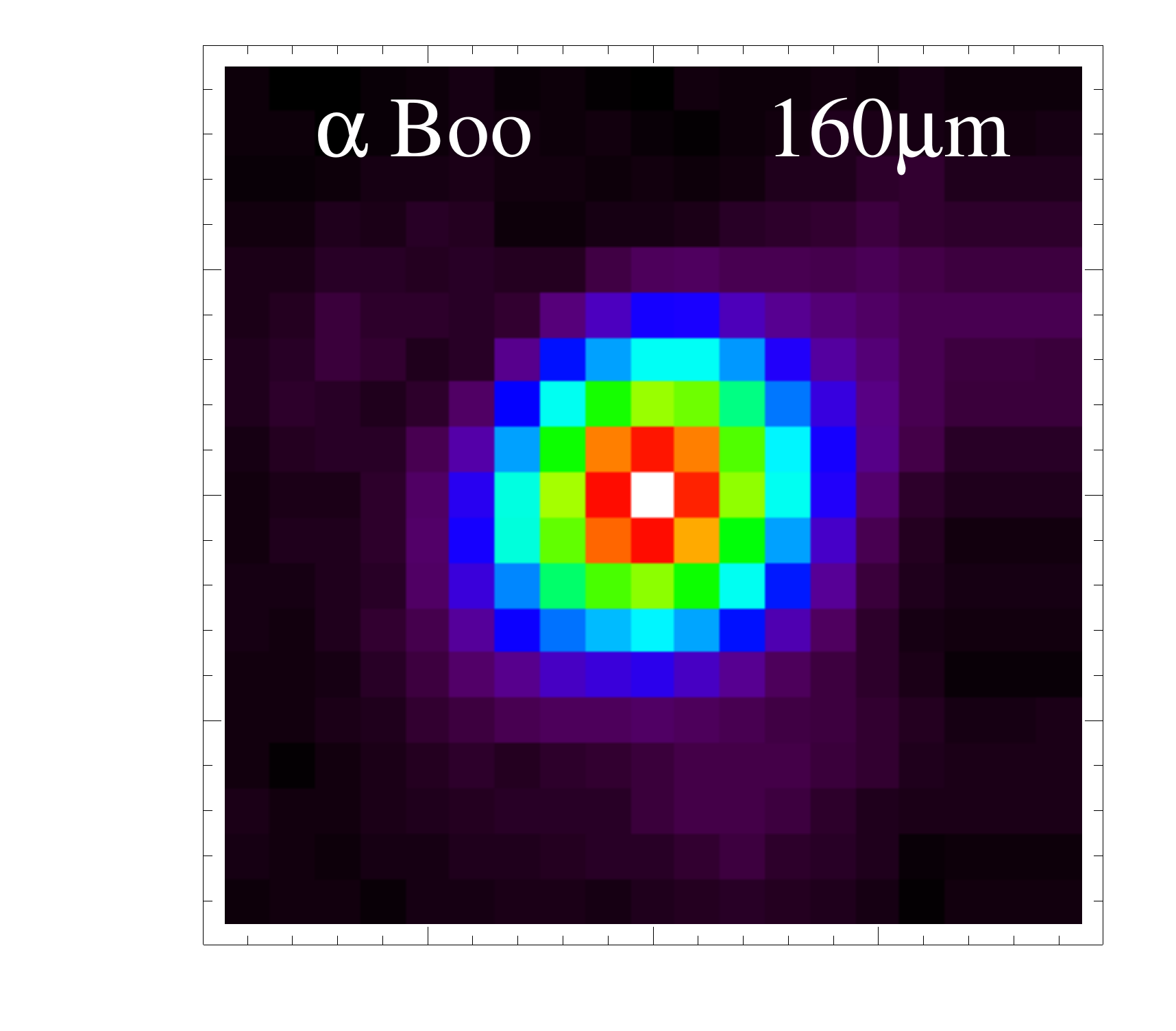} \\
			\vspace{-0.2in}
      \hspace{-0.5in} \includegraphics[width=0.31\textwidth]{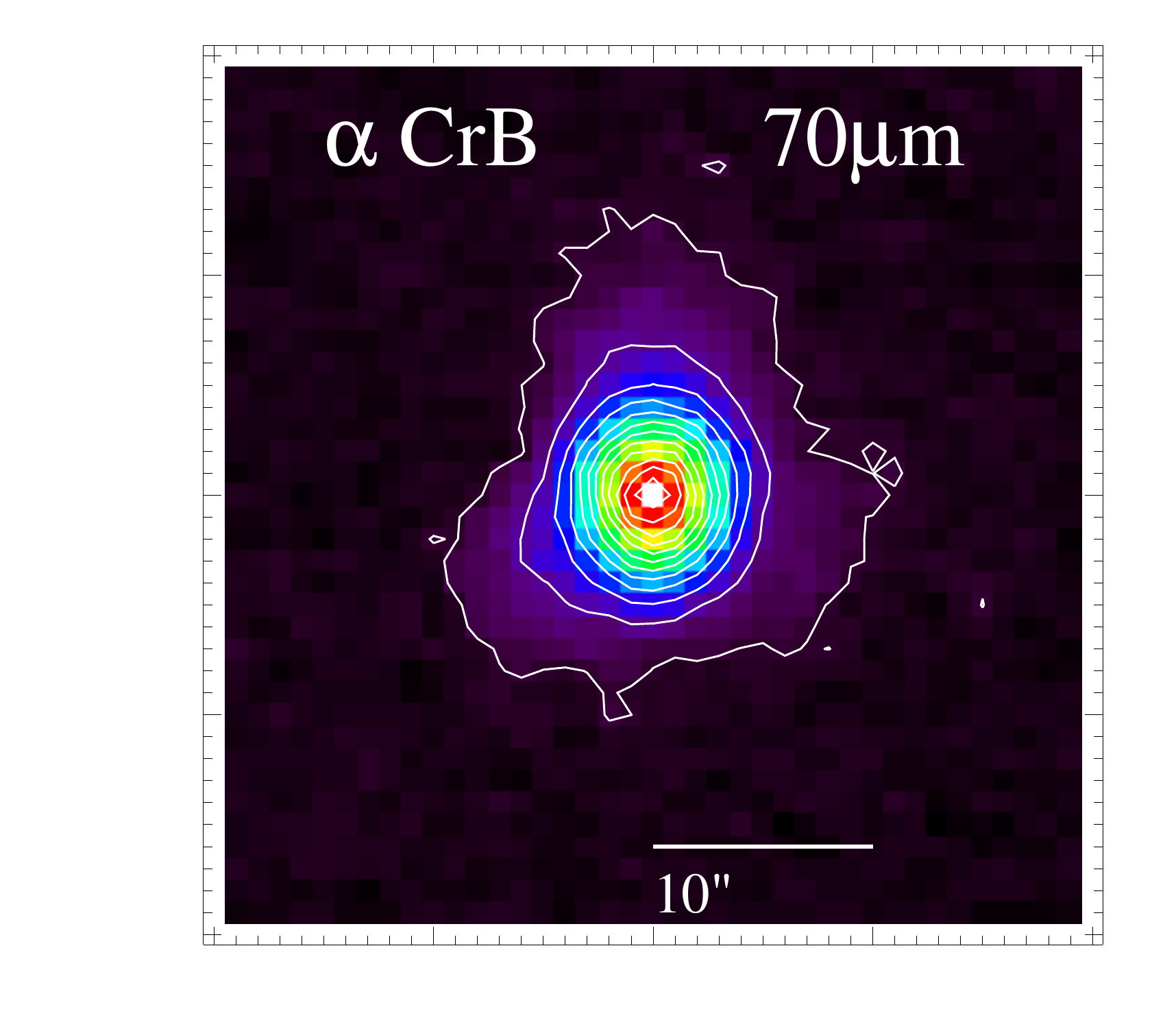} &
      \hspace{-0.5in} \includegraphics[width=0.31\textwidth]{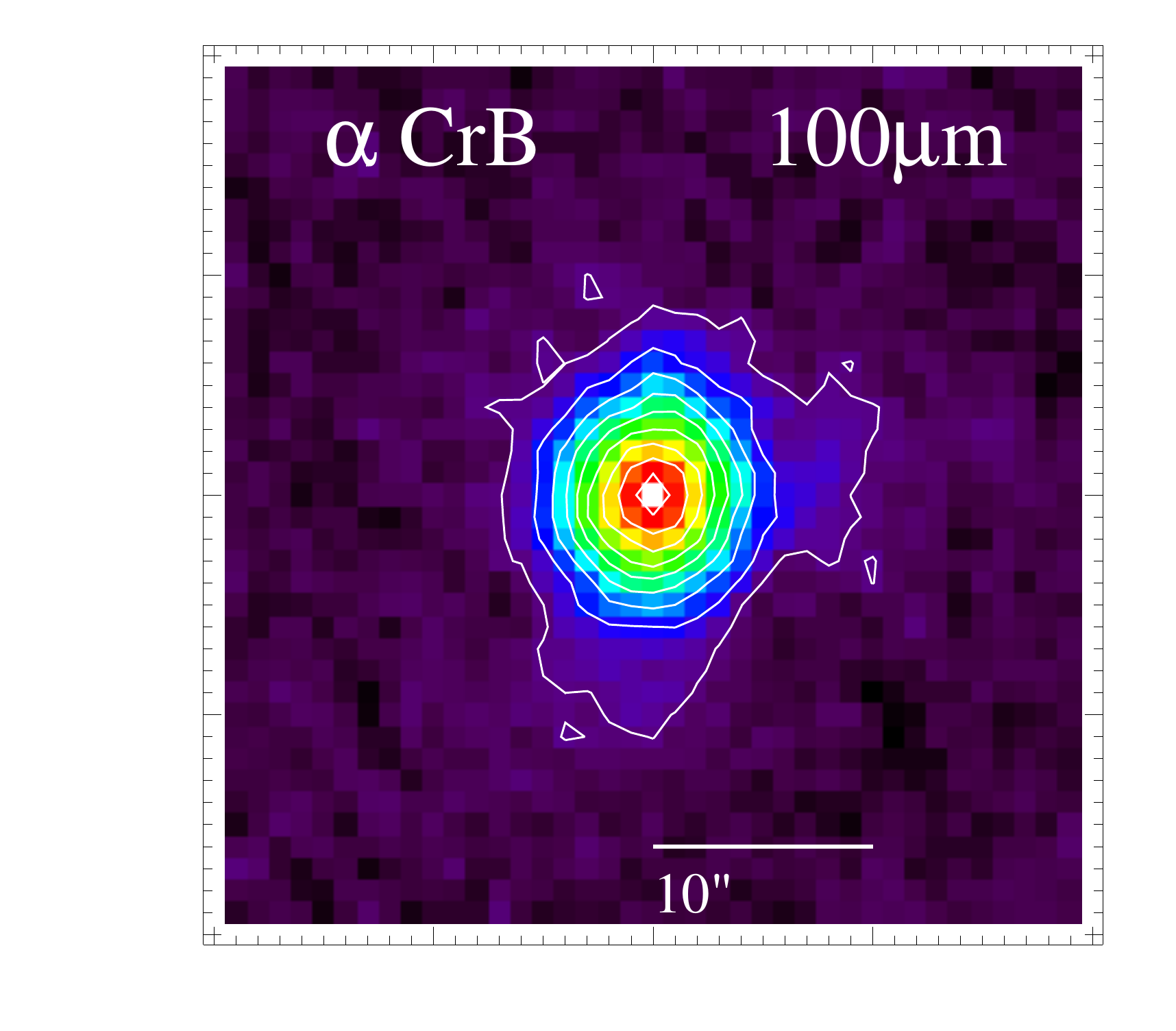} &
      \hspace{-0.5in} \includegraphics[width=0.31\textwidth]{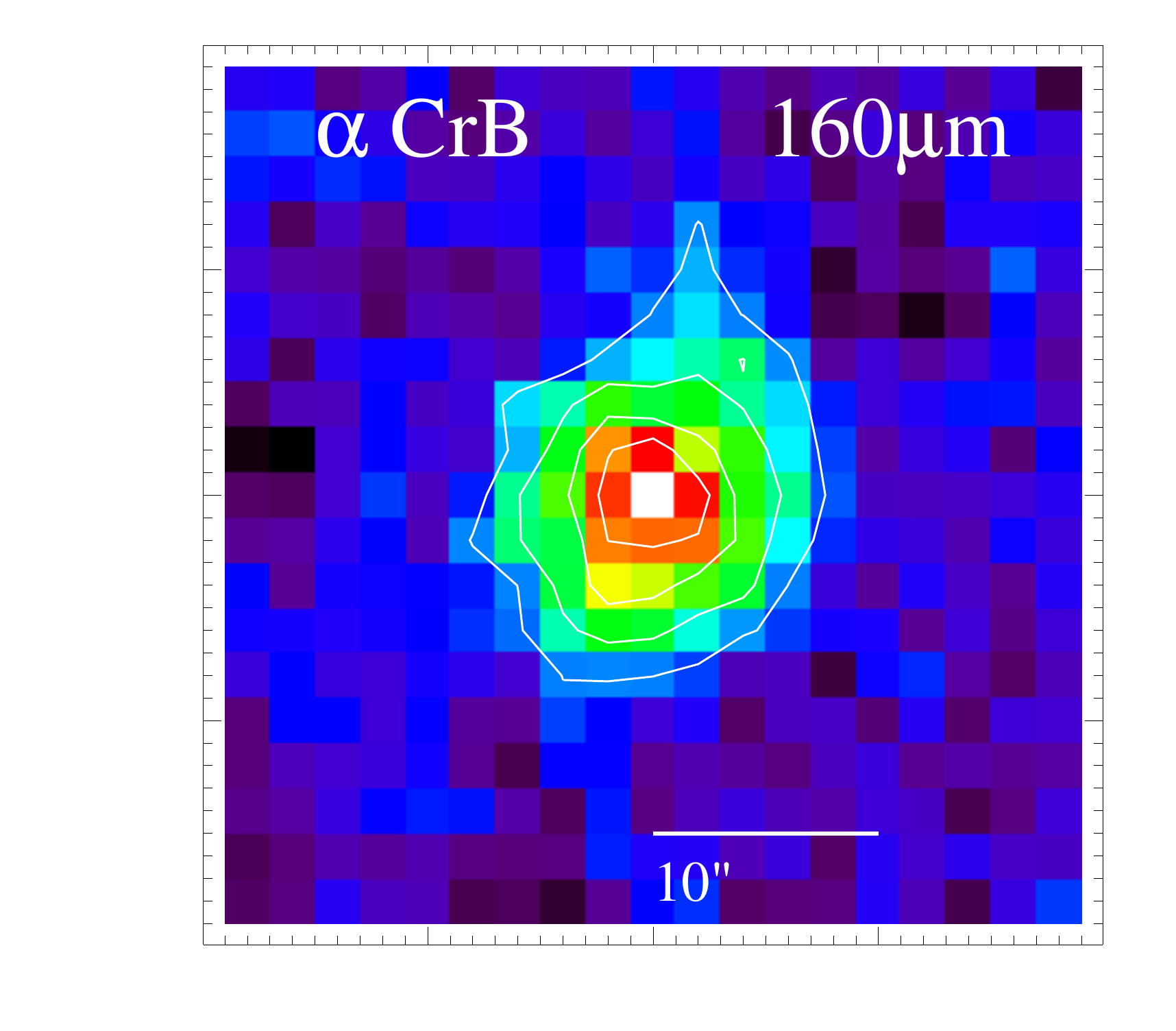} \\
			\vspace{-0.2in}

      \hspace{-0.5in} \includegraphics[width=0.31\textwidth]{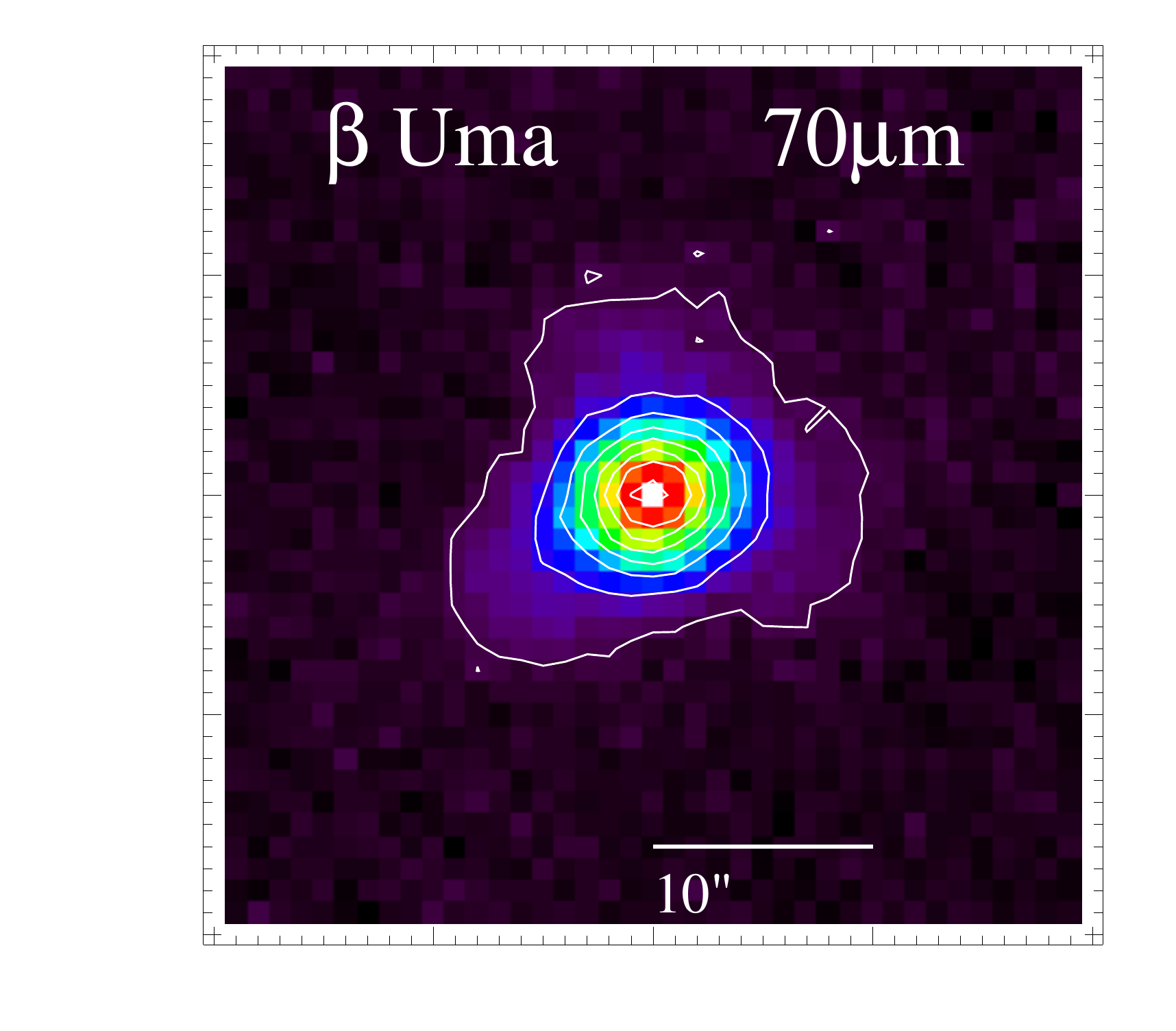} &
      \hspace{-0.5in} \includegraphics[width=0.31\textwidth]{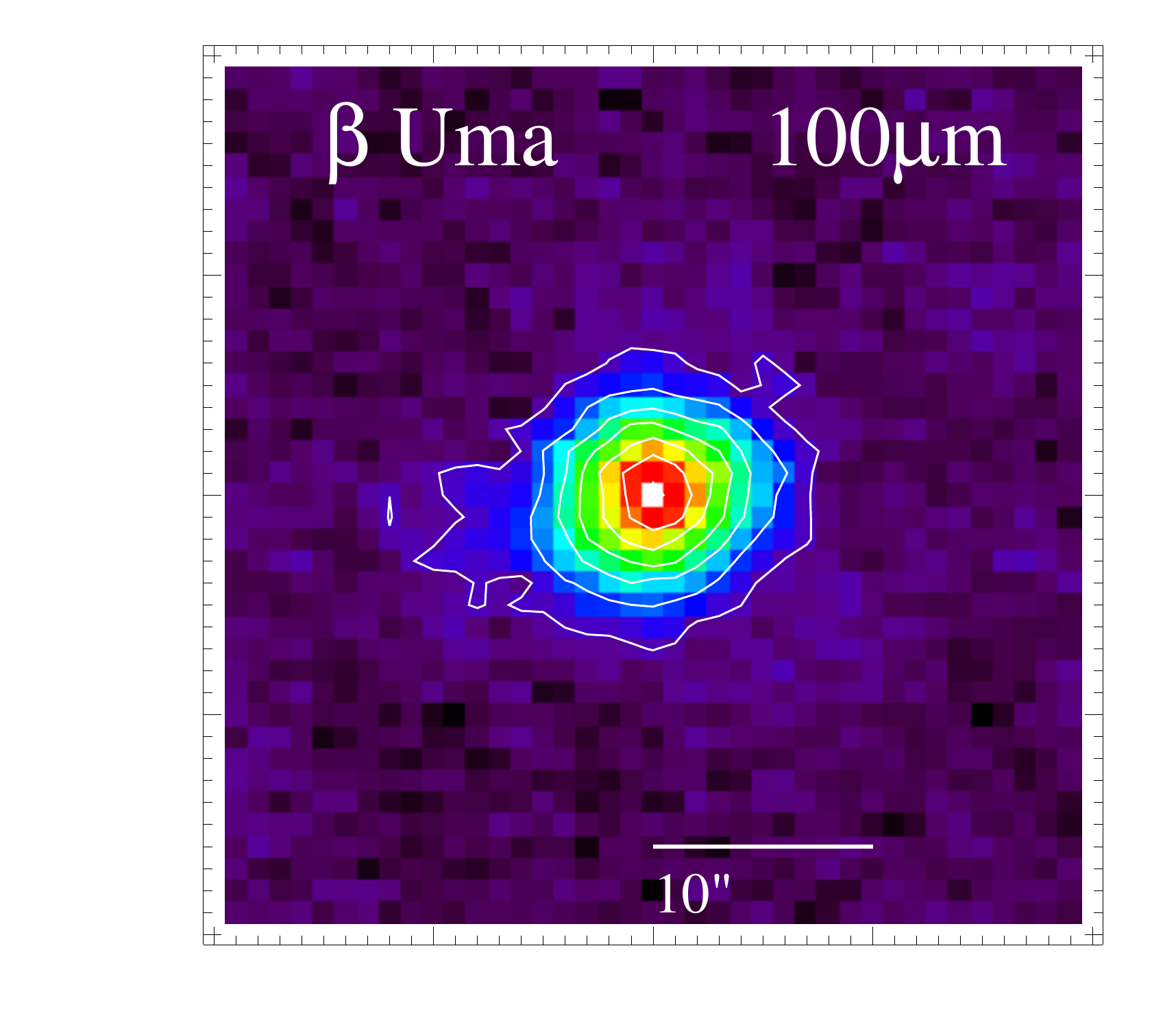} &
      \hspace{-0.5in} \includegraphics[width=0.31\textwidth]{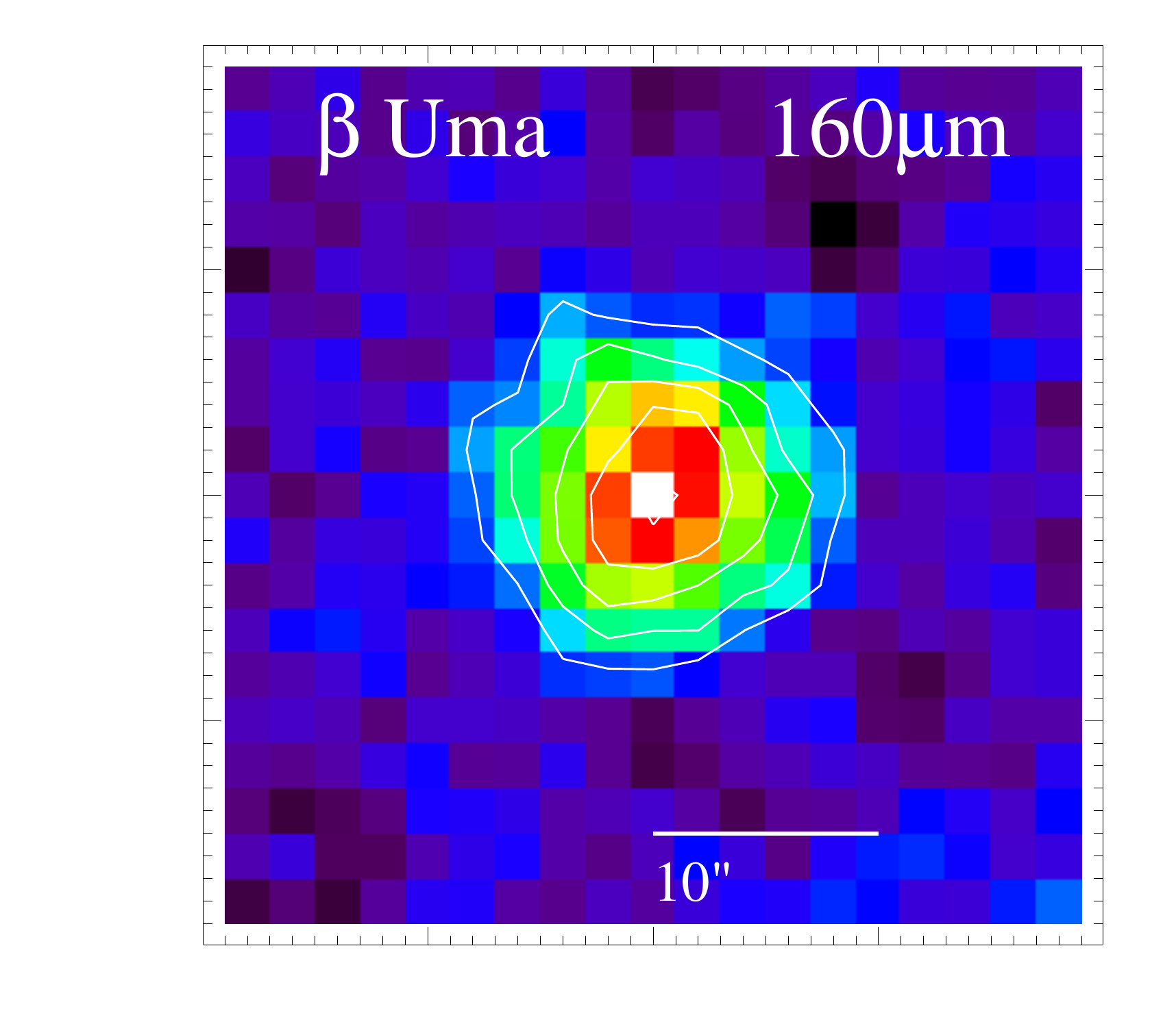} \\
			\vspace{-0.2in}
      
      \hspace{-0.5in} \includegraphics[width=0.31\textwidth]{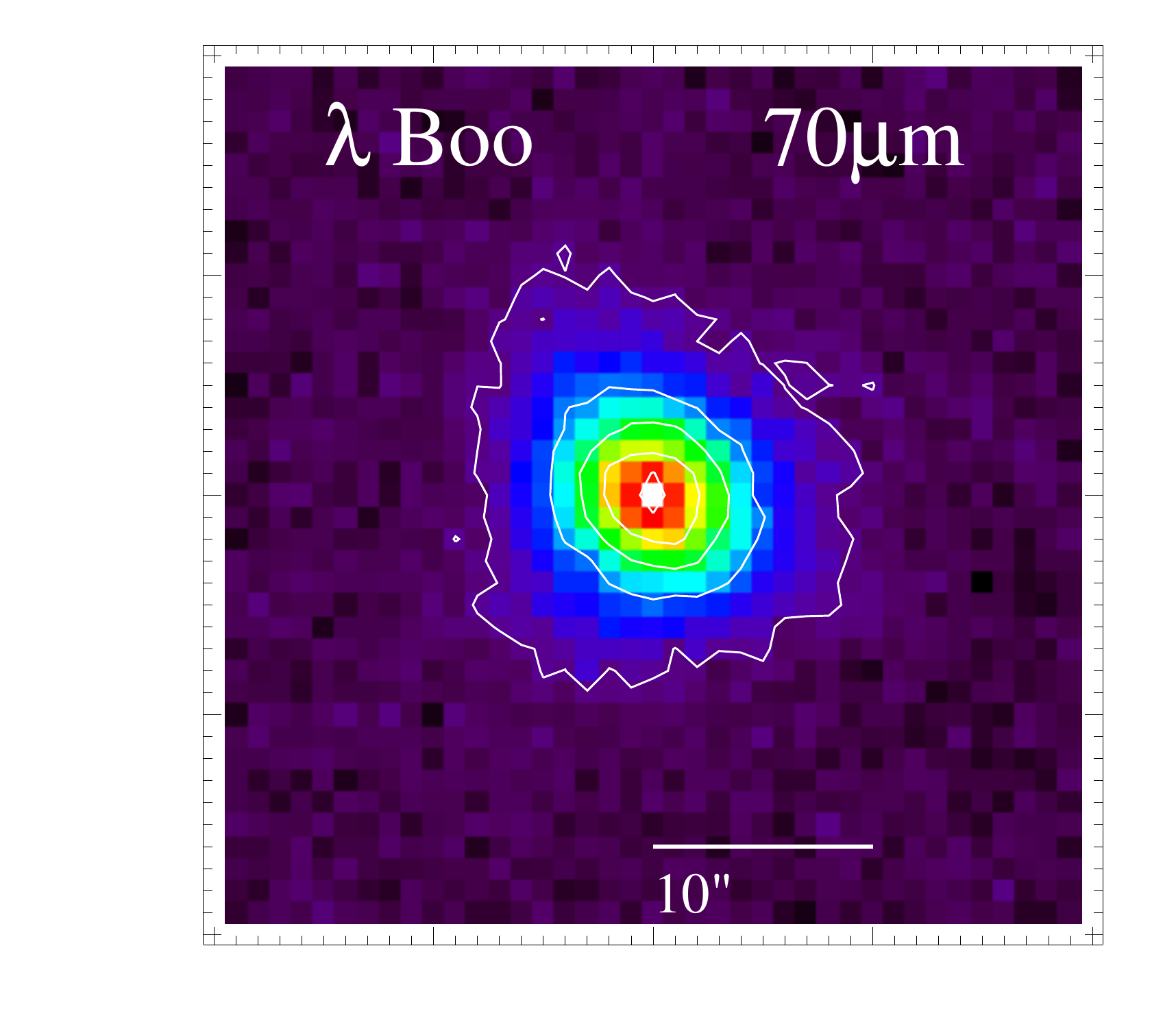} &
      \hspace{-0.5in} \includegraphics[width=0.31\textwidth]{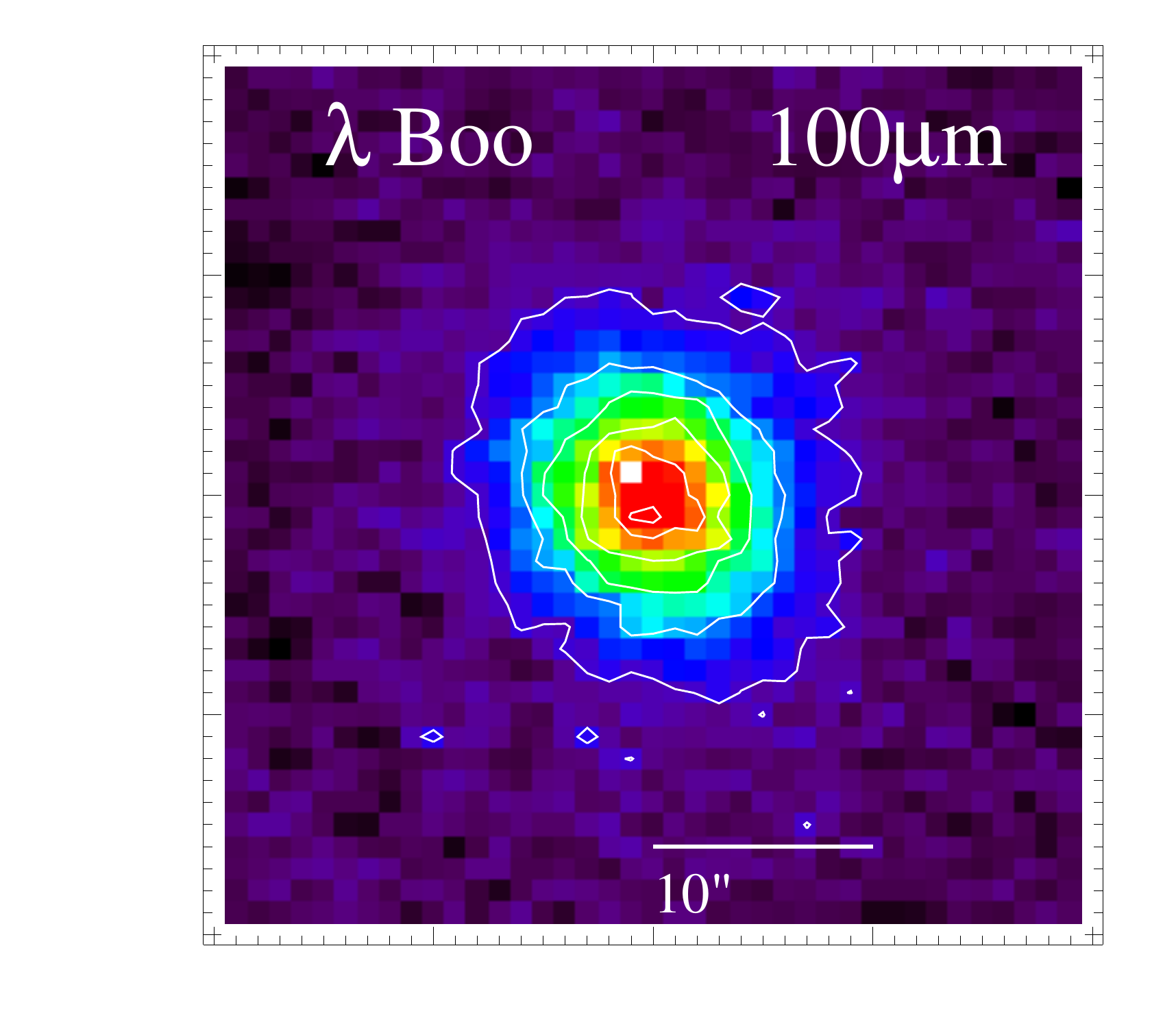} &
      \hspace{-0.5in} \includegraphics[width=0.31\textwidth]{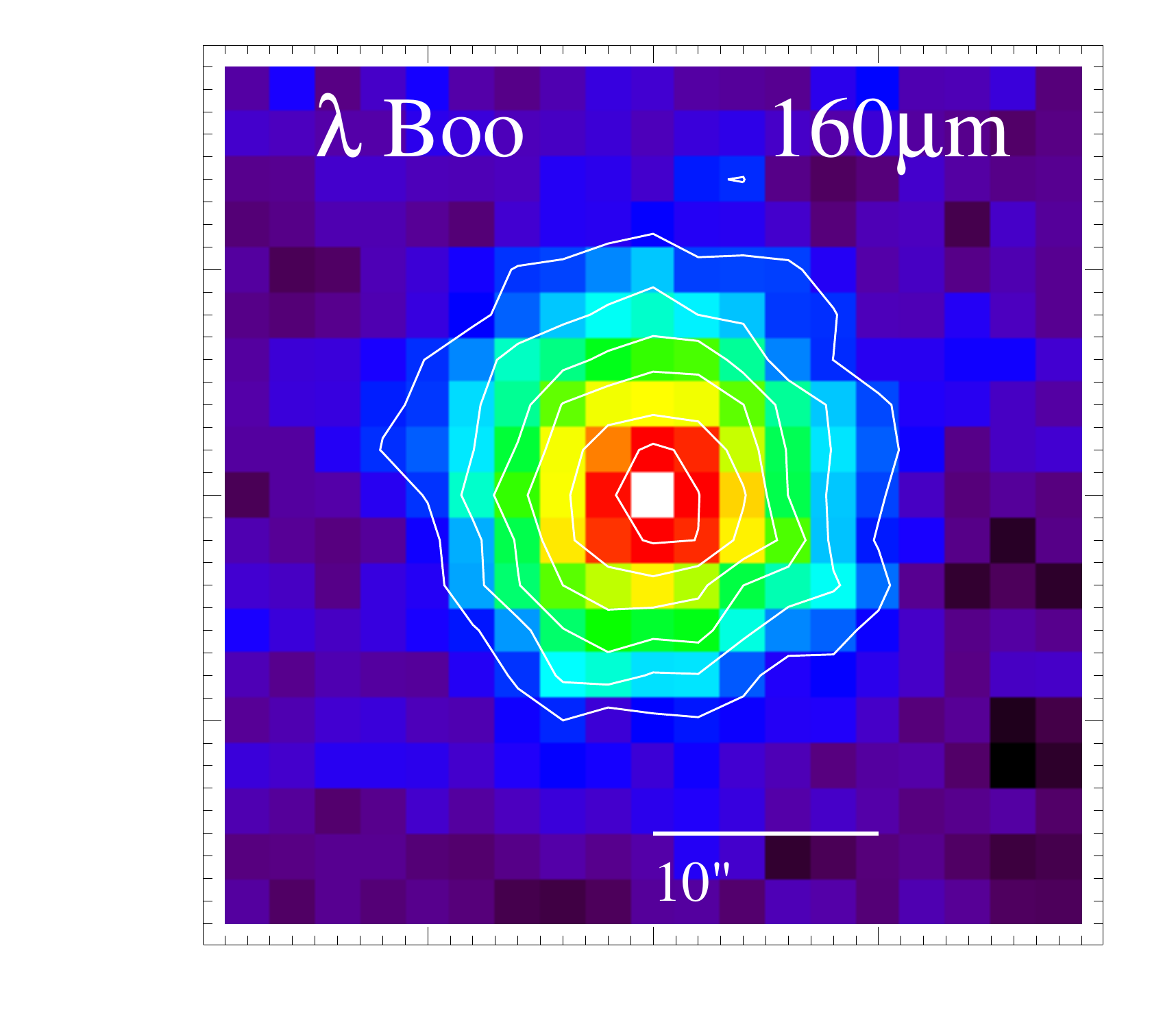} \\
			\vspace{-0.2in}
      
       \hspace{-0.5in} \includegraphics[width=0.31\textwidth]{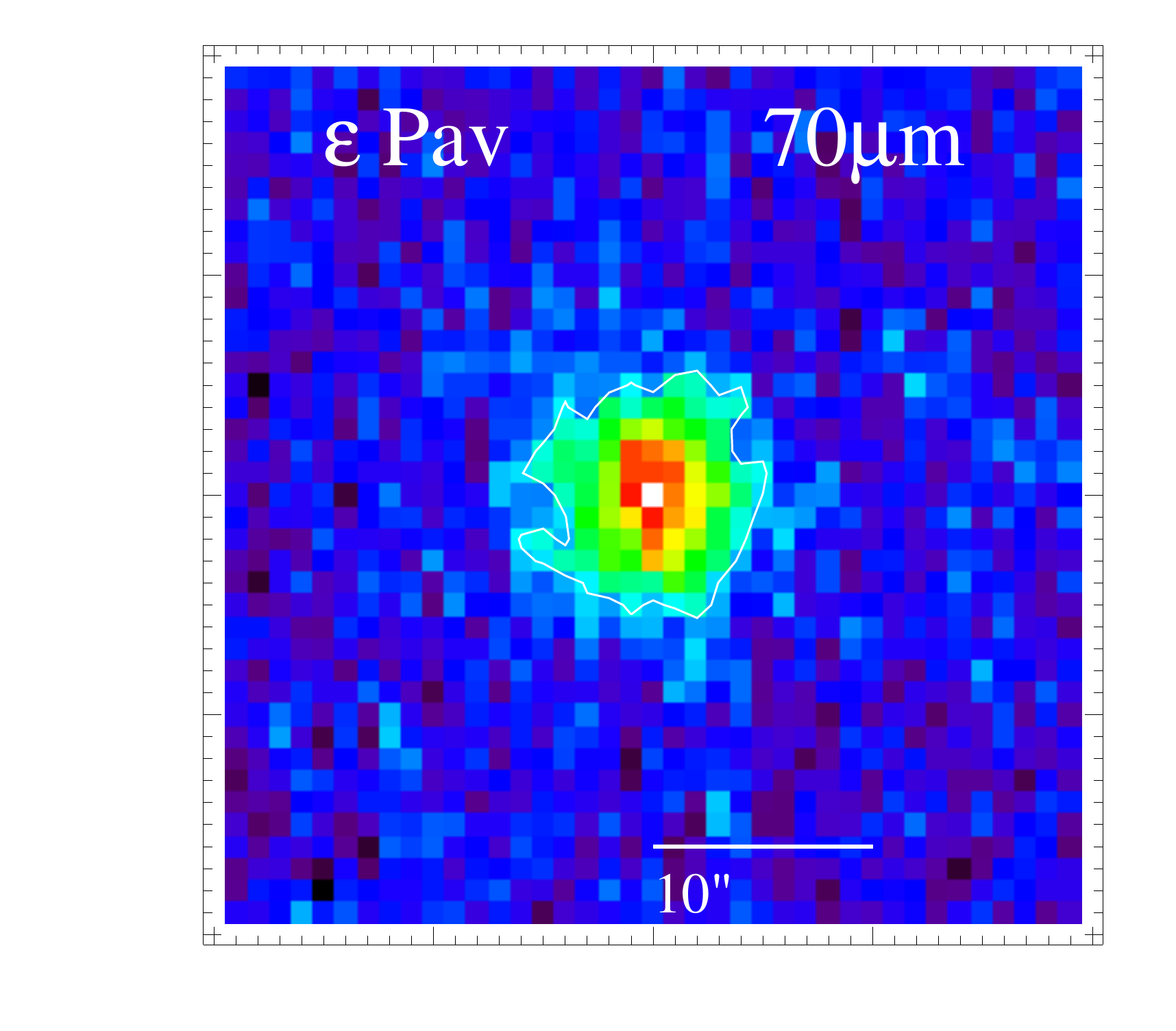} &
      \hspace{-0.5in} \includegraphics[width=0.31\textwidth]{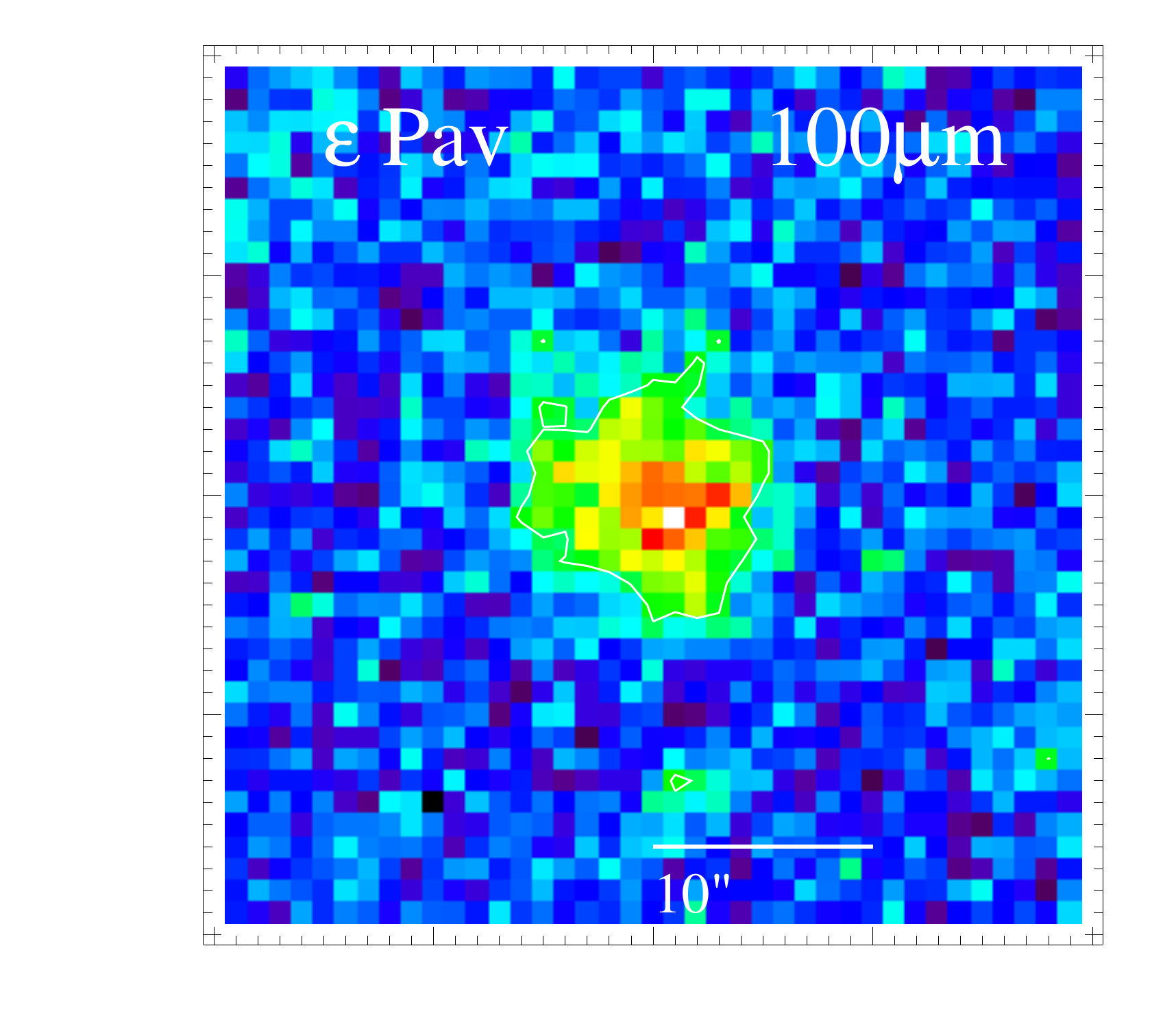} &
      \hspace{-0.5in} \includegraphics[width=0.31\textwidth]{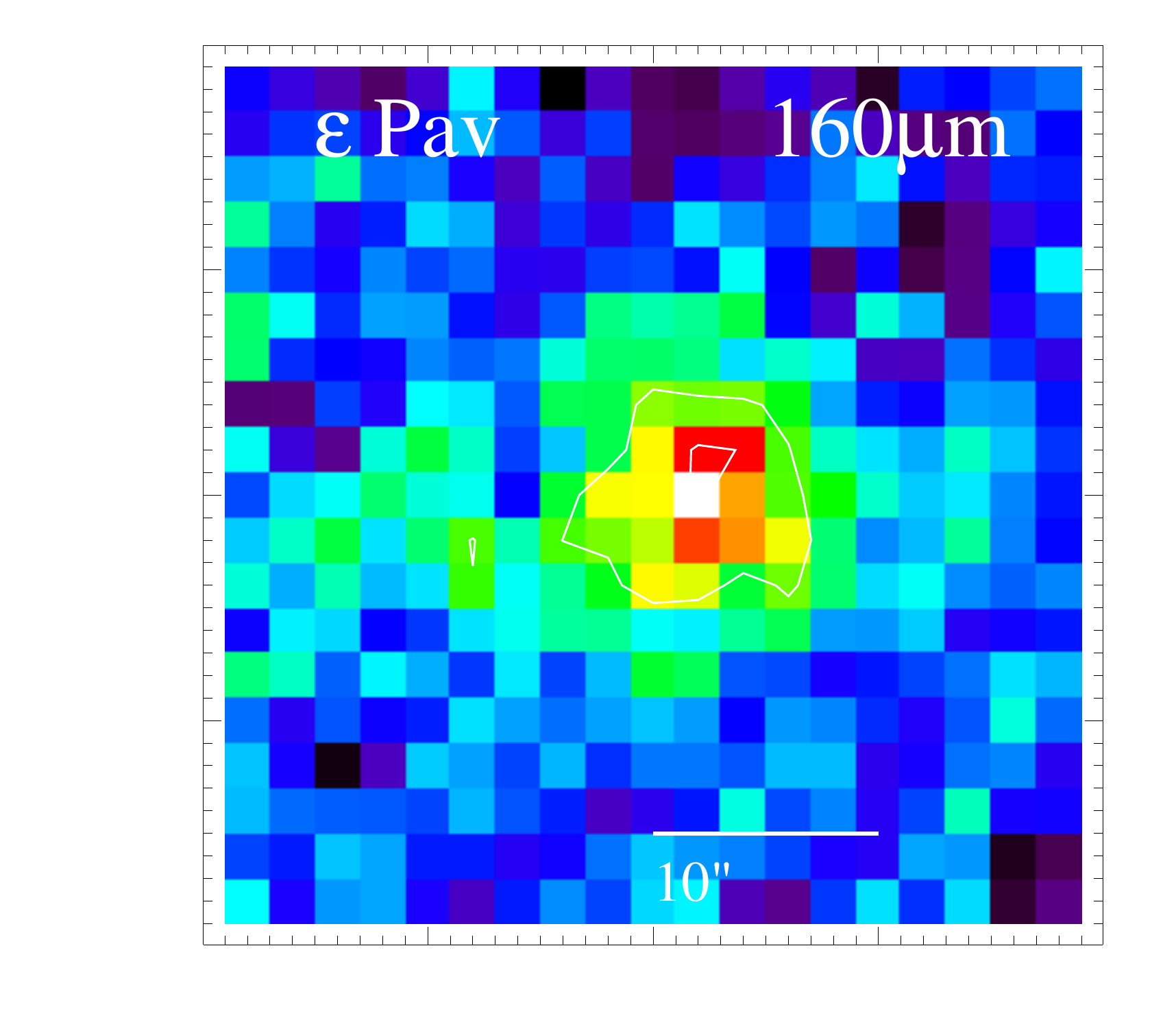} \\
      
    \end{tabular}

    \caption{PACS observations for the stars presented in this paper. 70$\mu$m images on the left, 100$\mu$m images in the middle and 160$\mu$m on the right. The top row shows the PSF star $\alpha$ Boo. The field of view of each image is 39{\arcsec}x39{\arcsec}. N is up and E is left. The lowest contour represents 3-$\sigma$ significance and the contours rise in increments of 12-$\sigma$ for 70$\mu$m, 6-$\sigma$ for 100$\mu$m and 3-$\sigma$ for 160$\mu$m.} 

   \label{fobs100}

  \end{center}

\end{figure*}

\begin{figure*}
  \begin{center}

    \begin{tabular}{ccc}
			\vspace{-0.2in}
      
      \hspace{-0.5in} \includegraphics[width=0.31\textwidth]{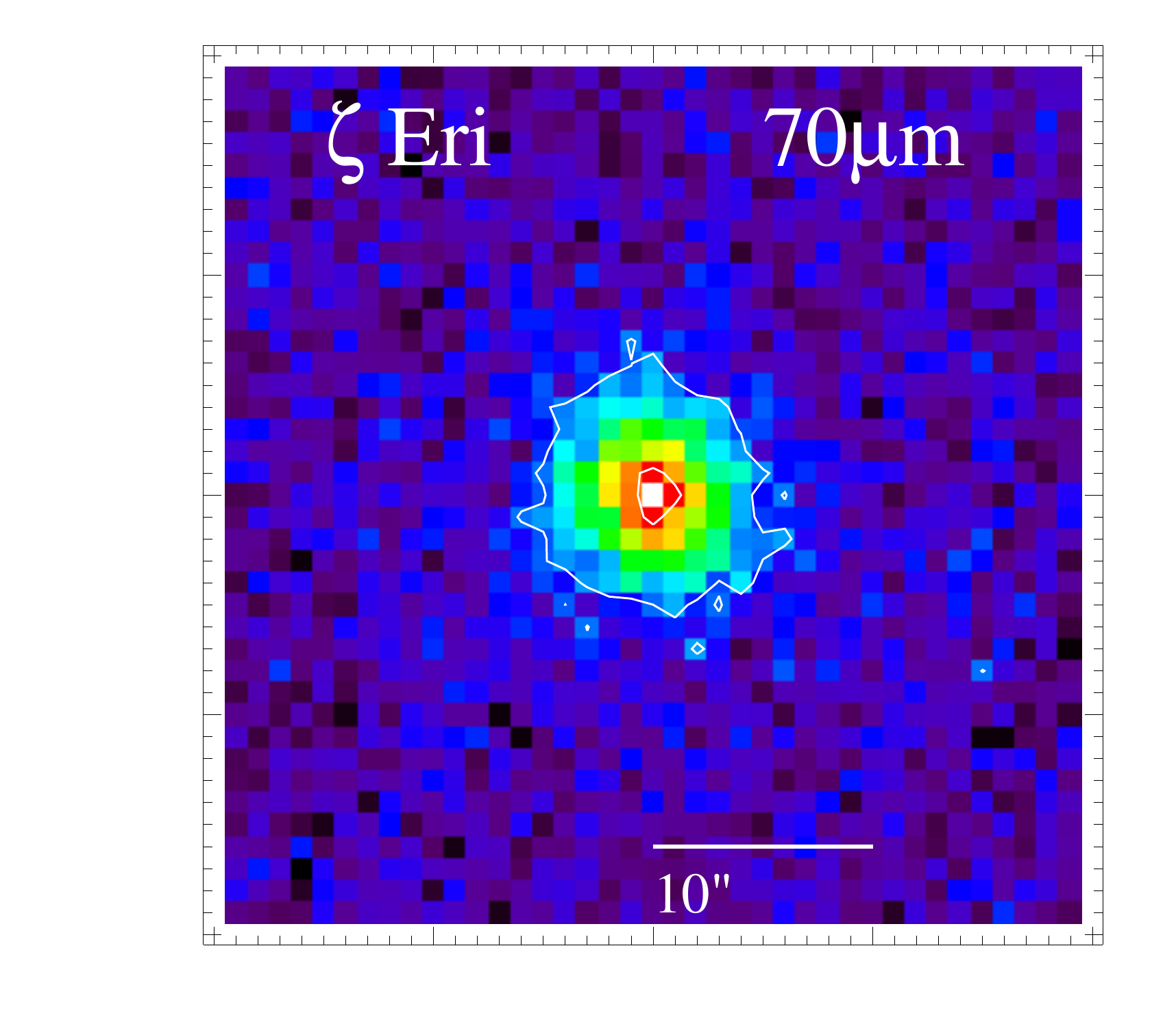} &
      \hspace{-0.5in} \includegraphics[width=0.31\textwidth]{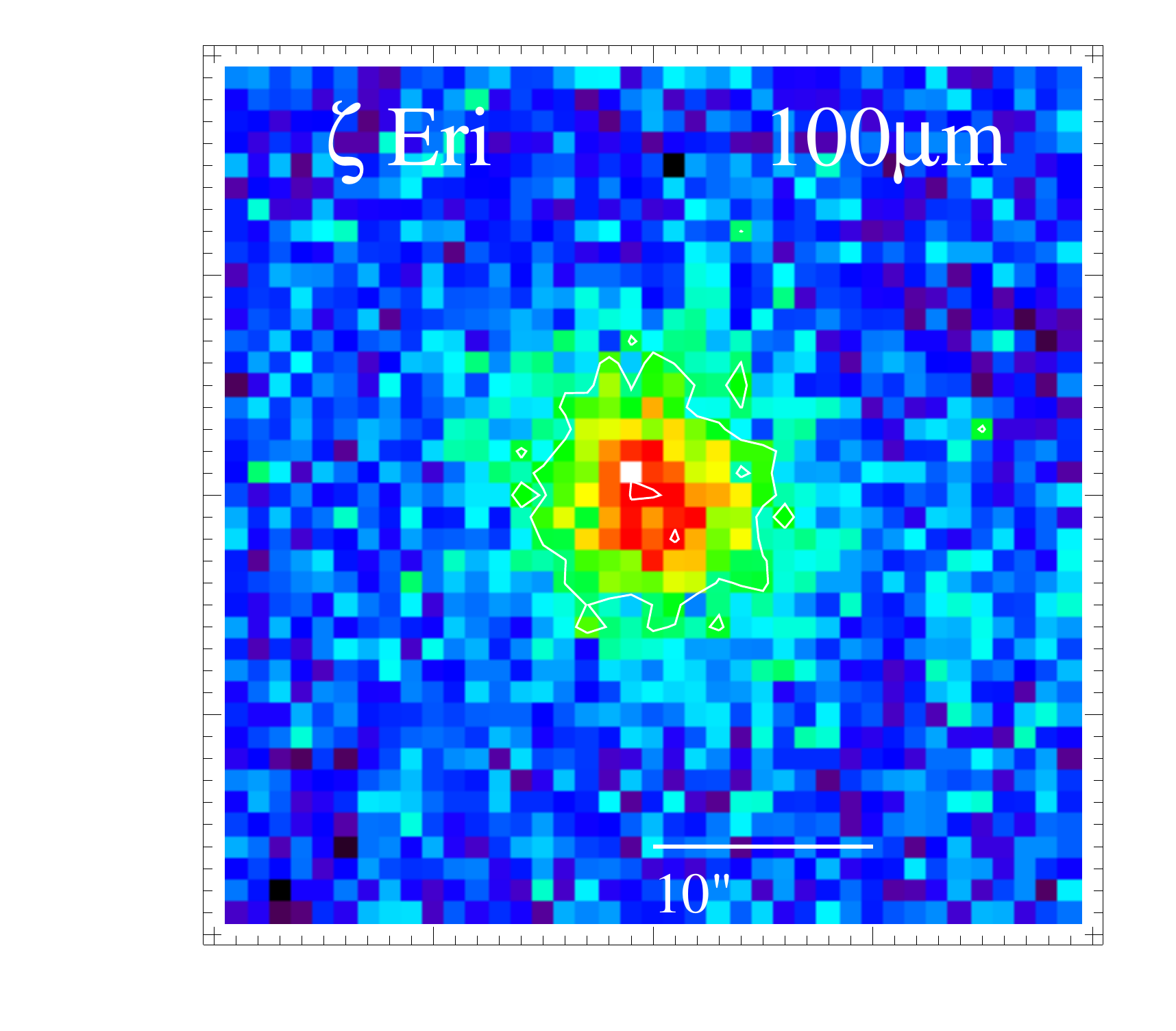} &
      \hspace{-0.5in} \includegraphics[width=0.31\textwidth]{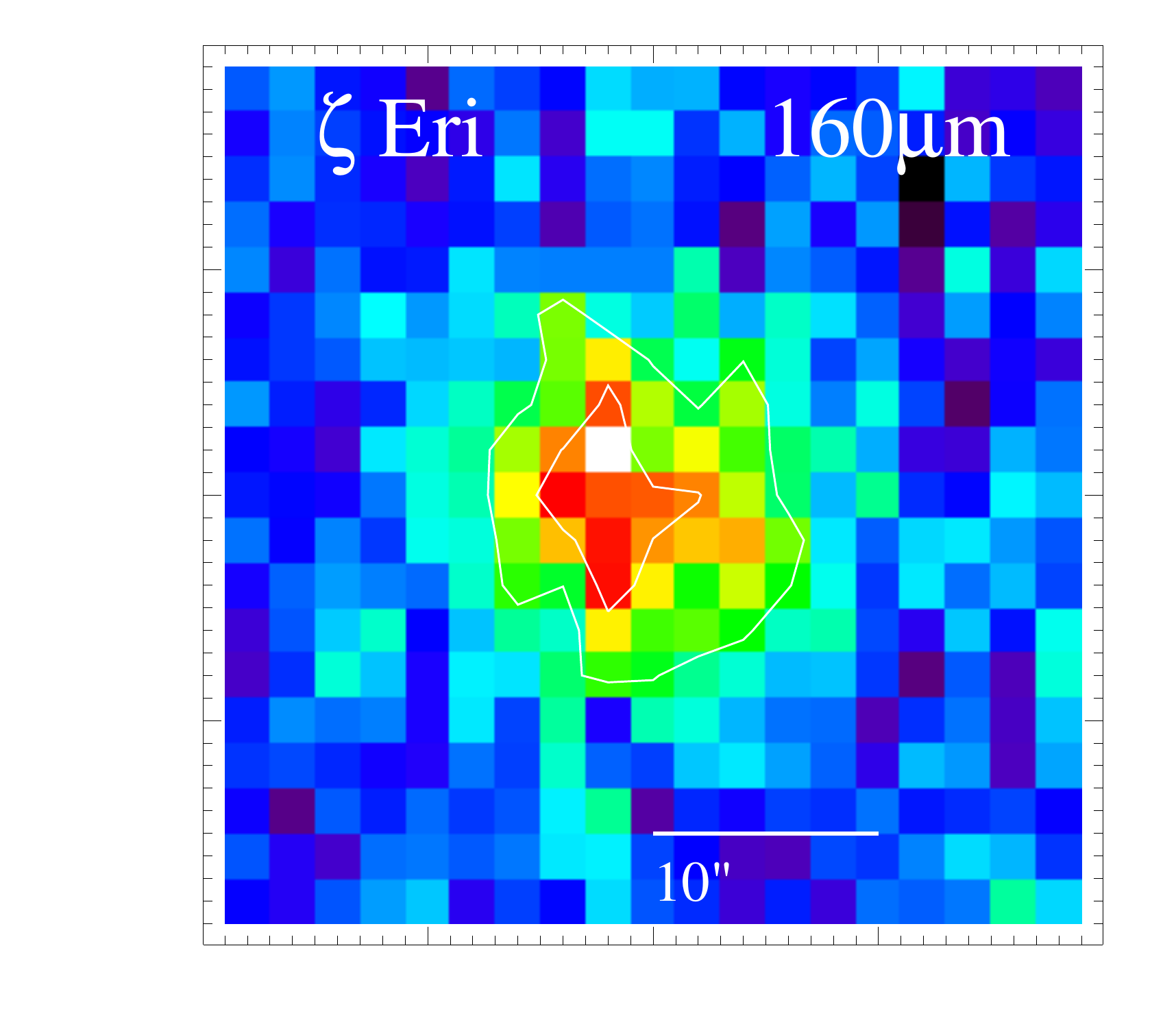} \\
      
			\vspace{-0.2in}
      \hspace{-0.5in} \includegraphics[width=0.31\textwidth]{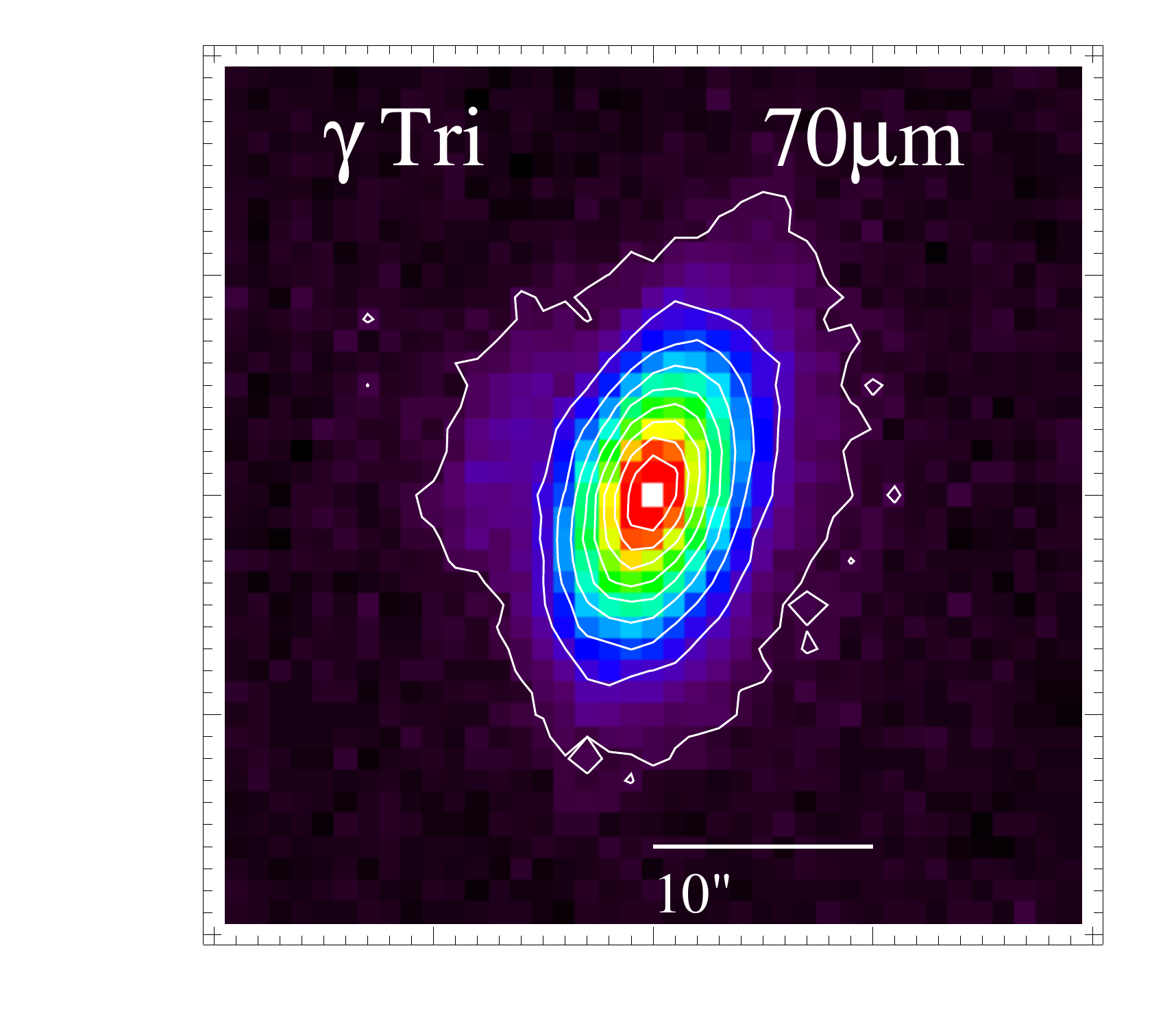} &
      \hspace{-0.5in} \includegraphics[width=0.31\textwidth]{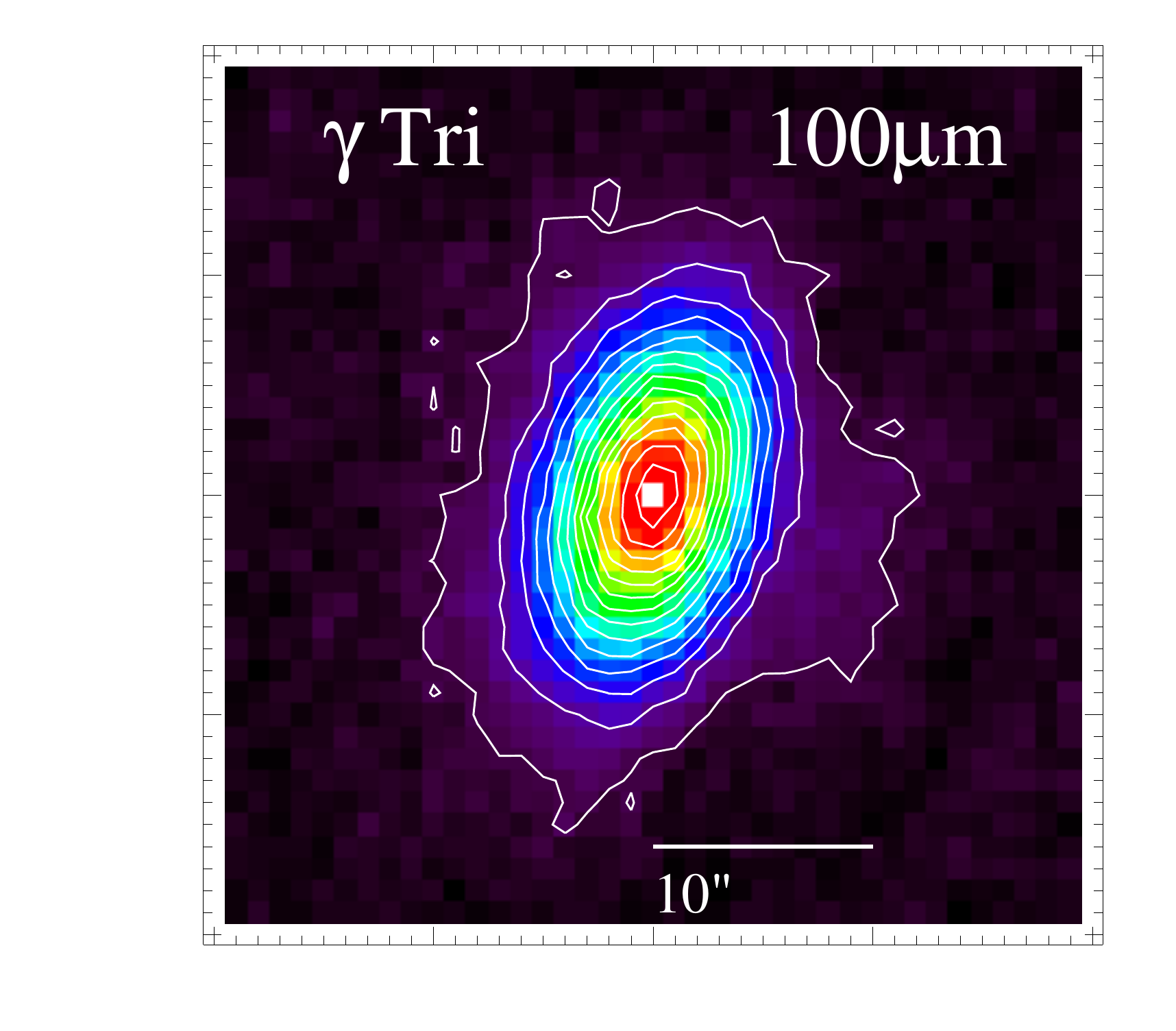} &
      \hspace{-0.5in} \includegraphics[width=0.31\textwidth]{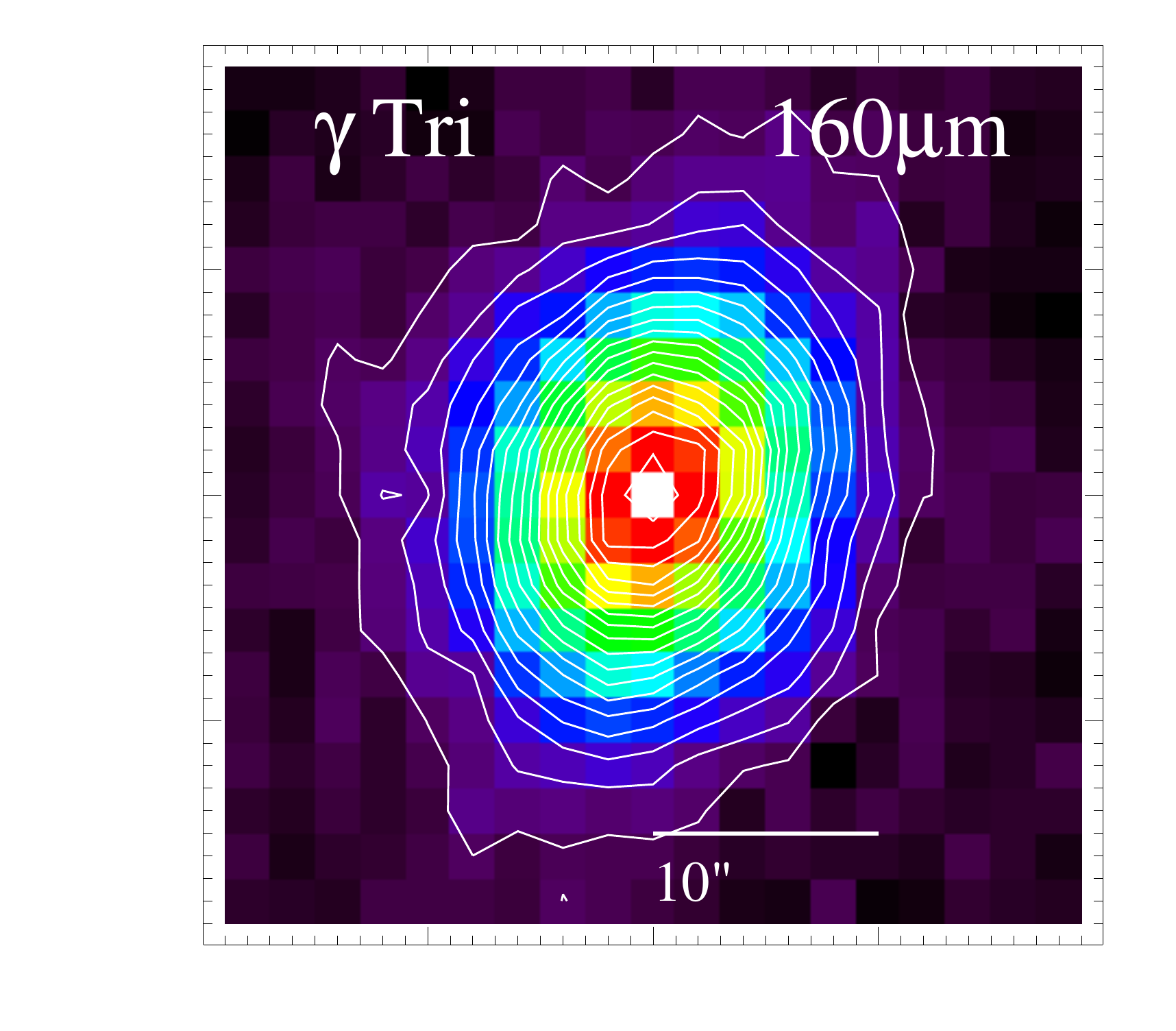} \\
      
 			\vspace{-0.2in}
      \hspace{-0.5in} \includegraphics[width=0.31\textwidth]{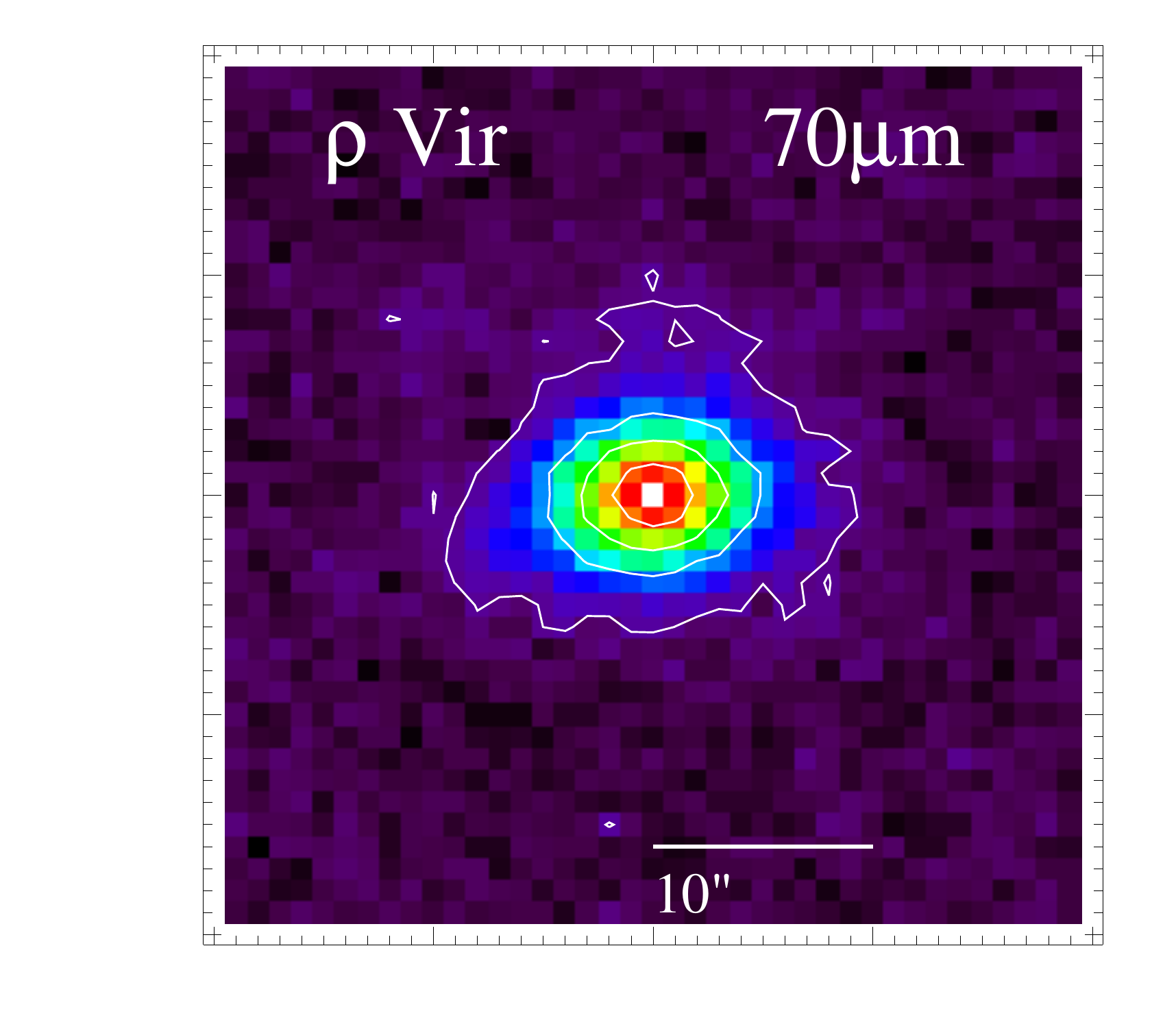} &
      \hspace{-0.5in} \includegraphics[width=0.31\textwidth]{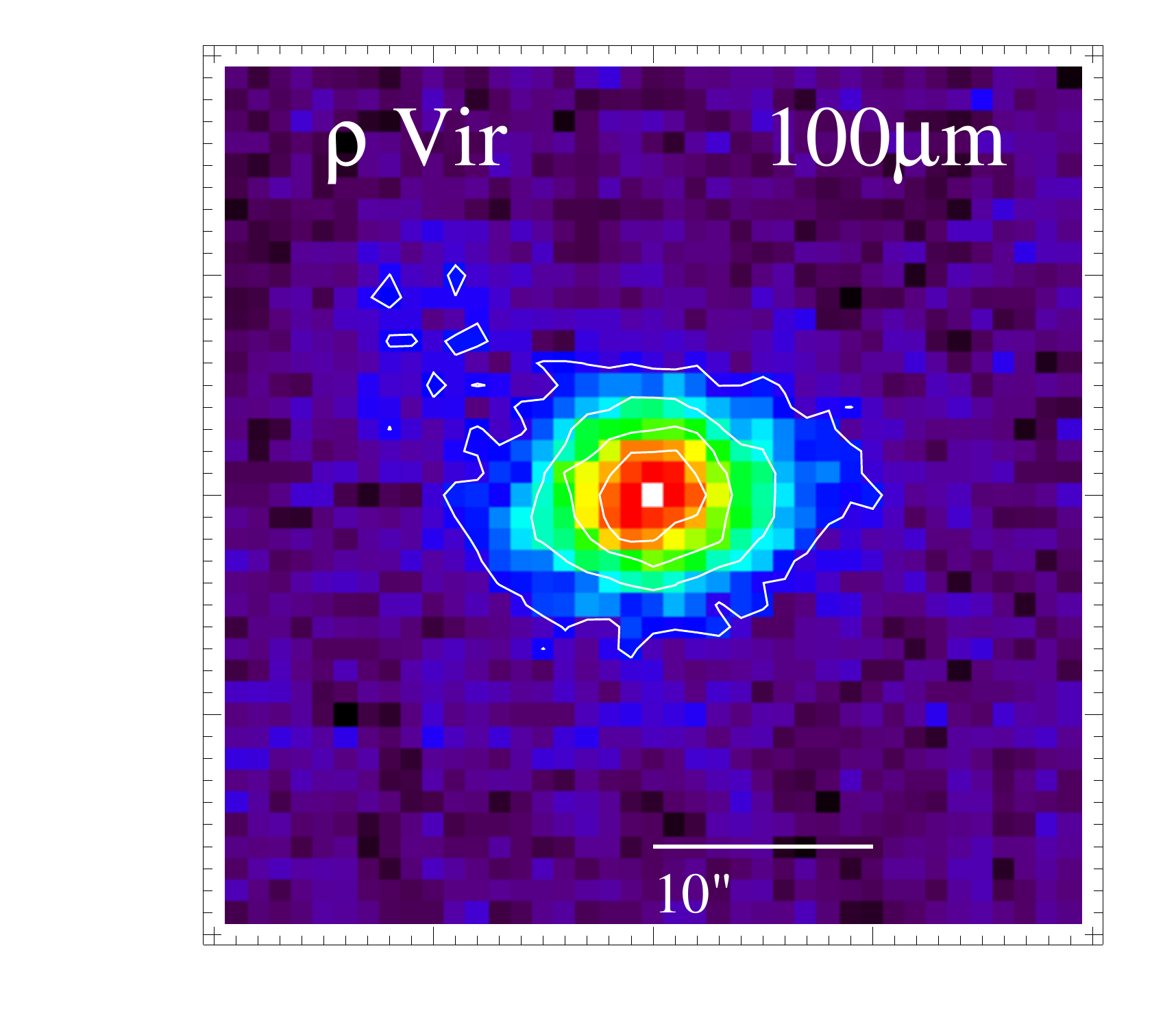} &
      \hspace{-0.5in} \includegraphics[width=0.31\textwidth]{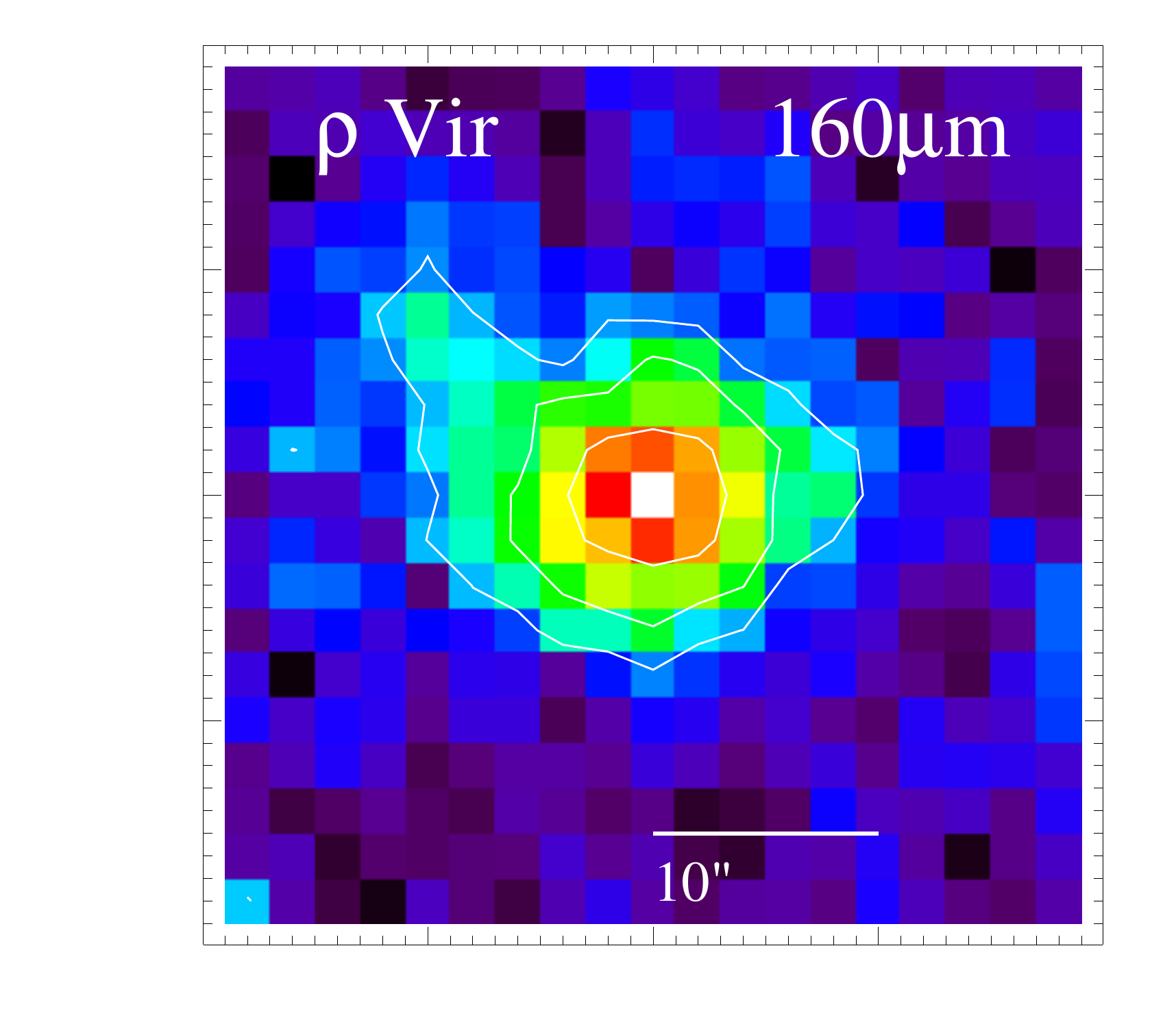} \\
      
			\vspace{-0.2in}
      \hspace{-0.5in} \includegraphics[width=0.31\textwidth]{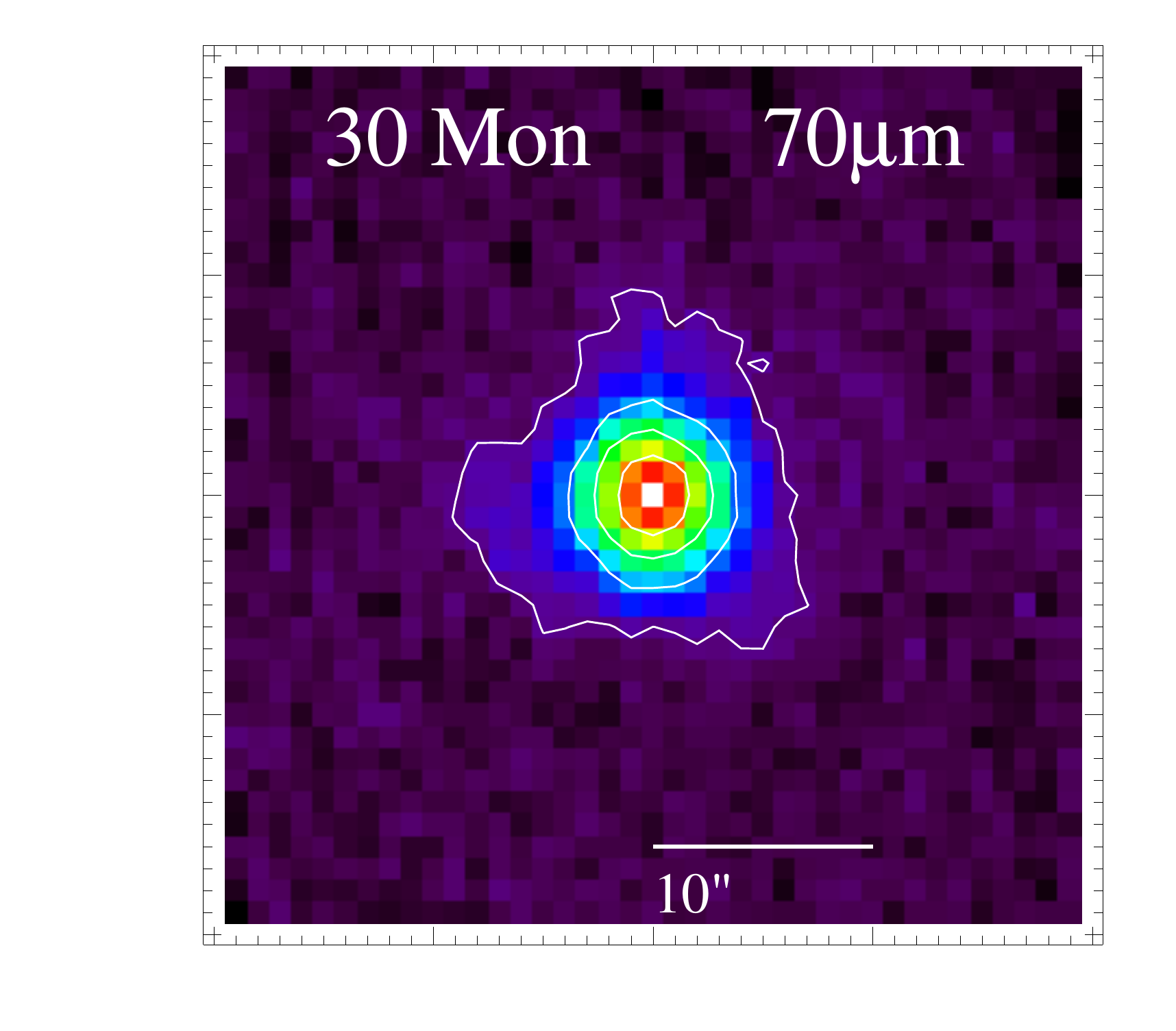} &
      \hspace{-0.5in} \includegraphics[width=0.31\textwidth]{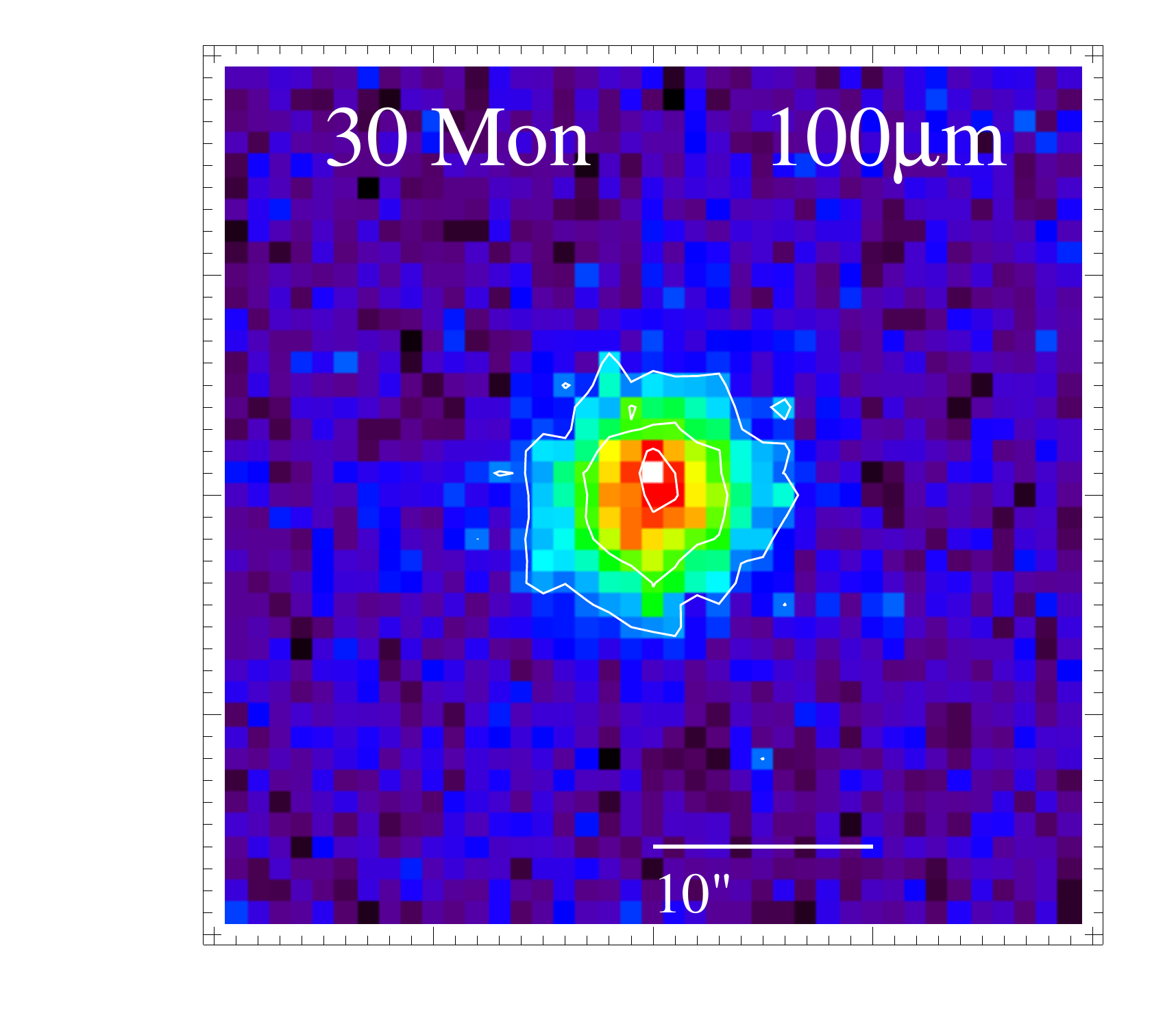} &
      \hspace{-0.5in} \includegraphics[width=0.31\textwidth]{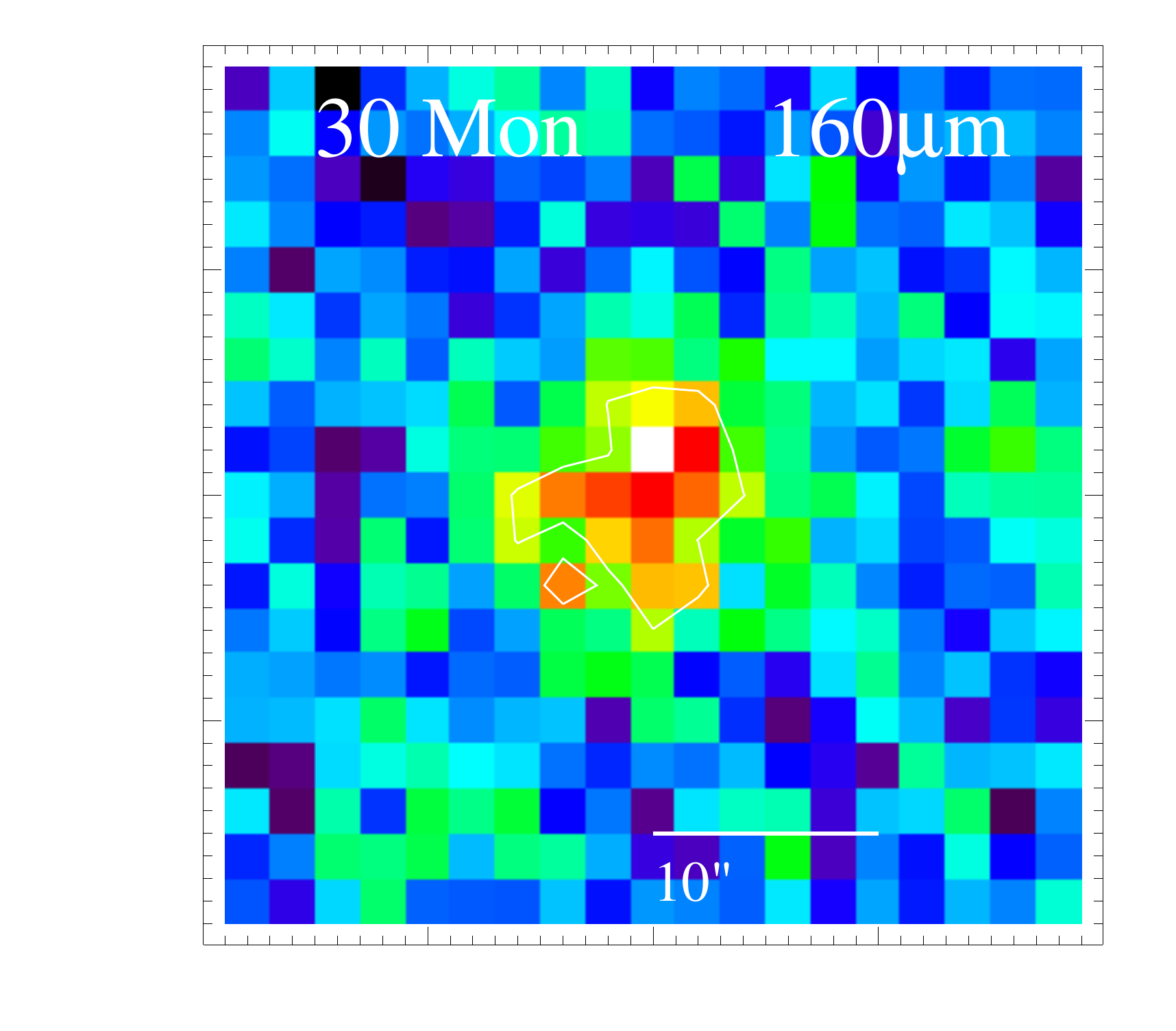} \\
      
			\vspace{-0.2in}
      \hspace{-0.5in} \includegraphics[width=0.31\textwidth]{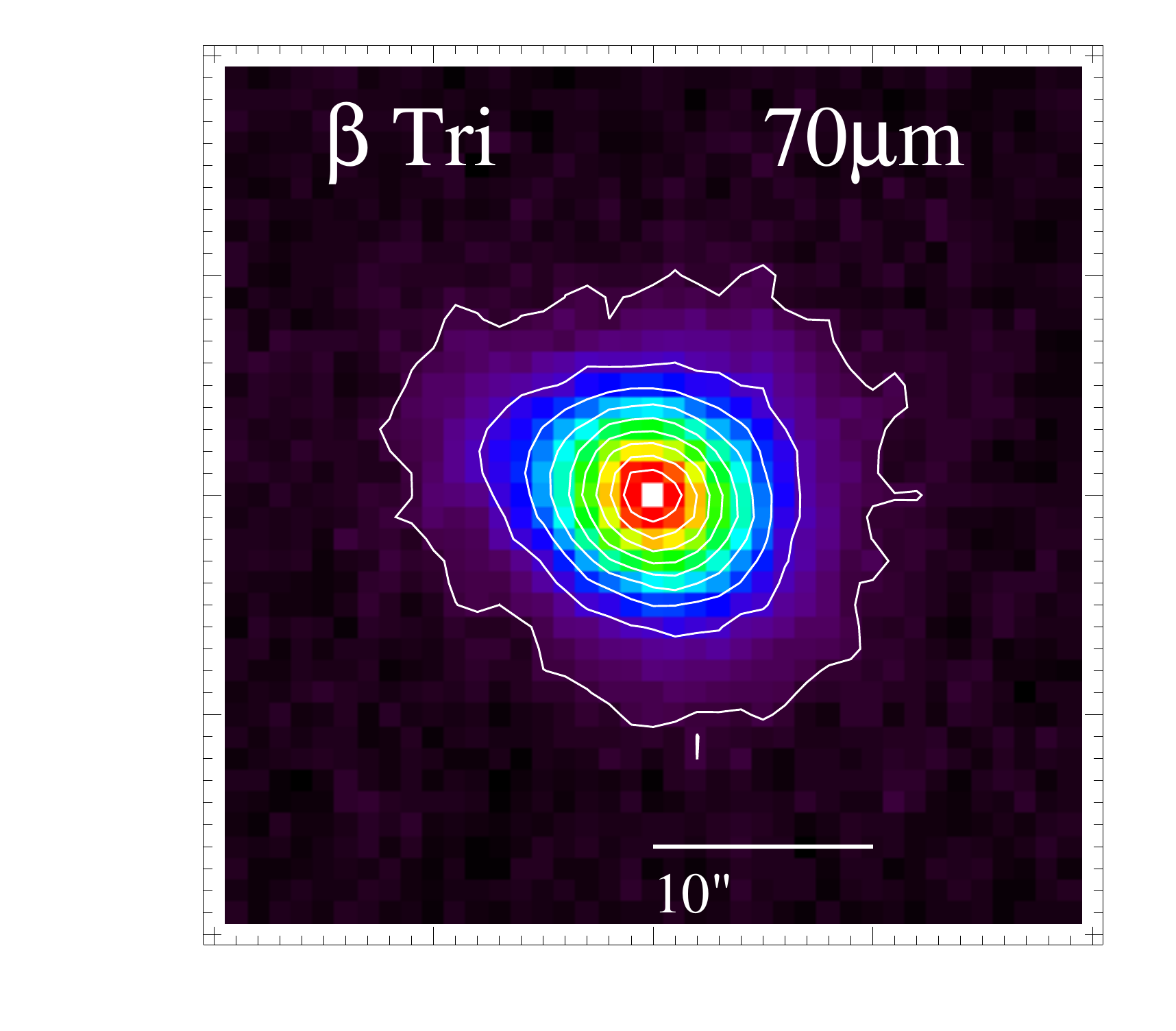} &
      \hspace{-0.5in} \includegraphics[width=0.31\textwidth]{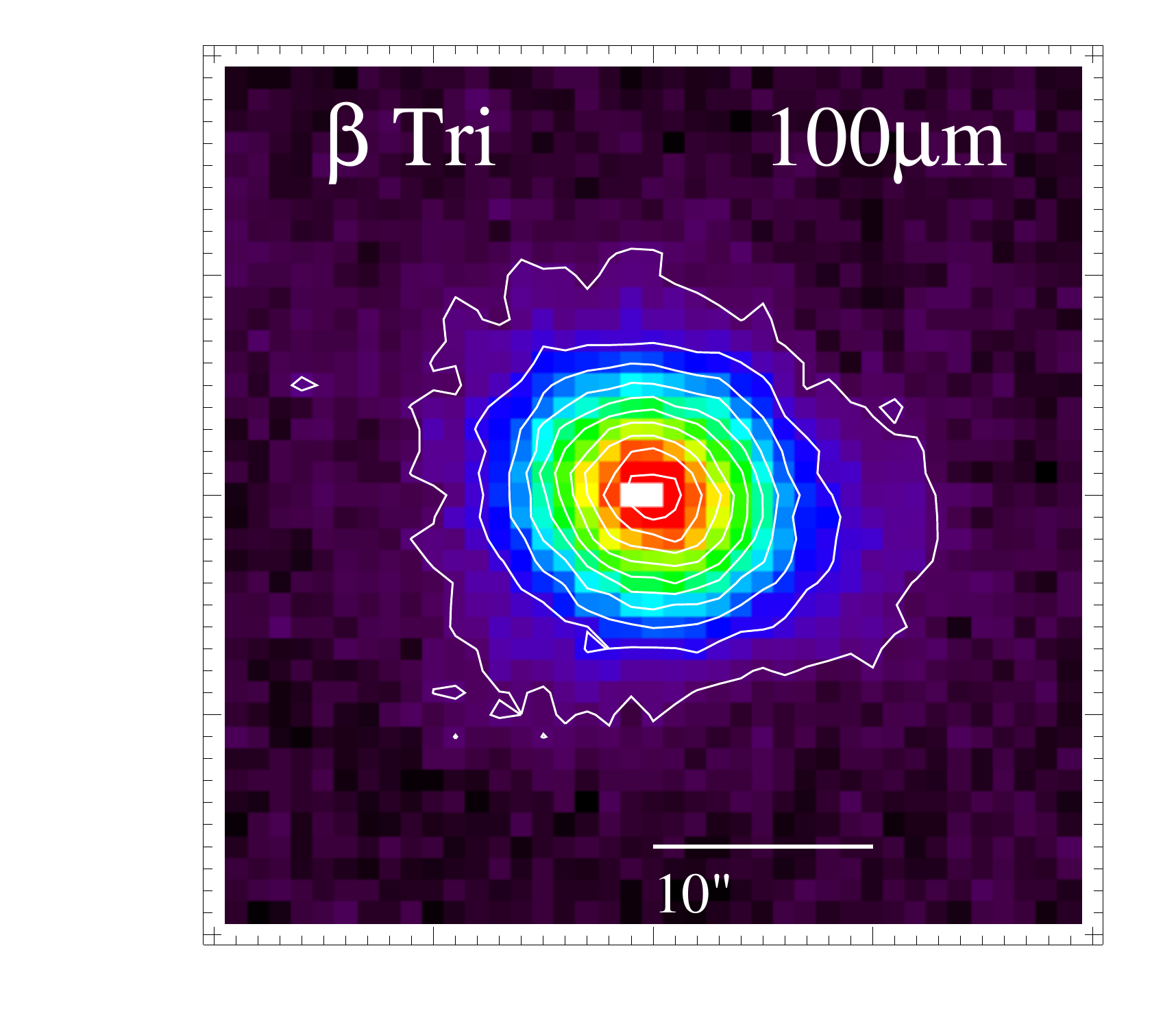} &
      \hspace{-0.5in} \includegraphics[width=0.31\textwidth]{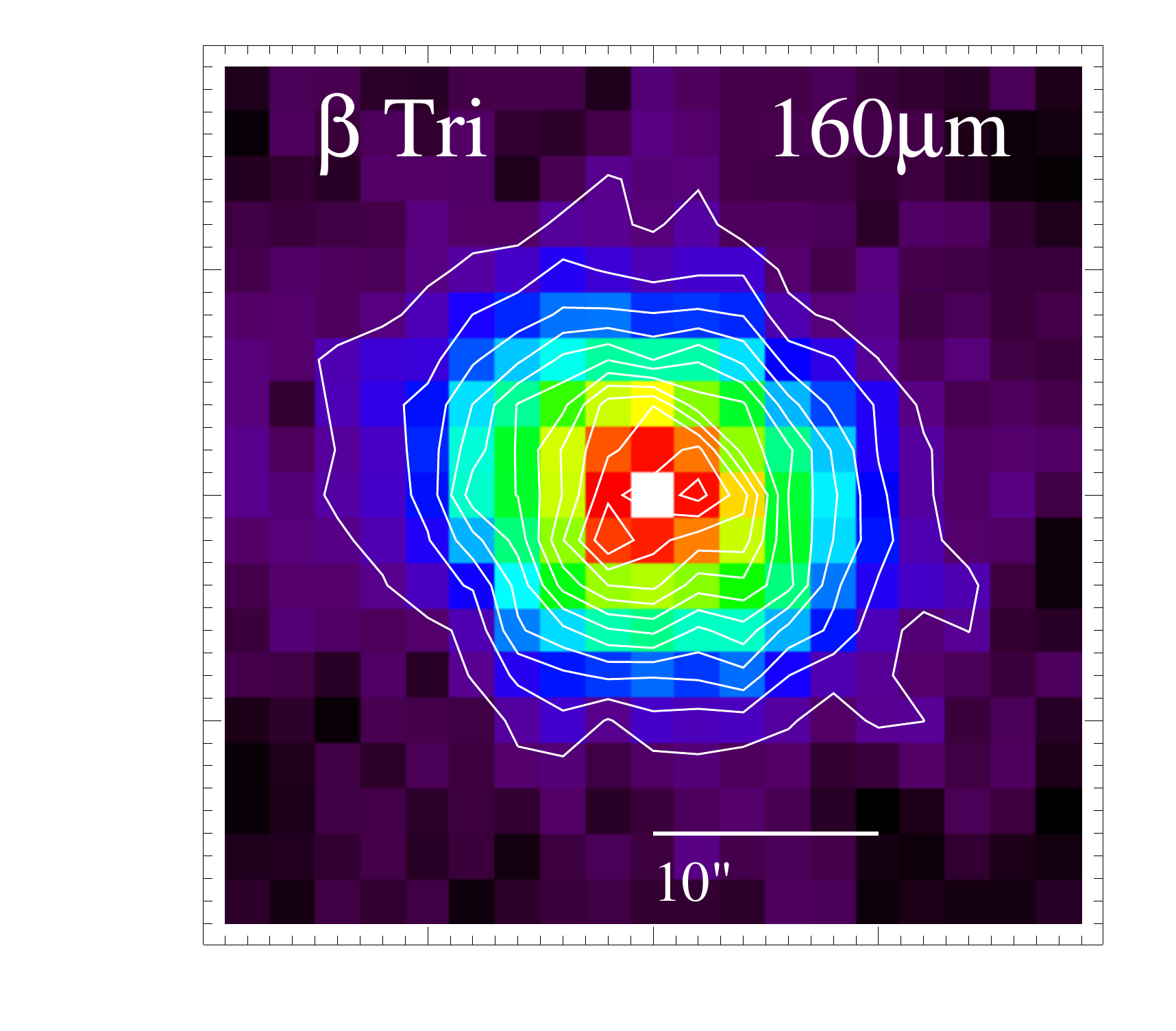}

    \end{tabular}
		\\ \vspace{0.1in}
		\textbf{Figure \ref{fobs100}.} (continued)

  \end{center}

\end{figure*}

\section{Observations}
\label{sobs}
\subsection{PACS}
Herschel's PACS \citep[Photodetecting Array Camera and Spectrometer][]{poglitsch10} instrument observes either at 100 and 160~$\mu$m simultaneously or at 70 and 160~$\mu$m simultaneously. The DEBRIS survey observed all target stars at 100~$\mu$m and 160~$\mu$m. PACS 70~$\mu$m provides little improvement in terms of sensitivity over MIPS \citep[The Multiband Imaging Photometer for Spitzer, ][]{rieke04} 70~$\mu$m and so was not used for most of the stars, however the increased resolution made it worthwhile observing at this wavelength for resolved systems. Doing so gave us an extra 160~$\mu$m image, thus making the 160~$\mu$m images of these discs twice as deep as the majority of the DEBRIS images. Observations were performed in scan-map mode at a scan rate of 20\arcsec/s for a total observing time of 890s (giving us a total observing time of 1780s at 160~$\mu$m). Eight, 3\arcmin-long scans are performed per observation in two directions that differ by 40\degr.

The data were reduced using HIPE \citep[Herschel Interactive Processing Environment, version 7.0, flux calibration FM6,][]{ott10} to a pixel scale of 1\arcsec/pix for 70~$\mu$m and 100~$\mu$m and 2\arcsec/pix for 160~$\mu$m. Nearest-neighbour mapping has been used throughout this paper rather than drizzle mapping to avoid the correlated error issues present in the drizzle method \citep{fruchter02,kennedy12}.

\begin{table}
	\caption{Observation data.}
\begin{tabular}{llllllll}
	\hline
  Star & Instrument & Obs ID & Obs Date  \\
	\hline
  $\alpha$ CrB & PACS 70/160 & 1342223846/7 & 2011-07-09 \\
       & PACS 100/160 & 1342213794/5 & 2011-02-07 \\
  \\
  $\beta$ Uma & PACS 70/160 & 1342220802/3 & 2011-04-29\\
       & PACS 100/160 & 1342197015/6 & 2010-05-24\\
      & SPIRE & 1342195687 & 2010-04-29 \\
  \\
  $\lambda$ Boo & PACS 70/160 & 1342221869/70 & 2011-05-29 \\
       & PACS 100/160 & 1342210928/9 & 2010-12-09 \\
      & SPIRE & 1342222905 & 2011-06-21 \\
  \\
  $\epsilon$ Pav & PACS 70/160 & 1342220576/7 & 2011-05-04\\
       & PACS 100/160 & 1342204276/7 & 2010-09-10  \\
      & SPIRE & 1342229226 & 	2011-09-22 \\
  \\
 $\zeta$ Eri & PACS 70/160 & 1342227302/3 & 2011-08-24 \\
      & PACS 100/160 & 1342216125/6 & 2011-03-05 \\
 \\
  $\gamma$ Tri & PACS 70/160 & 1342237466/7 & 2012-01-13 \\
      & PACS 100/160 & 1342223876/7 & 2011-07-10 \\
      & SPIRE & 1342237503 & 	2012-01-14 \\
 \\
 $\rho$ Vir & PACS 70/160 & 1342222452/3 & 2011-06-10 \\
       & PACS 100/160 & 1342212660/1 & 2011-01-15 \\
      & SPIRE & 1342222904 & 2011-06-21 \\
   \\
  30 Mon & PACS 70/160 & 1342220306/7 & 2011-05-10 \\
       & PACS 100/160 & 1342196127/8 & 2010-05-10 \\
       & SPIRE & 1342195715 & 2010-05-01 \\
   \\
  $\beta$ Tri & PACS 70/160 & 1342237390/1 & 2012-01-12 \\
      & PACS 100/160 & 1342223650/1 & 2011-07-04 \\
      & SPIRE & 1342237504 & 2012-01-14 \\
	\hline
\end{tabular}
	\label{tobsdata}
\end{table}

\begin{figure}
	\centering
	\includegraphics[width=0.47\textwidth]{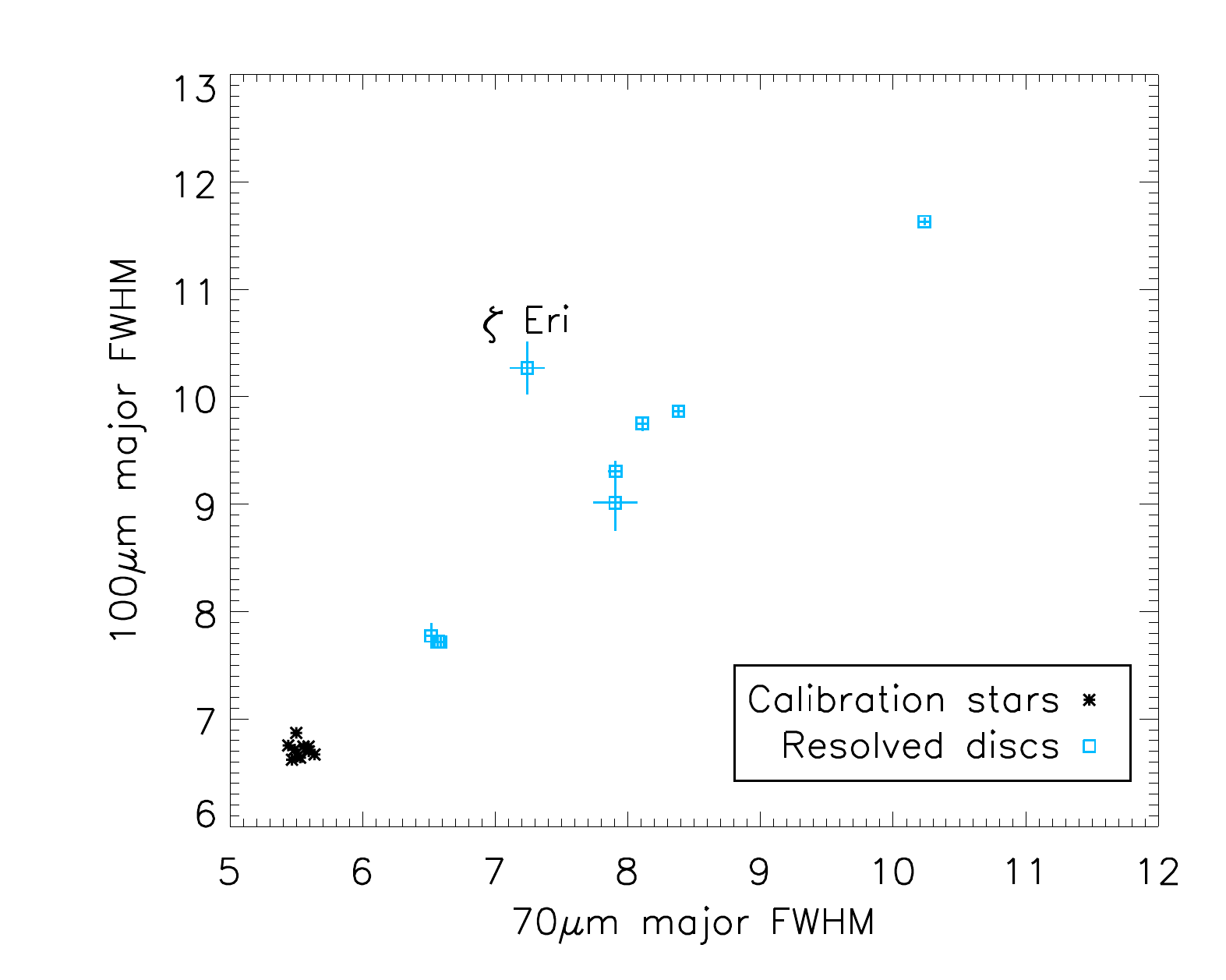}
	\caption{Gaussian major FWHM for the calibration stars and for the stars with resolved discs modelled in this paper. It is clear that these stars are greatly extended compared to the calibration stars.}
	\label{fmajflux}
\end{figure}

\begin{table*}
\begin{minipage}{166mm}
	\caption{Observed fluxes. F$_{obs}$ (mJy) is the observed flux measured by aperture photometry for resolved images and PSF fitting for unresolved images (see the aperture diameter column). F$_{\star}$ (mJy) is the flux of the star. F$_{disc}$ (mJy) is the excess flux. FWHM$_{\rm{maj}}$ is the Full Width Half Maximum of the major axis of a fitted Gaussian.}
\begin{tabular}{lllllll}
	\hline
  Star & $\lambda$ ($\mu$m)&  F$_{obs}$ (mJy) & F$_{\star}$ (mJy) & F$_{disc}$ (mJy) & Aperture Diameter (\arcsec) & FWHM$_{\rm{maj}}$ (\arcsec) \\
	\hline
$\alpha$ CrB & 70 & 515.0$\pm$25.2 & 105.50$\pm$1.73 & 409.5$\pm$25.3 & 30 & 6.57$\pm$0.03 \\
 & 100 & 235.1$\pm$12.6 & 51.21$\pm$0.84 & 183.8$\pm$12.6 & 30 & 7.72$\pm$0.04 \\
 & 160 & 67.6$\pm$2.4 & 19.70$\pm$0.32 & 47.9$\pm$2.5 & PSF fit & 10.83$\pm$0.29 \\
 \\
$\beta$ Uma & 70 & 393.0$\pm$19.4 & 93.65$\pm$1.58 & 299.4$\pm$19.5 & 30 & 6.59$\pm$0.03 \\
 & 100 & 189.2$\pm$9.6 & 45.45$\pm$0.77 & 143.7$\pm$9.6 & 30 & 7.72$\pm$0.06 \\
 & 160 & 58.3$\pm$10.5 & 17.49$\pm$0.30 & 40.8$\pm$10.5 & 30 & 10.19$\pm$0.21 \\
 & 250 & 18.3$\pm$3.9 & 7.05$\pm$0.12 & 11.2$\pm$3.9 & PSF fit \\
 & 350 & 14.1$\pm$4.3 & 3.55$\pm$0.06 & 10.5$\pm$4.3 & PSF fit \\
 & 500 & 9.1$\pm$4.7 & 1.72$\pm$0.03 & 7.4$\pm$4.7 & PSF fit \\ \\
$\lambda$ Boo & 70 & 345.3$\pm$17.3 & 22.61$\pm$0.40 & 322.6$\pm$17.3 & 40 & 8.11$\pm$0.05 \\
 & 100 & 272.1$\pm$15.4 & 11.01$\pm$0.19 & 261.1$\pm$15.4 & 40 & 9.75$\pm$0.07 \\
 & 160 & 142.4$\pm$12.1 & 4.25$\pm$0.07 & 138.1$\pm$12.1 & 40 & 12.90$\pm$0.19 \\
 & 250 & 50.7$\pm$5.1 & 1.72$\pm$0.03 & 48.9$\pm$5.1 & PSF fit \\
 & 350 & 21.3$\pm$5.3 & 0.87$\pm$0.02 & 20.5$\pm$5.3 & PSF fit \\
 & 500 & 4.2$\pm$4.9 & 0.42$\pm$0.01 & 3.8$\pm$4.9 & PSF fit \\ \\
$\epsilon$ Pav & 70 & 63.1$\pm$4.8 & 19.55$\pm$0.29 & 43.5$\pm$4.8 & 30 & 7.90$\pm$0.17 \\
 & 100 & 41.9$\pm$4.6 & 9.48$\pm$0.14 & 32.4$\pm$4.6 & 30 & 9.01$\pm$0.26 \\
 & 160 & 23.6$\pm$3.2 & 3.65$\pm$0.05 & 19.9$\pm$3.2 & PSF fit & 13.06$\pm$0.76 \\
 & 250 & 5.0$\pm$3.9 & 1.47$\pm$0.02 & 3.5$\pm$3.9 & PSF fit \\
 & 350 & 0.7$\pm$4.8 & 0.74$\pm$0.01 & -0.0$\pm$4.8 & PSF fit \\
 & 500 & 0.0$\pm$4.7 & 0.36$\pm$0.01 & -0.4$\pm$4.7 & PSF fit \\ \\
$\zeta$ Eri & 70 & 93.9$\pm$5.8 & 17.40$\pm$0.43 & 76.5$\pm$5.8 & 30 & 7.24$\pm$0.13 \\
 & 100 & 84.1$\pm$5.9 & 8.46$\pm$0.21 & 75.6$\pm$5.9 & 30 & 10.27$\pm$0.24 \\
 & 160 & 42.1$\pm$0.8 & 3.26$\pm$0.08 & 38.8$\pm$0.8 & 30 & 13.38$\pm$0.52 \\
 \\
$\gamma$ Tri & 70 & 777.6$\pm$38.8 & 21.20$\pm$0.40 & 756.4$\pm$38.8 & 40 & 10.24$\pm$0.04 \\
 & 100 & 718.8$\pm$35.5 & 10.29$\pm$0.20 & 708.5$\pm$35.5 & 40 & 11.63$\pm$0.04 \\
 & 160 & 444.3$\pm$10.4 & 3.96$\pm$0.08 & 440.4$\pm$10.4 & 40 & 14.09$\pm$0.08 \\
 & 250 & 186.6$\pm$13.6 & 1.60$\pm$0.03 & 185.0$\pm$13.6 & PSF fit \\
 & 350 & 78.1$\pm$7.2 & 0.80$\pm$0.02 & 77.3$\pm$7.2 & PSF fit \\
 & 500 & 21.8$\pm$5.1 & 0.39$\pm$0.01 & 21.4$\pm$5.1 & PSF fit \\ \\
$\rho$ Vir & 70 & 230.1$\pm$4.3 & 11.25$\pm$0.20 & 218.8$\pm$4.3 & 40 & 7.91$\pm$0.06 \\
 & 100 & 154.2$\pm$7.0 & 5.48$\pm$0.10 & 148.7$\pm$7.0 & 40 & 9.31$\pm$0.10 \\
 & 160 & 67.3$\pm$7.0 & 2.12$\pm$0.04 & 65.2$\pm$7.0 & 40 & 13.62$\pm$0.39 \\
 & 250 & 37.9$\pm$0.8 & 0.86$\pm$0.02 & 37.0$\pm$0.8 & PSF fit \\
 & 350 & 22.7$\pm$0.5 & 0.43$\pm$0.01 & 22.2$\pm$0.5 & PSF fit \\
 & 500 & 20.3$\pm$0.4 & 0.21$\pm$0.00 & 20.0$\pm$0.4 & PSF fit \\ \\
30 Mon & 70 & 206.2$\pm$10.5 & 22.08$\pm$0.46 & 184.1$\pm$10.5 & 30 & 6.52$\pm$0.05 \\
 & 100 & 86.5$\pm$5.9 & 10.71$\pm$0.22 & 75.8$\pm$5.9 & 30 & 7.77$\pm$0.12 \\
 & 160 & 25.5$\pm$0.7 & 4.12$\pm$0.09 & 21.3$\pm$0.7 & PSF fit & 11.45$\pm$0.76 \\
 & 250 & 7.0$\pm$3.4 & 1.66$\pm$0.03 & 5.4$\pm$3.4 & PSF fit \\
 & 350 & 4.2$\pm$4.5 & 0.84$\pm$0.02 & 3.4$\pm$4.5 & PSF fit \\
 & 500 & 0.7$\pm$4.6 & 0.41$\pm$0.01 & 0.3$\pm$4.6 & PSF fit \\ \\
$\beta$ Tri & 70 & 641.2$\pm$31.5 & 73.01$\pm$1.42 & 568.2$\pm$31.6 & 50 & 8.38$\pm$0.04 \\
 & 100 & 481.1$\pm$23.4 & 35.47$\pm$0.69 & 445.6$\pm$23.4 & 50 & 9.87$\pm$0.05 \\
 & 160 & 263.6$\pm$0.3 & 13.68$\pm$0.27 & 249.9$\pm$0.4 & 40 & 13.81$\pm$0.12 \\
 & 250 & 87.1$\pm$7.3 & 5.53$\pm$0.11 & 81.6$\pm$7.3 & PSF fit \\
 & 350 & 34.6$\pm$5.6 & 2.79$\pm$0.05 & 31.8$\pm$5.6 & PSF fit \\
 & 500 & 5.1$\pm$4.9 & 1.35$\pm$0.03 & 3.7$\pm$4.9 & PSF fit \\
 	\hline
\end{tabular}
	\\
	\label{tfluxdata}
\end{minipage}
\end{table*}

As the observations are performed in scan-map mode, the coverage of each point in the map varies. This is especially true of data in the DEBRIS survey where a non-standard technique has been developed to allow us to include data taken whilst the telescope was turning around. Such variation in coverage also means that the noise varies greatly over the map. Due to this, the pixel-pixel noise in the sky background cannot be measured by measuring the variation in pixels in an annulus around the source as this will include pixels of different coverage. Instead we separate the background pixels into 10 bins of coverage and calculate the pixel-pixel noise for each bin.

The FWHM sizes for PACS at 70, 100 and 160~$\mu$m are 5.6\arcsec, 6.8{\arcsec} and 11.4{\arcsec} respectively. To discover which of the DEBRIS targets have discs larger than these sizes and so can be considered to be resolved, Gaussians can be fit to all the systems and the FWHMs are compared to the PSF FWHMs. As a secondary check we also fit PSFs to all systems and check for significant residuals after PSF subtraction. It is important to also consider the possibility of background confusion in these maps, which is discussed in more detail in section \ref{sconf}. 

Amongst the 83 A stars in the DEBRIS sample, 7 are within 20~pc, one of which is resolved -- $\beta$ Leo -- and is modelled in detail by \citet{churcher11}. 57 of the stars are between 20-40~pc and 9 of these host resolved discs. These 9 stars form the basis of this paper. Images of the nine systems presented here as well as one of the PSF stars ($\alpha$ Boo) are shown in figure \ref{fobs100}. These have been cropped to a region of 39{\arcsec}$\times$39{\arcsec} and centred on the peak of the emission. A comparison between the FWHMs of these resolved discs and the PSF stars is shown in figure \ref{fmajflux}. Details of these stars are given in table \ref{tsparam} and details of the observations are given in table \ref{tobsdata}.

The fluxes are calculated through aperture photometry and are given in table \ref{tfluxdata}. PACS data are subject to fairly strong low-frequency ($1/f$) noise and so the data were pre-filtered to remove this. However, the PACS beam has been found to extend to large angular scales with $\approx$10\% of the energy beyond 1\arcmin. Some of the flux is therefore filtered out during the high-pass filtering. To account for this, the fluxes need to be increased by 16$\pm$5\%, 19$\pm$5\% and 21$\pm$5\% at 70, 100 and 160~$\mu$m respectively \citep{kennedy12}. Although this correction factor has been derived for point sources, it should also apply to resolved sources as long as the source extent is less than the size scales from which flux is removed from the beam due to filtering. A further issue with PACS data is that the PACS beam exhibits significant variation at 70~$\mu$m as described in more detail in Kennedy et al. (in prep). Because of this, we have five different calibration stars that we can use when comparing to our observations.

\subsection{SPIRE}
Those systems where a blackbody fit to the spectrum predicted sub-mm fluxes that we expected to be detectable by the SPIRE instrument \citep{griffin10} were followed up with observations at 250, 350 and 500~$\mu$m. The larger beam sizes of 18.2\arcsec, 24.9{\arcsec} and 36.3{\arcsec} at these wavelengths meant that the systems were not resolved at these wavelengths, however these data are still useful for constraining the SED. The fluxes are measured by fitting PSFs to the images and are given in table \ref{tfluxdata}.

\subsection{Confusion Issues}
\label{sconf}
At the depths of the DEBRIS survey a large number of background galaxies become visible. Analysis of all DEBRIS maps shows that the probability of a background source of 3-$\sigma$ significance showing up within 10{\arcsec} of the target is about 36\% for the 160~$\mu$m maps based on \citet{berta11}. Therefore, it is likely that some of our discs are in fact confused with background sources. This does not necessarily mean that what was thought to be disc emission is all background emission, just that care has to be taken when modelling the disc as has been shown to be the case for $\beta$ Pic \citep{regibo12} and 61 Vir \citep{wyatt12} for example. 

Careful analysis of the SED and images for each disc shows that one of the discs in our sample is likely affected by background sources. The images of $\rho$ Vir (figure \ref{fobs100}) show significant emission to the NE at an RA offset of +10\arcsec and a DEC offset of +6\arcsec. In this case the background source can be fit with fluxes of 7.2, 11.3, and 17.7~mJy at 70, 100, and 160~$\mu$m, which are consistent with a submillimetre galaxy of redshift, $z\approx1$ \citep{blain02}. The aperture for the photometry is increased to include both the disc and background emission with the background emission then subtracted from the aperture flux.

\section{Spectral Energy Distributions}
\label{ssedfit}
Debris discs are usually found by excess IR emission above the level expected from the stellar photosphere. Thus, the first step is to make a model of the stellar spectral energy distribution (SED). To do this we start by collating optical and near-IR photometry for all of our stars. Because many of our stars are very bright, and therefore saturated in the 2MASS catalogue \citep{cutri03}, we use ``heritage'' UBV \citep{mermilliod06} and ubvy \citep{hauck98} photometry, as well as more recent optical and near-IR photometry from Tycho 2 \citep{hog00}, Hipparcos \citep{perryman97a}, AKARI 9~$\mu$m \citep{ishihara10}, and IRAS 12~$\mu$m \citep{moshir90}. For stars that show excess emission at near-IR wavelengths, we do not include AKARI and IRAS fluxes in the stellar spectrum fitting. 

The stellar photosphere model is then found by least squares minimisation of the GAIA grid of PHOENIX models \citep{brott05}. Interstellar reddening is not taken into account for the photosphere fitting as it is not expected to be important for stars within the Local Bubble \citep{su05}. This fitting process gives us the effective temperature and luminosity of the stars (table \ref{tstfit}). These parameters are also used to interpolate the mass from the tables of \citet{schmidt82}. Comparison of the predicted photospheric fluxes with MIPS 24~$\mu$m photometry for stars in the DEBRIS survey that do not exhibit a detectable excess at 24~$\mu$m shows that the photospheric predictions have a 1-$\sigma$ uncertainty of about 2\%. Once modelled, the photospheric flux can be subtracted from the observed fluxes, yielding the disc spectrum.

\begin{table}
	\caption{Stellar parameters from the photospheric fitting.}
\begin{tabular}{llll}
	\hline
  Star & $T_{\rm{eff}}$ (K) & $M_\star$ ($M_\odot$) & $L_\star$ ($L_\odot$)  \\
	\hline
$\alpha$ CrB$^{a}$ & 9302$\pm$61 & 2.7 & 59.8$\pm$1.0 \\
$\beta$ Uma & 9291$\pm$61 & 2.7 & 59.7$\pm$1.0 \\
$\lambda$ Boo & 8608$\pm$50 & 2.3 & 17.1$\pm$0.3 \\
$\epsilon$ Pav & 9794$\pm$61 & 3.0 & 25.6$\pm$0.4 \\
$\zeta$ Eri & 7516$\pm$44 & 1.7 & 10.9$\pm$0.3 \\
$\gamma$ Tri & 9127$\pm$111 & 2.6 & 25.2$\pm$0.5 \\
$\rho$ Vir & 8817$\pm$55 & 2.4 & 13.2$\pm$0.2 \\
30 Mon & 9401$\pm$84 & 2.8 & 34.5$\pm$0.7 \\
$\beta$ Tri$^{a}$ & 8011$\pm$42 & 1.9 & 73.4$\pm$1.4 \\

	\hline
\end{tabular}
        \medskip \\$^{a}${These systems are close binaries and so the fitted parameters are representative of the combined spectra rather than being real properties of the stars.}
	\label{tstfit}
\end{table}

With flux density measurements of an excess at just two wavelengths a blackbody can be fit to give the colour temperature of the excess spectrum. As many of our stars have been observed before, more data than just the Herschel observations can be used to more tightly constrain the blackbody fit. Photometry from all available sources was collected. This includes (where available) IRAS \citep[Faint Source Catalog v2;][]{moshir90}, ISO \citep{habing01}, Spitzer IRS \citep{lebouteiller11}, Spitzer MIPS (re-reduced data from an updated pipeline), WISE preliminary release \citep{wright10}, AKARI \citep{ishihara10} and SCUBA \citep{williams06}.

\begin{figure*}
	\begin{center}
	\includegraphics[width=\textwidth]{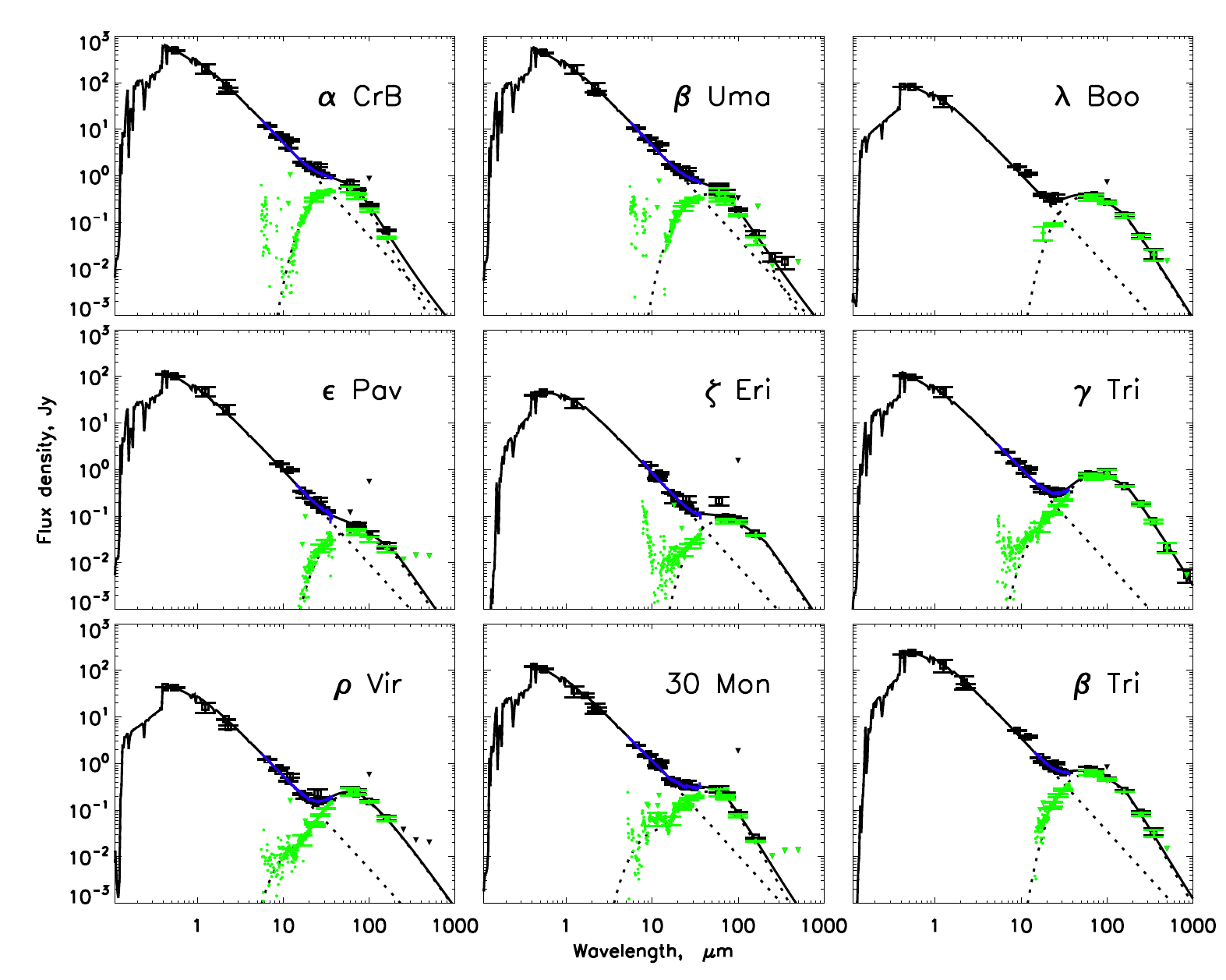}
	\caption{SED fits to each of the stars. The solid line represents the combined star and disc fit, whereas the dashed lines represent separate fits to the star and disc. The black squares represent the observed photometry and the green diamonds represent the star subtracted photometry. The black and green downward triangles represent upper limits for the observed and star subtracted photometry. The blue line represents the IRS spectrum. The green dots represent the star subtracted IRS data. No error bars have been plotted for the IRS data for clarity, but the uncertainties do increase greatly at shorter wavelengths.}
	\label{fseds}
	\end{center}
\end{figure*}

From observations at sub-millimetre wavelengths it is known that the spectrum falls off more steeply than blackbody at wavelengths much larger than the grain size due to the absorption and emission inefficiencies. To account for this, we use a modified blackbody where the standard blackbody equation is modified by dividing by a factor $X_\lambda$ as given by
\begin{align}
X_\lambda=\left\{
	\begin{array}{lr}
		1 & \lambda < \lambda_0 \\
		(\lambda/\lambda_0)^{\beta} & \lambda > \lambda_0 
	\end{array} \right.
\end{align}
where $\lambda_0$ and $\beta$ are free parameters. In some cases there is not enough long wavelength data to constrain the fall-off in the SED. In these cases values of $\lambda_0=210$~$\mu$m and $\beta=1$ are used based on the average values for A star discs \citep{wyatt07a}. The SED fits for the stars are shown in figure \ref{fseds}. The results of the SED fits are listed in table \ref{tsedfit}. 

The wavelengths observed using Herschel are dominated by cold dust. Due to the inclusion of shorter wavelength data in the SED, it is important to also consider the possibility of emission from a warm disc that may be analogous to the asteroid belt but could also be evidence of emission from dust at a single radial location at multiple temperatures. Three systems show evidence for warm emission based on their SEDs: $\rho$ Vir, $\gamma$ Tri and 30 Mon.

\citet{morales11} showed that the Spitzer data for $\rho$ Vir is best fit by a two ring system with belts at temperatures of 237 and 74~K. With our increased wavelength range for the SED we find this to be broadly correct, with our best fit giving temperatures of 279$\pm$107 and 81$\pm$3~K. The disc around $\gamma$ Tri has previously been reported to have temperatures ranging from 65~K \citep{zuckerman04,williams06} to 120~K \citep{chen06}. Clearly the dust in this system is at more than one temperature so here we choose to fit it with two ring components, although we will see later that narrow rings are not appropriate for this system. Note that the inner components for these stars are poorly constrained and improved mid-IR photometry, spectra and/or imaging will be required to get a better idea of the properties of this warm component. Such mid-IR imaging allowed \citet{smith10} to show that 30 Mon (HD 71155) most likely has two components and the inner component has been resolved by \citet{moerchen10} at about 2~AU, which our value of 3$\pm$4~AU is consistent with.

\begin{table*}
\begin{minipage}{170mm}
	\caption{Parameters from a blackbody fit to the SED. Columns 2 to 7 give the temperature, radius and fractional luminosity for the inner belt (subscript $i$), where this is included in the fit, and for the outer belt (subscript $o$). $\lambda_0$ and $\beta$ are the parameters for the modified part of the blackbody (see section \ref{ssedfit}). $\chi^2_{red}$ gives the reduced chi-squared of the fit.}
\begin{tabular}{llllllllll}
	\hline
  Star & T$_i$ (K) & R$_i$ (AU) & f$_i$ ($\times 10^{-5}$) & T$_o$ (K) & R$_o$ (AU) & f$_o$ ($\times 10^{-5}$) & $\lambda_0$ & $\beta$ & $\chi^2_{red}$ \\
	\hline
$\alpha$ CrB &  &  &  & 121.8$\pm$3.4 & 40.4$\pm$2.3 & 1.7$\pm$0.2 & 74.6$\pm$6.7 & 1.6$\pm$0.2 & 0.241 \\
$\beta$ Uma &  &  &  & 119.1$\pm$3.6 & 42.2$\pm$2.6 & 1.3$\pm$0.2 & 72.1$\pm$9.3 & 1.3$\pm$0.3 & 0.664 \\
$\lambda$ Boo &  &  &  & 91.1$\pm$1.1 & 38.6$\pm$1.0 & 5.1$\pm$0.3 & 162.2$\pm$41.9 & 1.1$\pm$0.6 & 0.722 \\
$\epsilon$ Pav &  &  &  & 87.0$\pm$3.1 & 51.9$\pm$3.7 & 0.5$\pm$0.1 & 210.0$\pm$0.0 & 1.0$\pm$0.0 & 1.313 \\
$\zeta$ Eri &  &  &  & 80.1$\pm$1.1 & 39.7$\pm$1.2 & 2.0$\pm$0.1 & 210.0$\pm$0.0 & 1.0$\pm$0.0 & 1.434 \\
$\gamma$ Tri & 182.5$\pm$38.4 & 11.7$\pm$4.9 & 2.15$\pm$1.93 & 65.6$\pm$2.1 & 90.2$\pm$5.8 & 6.4$\pm$0.5 & 175.2$\pm$21.3 & 1.1$\pm$0.2 & 1.164 \\
$\rho$ Vir & 278.6$\pm$107.1 & 3.6$\pm$2.8 & 1.48$\pm$1.99 & 80.7$\pm$2.6 & 43.1$\pm$2.8 & 5.0$\pm$0.6 & 71.8$\pm$16.4 & 0.7$\pm$0.3 & 1.124 \\
30 Mon & 391.4$\pm$249.3 & 3.0$\pm$3.8 & 1.73$\pm$3.81 & 107.8$\pm$9.3 & 39.2$\pm$6.7 & 2.9$\pm$0.9 & 64.7$\pm$10.1 & 1.5$\pm$0.1 & 0.753 \\
$\beta$ Tri &  &  &  & 86.4$\pm$1.0 & 89.0$\pm$2.3 & 3.1$\pm$0.2 & 162.6$\pm$21.2 & 1.3$\pm$0.3 & 0.982 \\	\hline
\end{tabular}
        \\
        $^{a}$ In these cases the fall off in emission at long wavelengths is unconstrained by the data and so $\lambda_0$ and $\beta$ are fixed to values believed to be typical for A stars (see text). 
	\label{tsedfit}
\end{minipage}
\end{table*}

\begin{figure*}
	\begin{tabular}{cc}
	\includegraphics[width=0.47\textwidth]{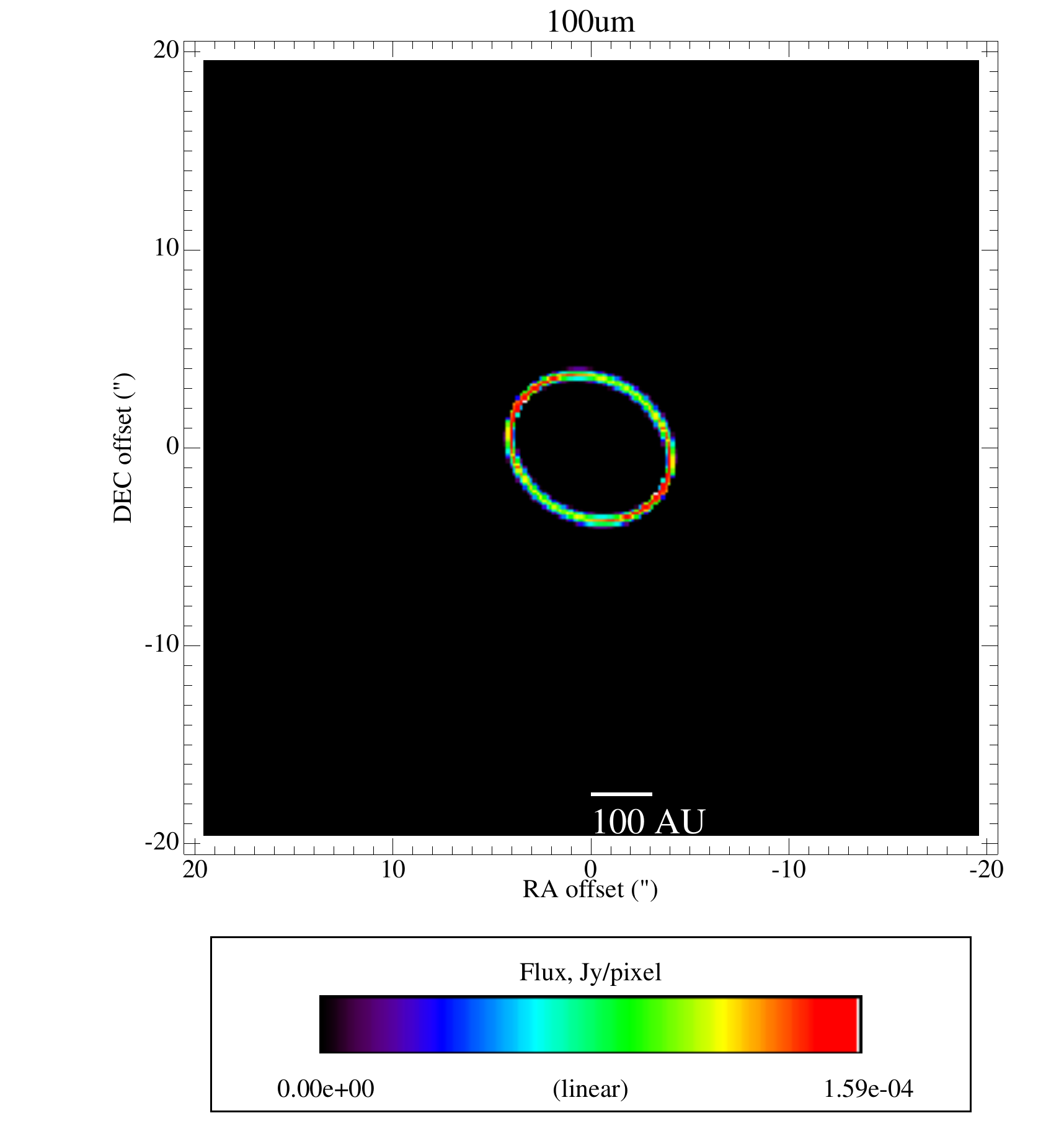} &
	\includegraphics[width=0.47\textwidth]{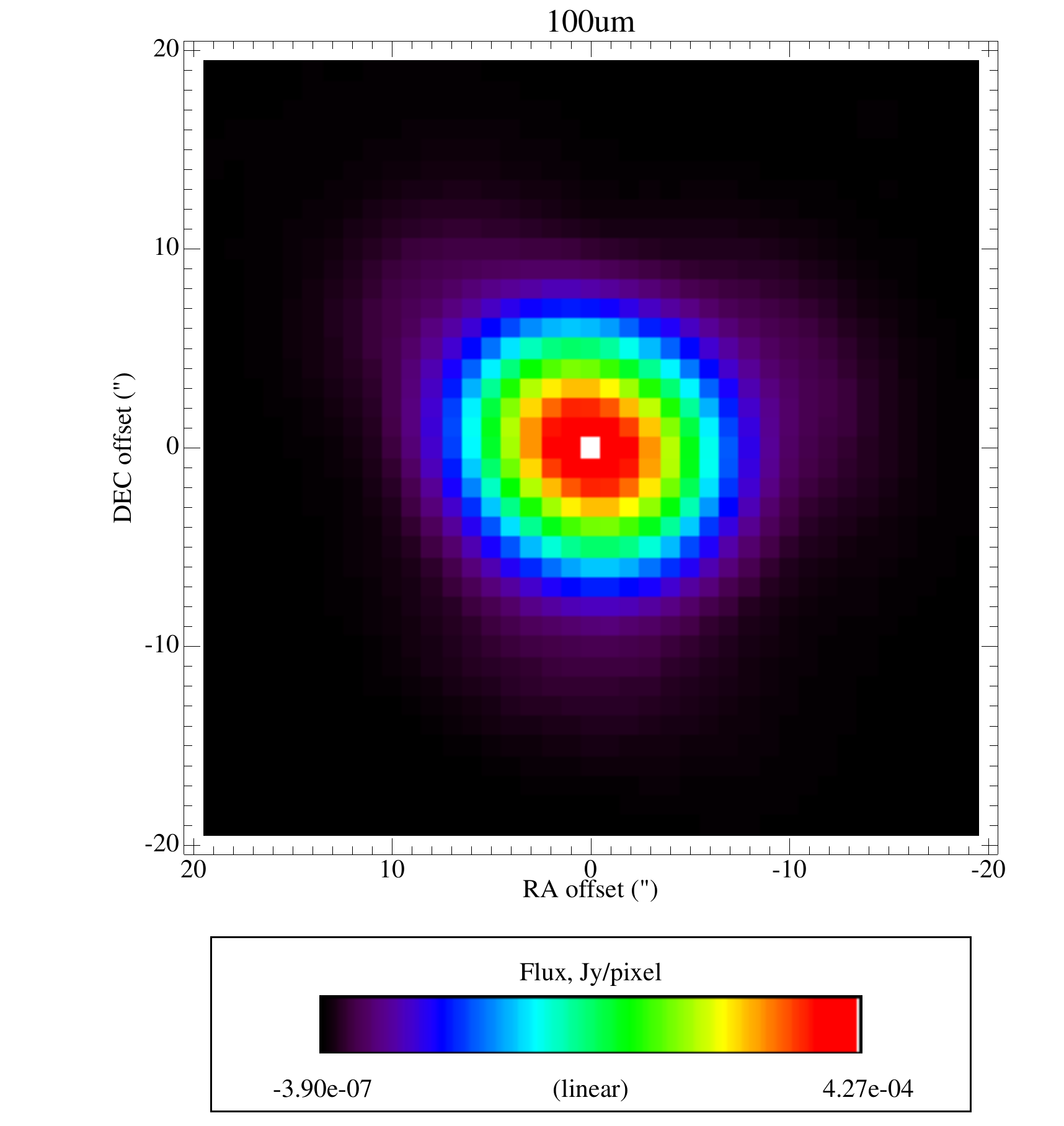}\\
	\end{tabular}

	\caption{Left: Best-fit model of the $\epsilon$ Pav disc at 100~$\mu$m. Right: Best-fit model at 100~$\mu$m of $\epsilon$ Pav combined with the stellar photosphere and convolved with the PSF.}
	\label{fa076ring}

\end{figure*}

\section{Resolved Modelling}
\label{smod}
\subsection{Image preparation}
Herschel has a nominal 1-$\sigma$ pointing error of 2.3\arcsec in scan-map mode\footnote{http://herschel.esac.esa.int/Docs/Herschel/html/ch02.html}, although errors as high as 4\arcsec or more have been reported\footnote{http://herschel.esac.esa.int/twiki/bin/view/Public/DpKnownIssues}. So, rather than centring the images on where we expect the star to be, we centre them (to sub-pixel resolution) on the peak of the emission. Although this does have the potential to introduce correlated noise as we are regridding the data, the difference to the final results is insignificant and this technique has the benefit of allowing us to combine the 160~$\mu$m images as noted below. Offsets larger than the average pointing error may be indicative of asymmetric discs or confusion. The RMS value for the offsets of these stars is 1.9{\arcsec} and the largest offset is 3.0{\arcsec} and so we do not see evidence that the emission is not centred on the star's position for any of our targets.

Since the 70 and 100~$\mu$m data are observed at different times and each comes with a simultaneous 160~$\mu$m observation, the 160~$\mu$m images are co-added based on the centre location of their respective 70 or 100~$\mu$m image.

\begin{figure*}

  \begin{center}


    \begin{tabular}{ccc}

			\vspace{-0.2in}
      \hspace{-0.5in} \includegraphics[width=0.31\textwidth]{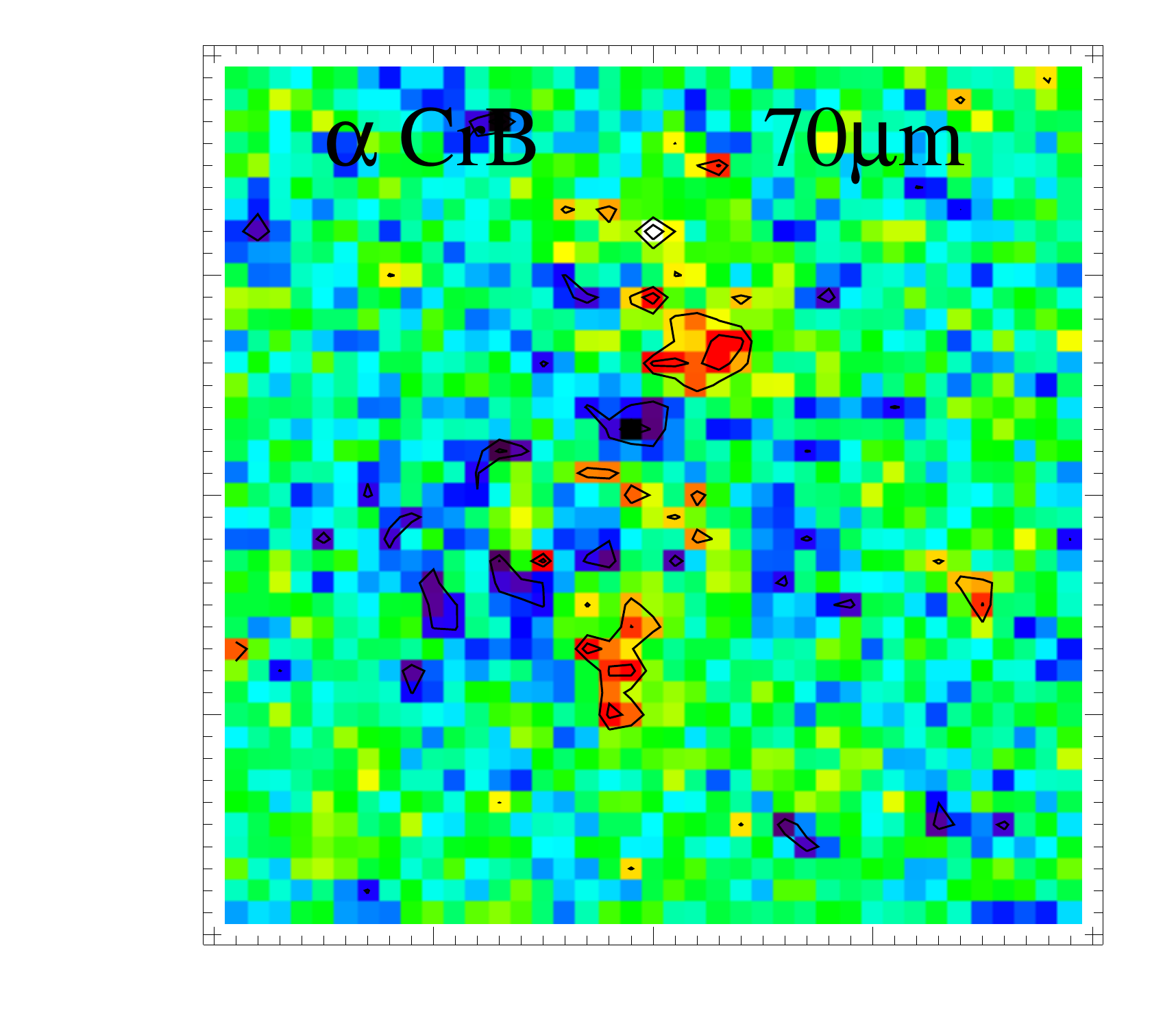} &
      \hspace{-0.5in} \includegraphics[width=0.31\textwidth]{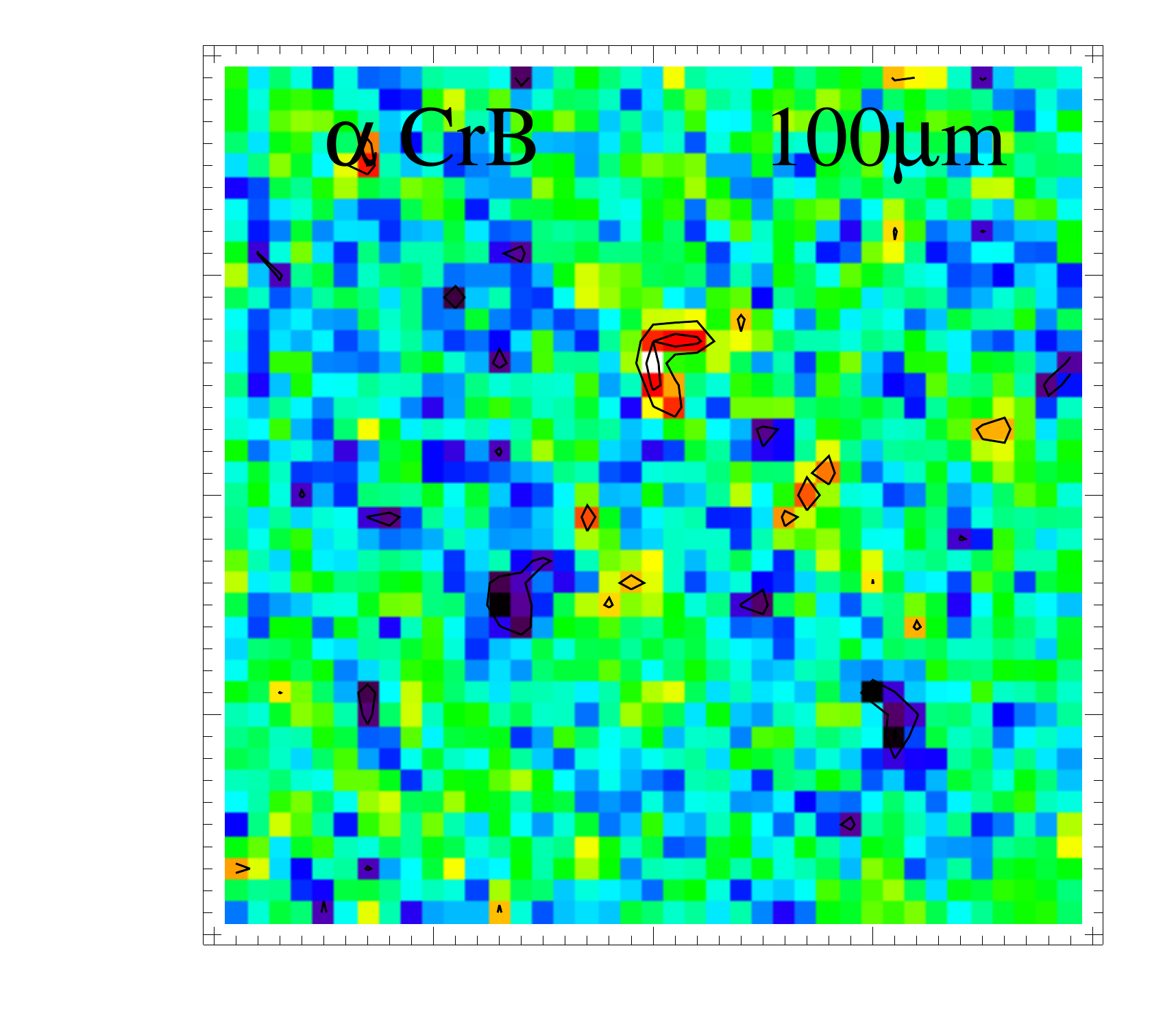} &
      \hspace{-0.5in} \includegraphics[width=0.31\textwidth]{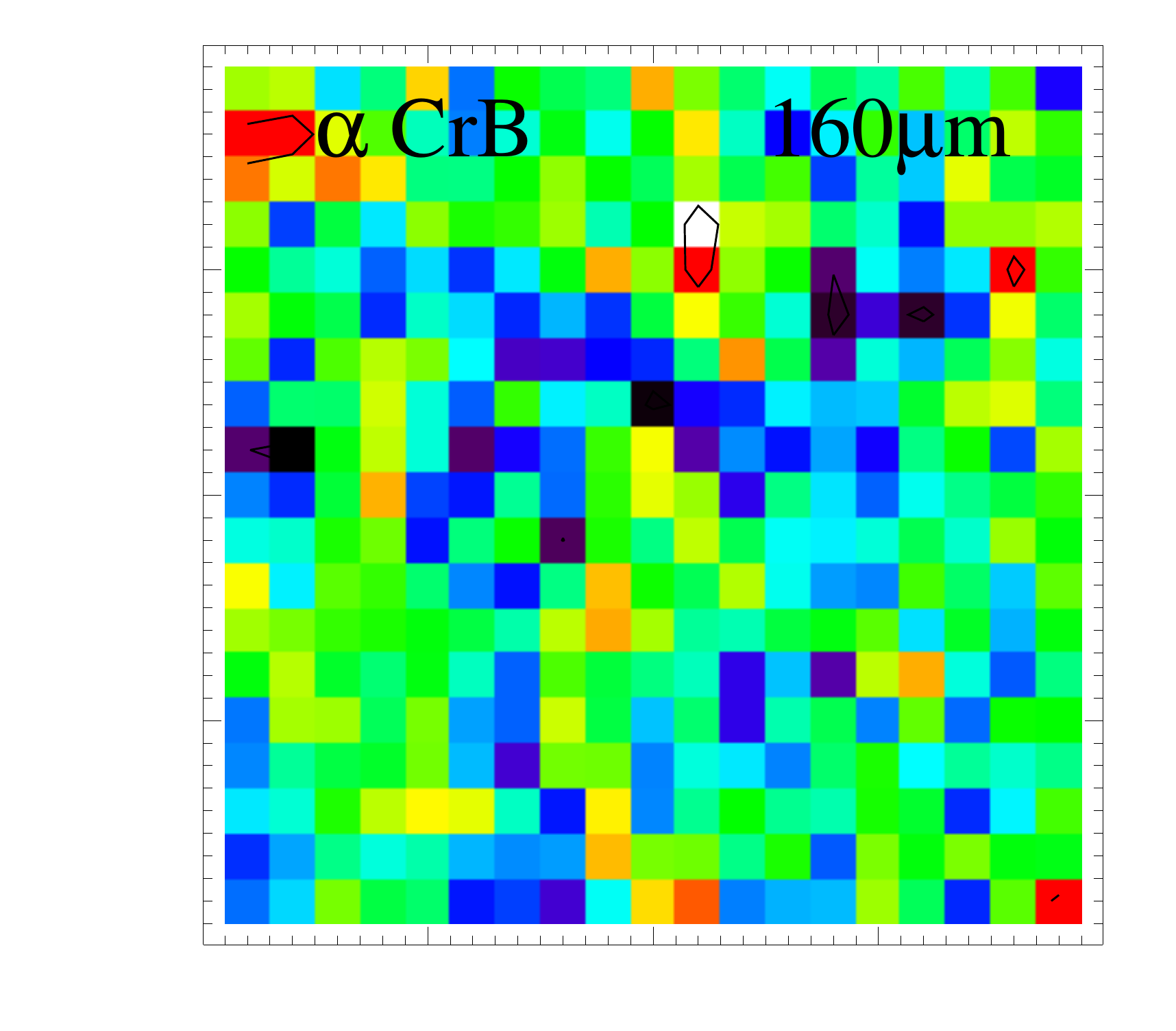} \\

			\vspace{-0.2in}
      \hspace{-0.5in} \includegraphics[width=0.31\textwidth]{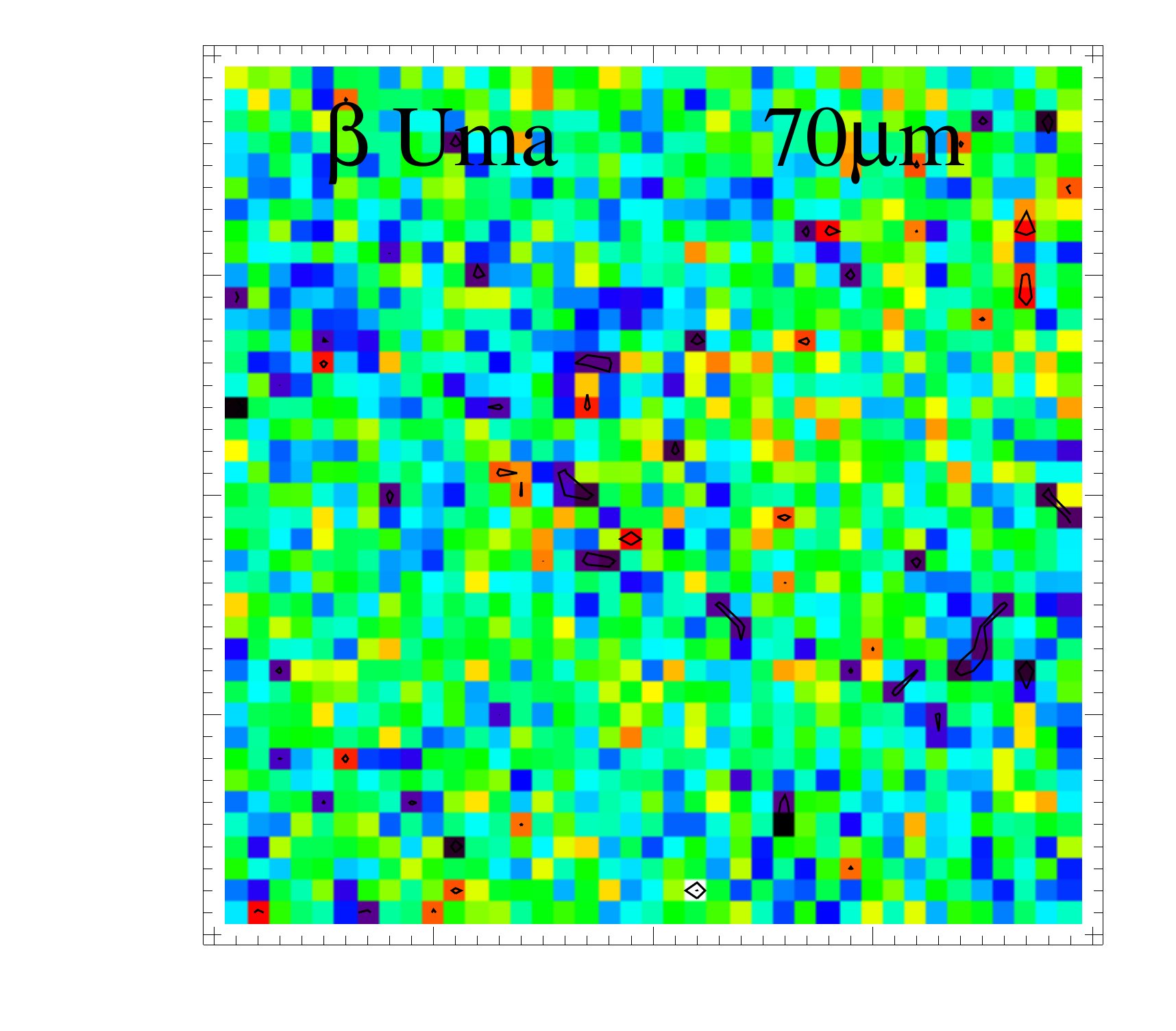} &
      \hspace{-0.5in} \includegraphics[width=0.31\textwidth]{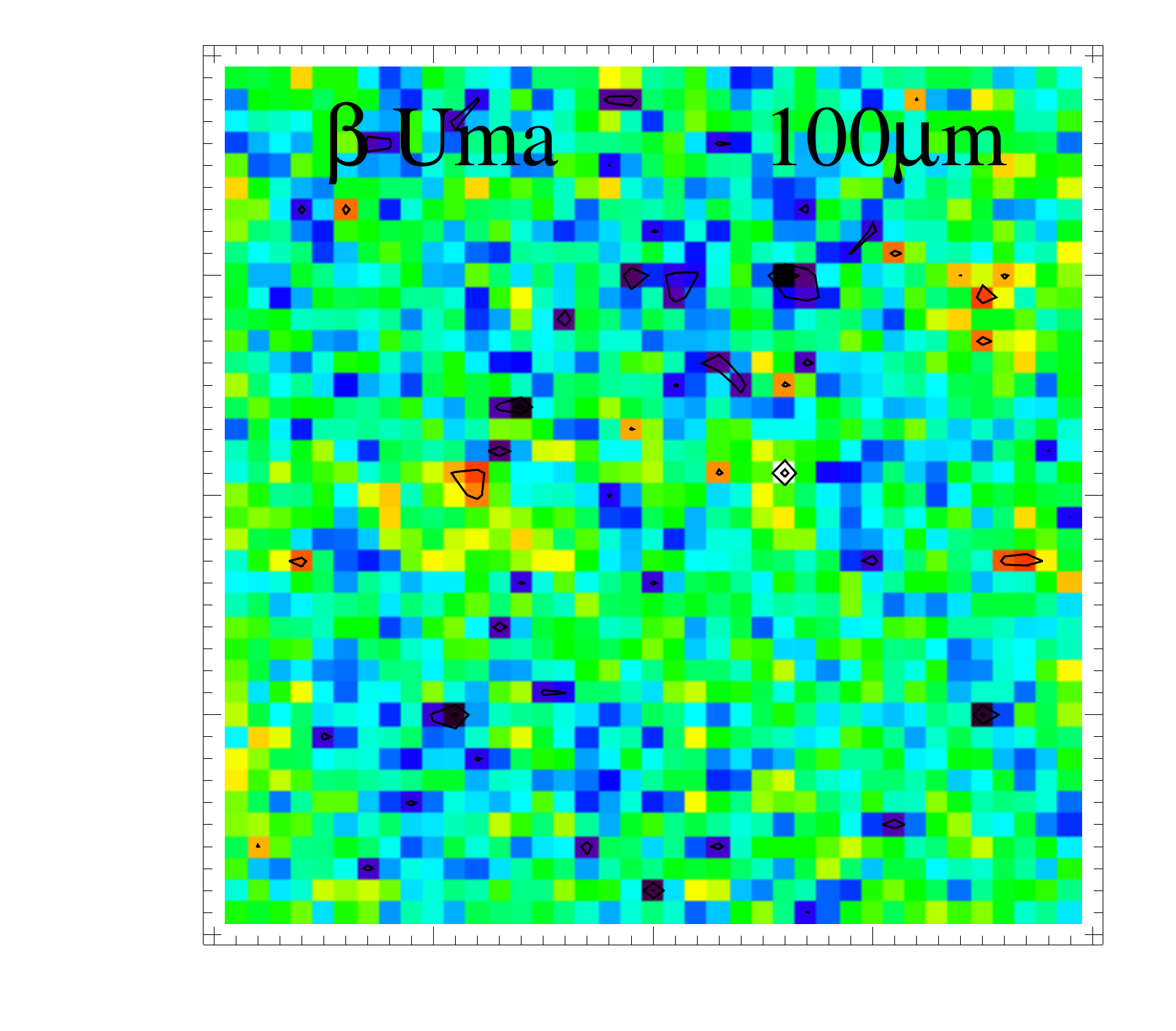} &
      \hspace{-0.5in} \includegraphics[width=0.31\textwidth]{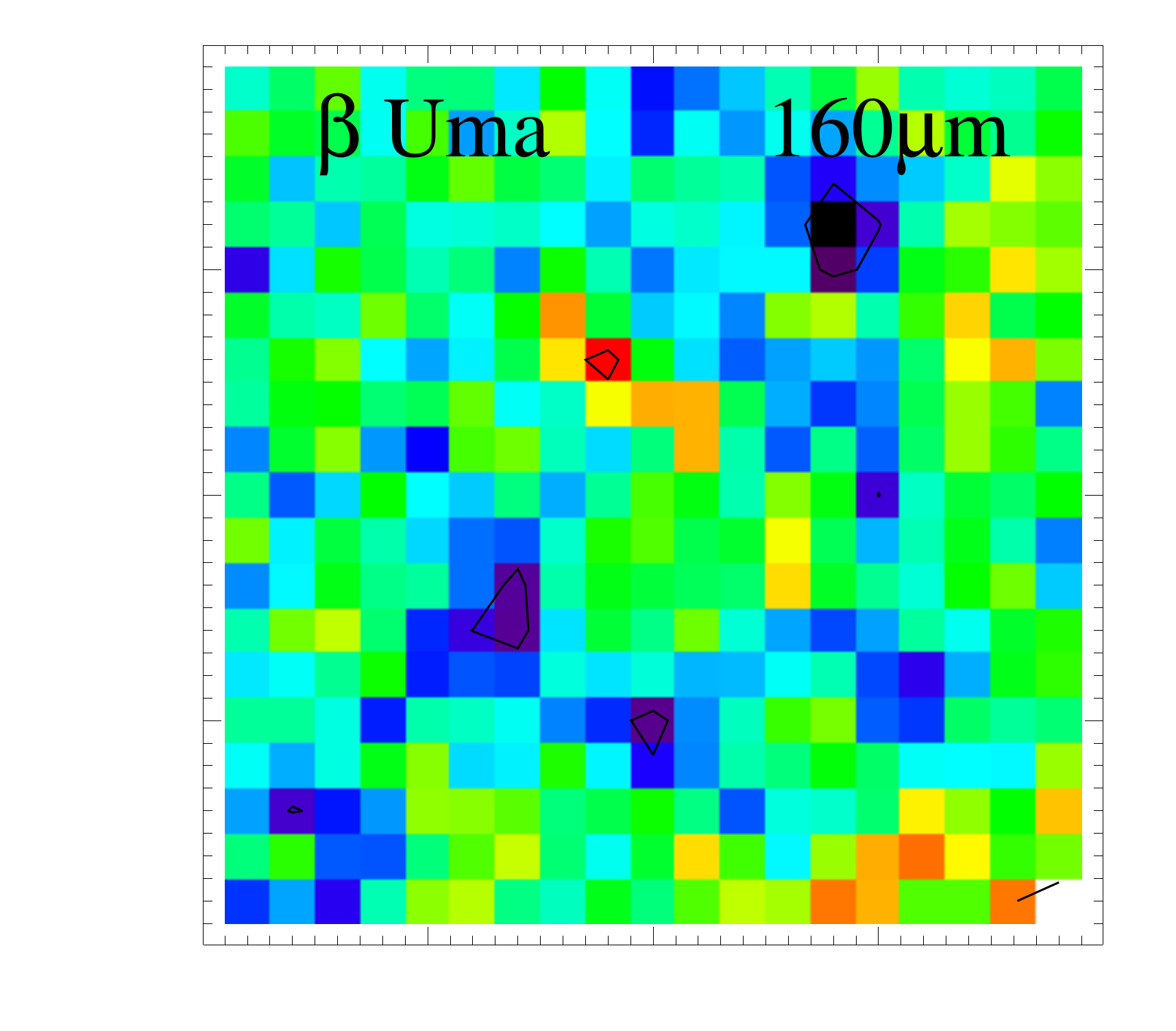} \\
      
			\vspace{-0.2in}
      \hspace{-0.5in} \includegraphics[width=0.31\textwidth]{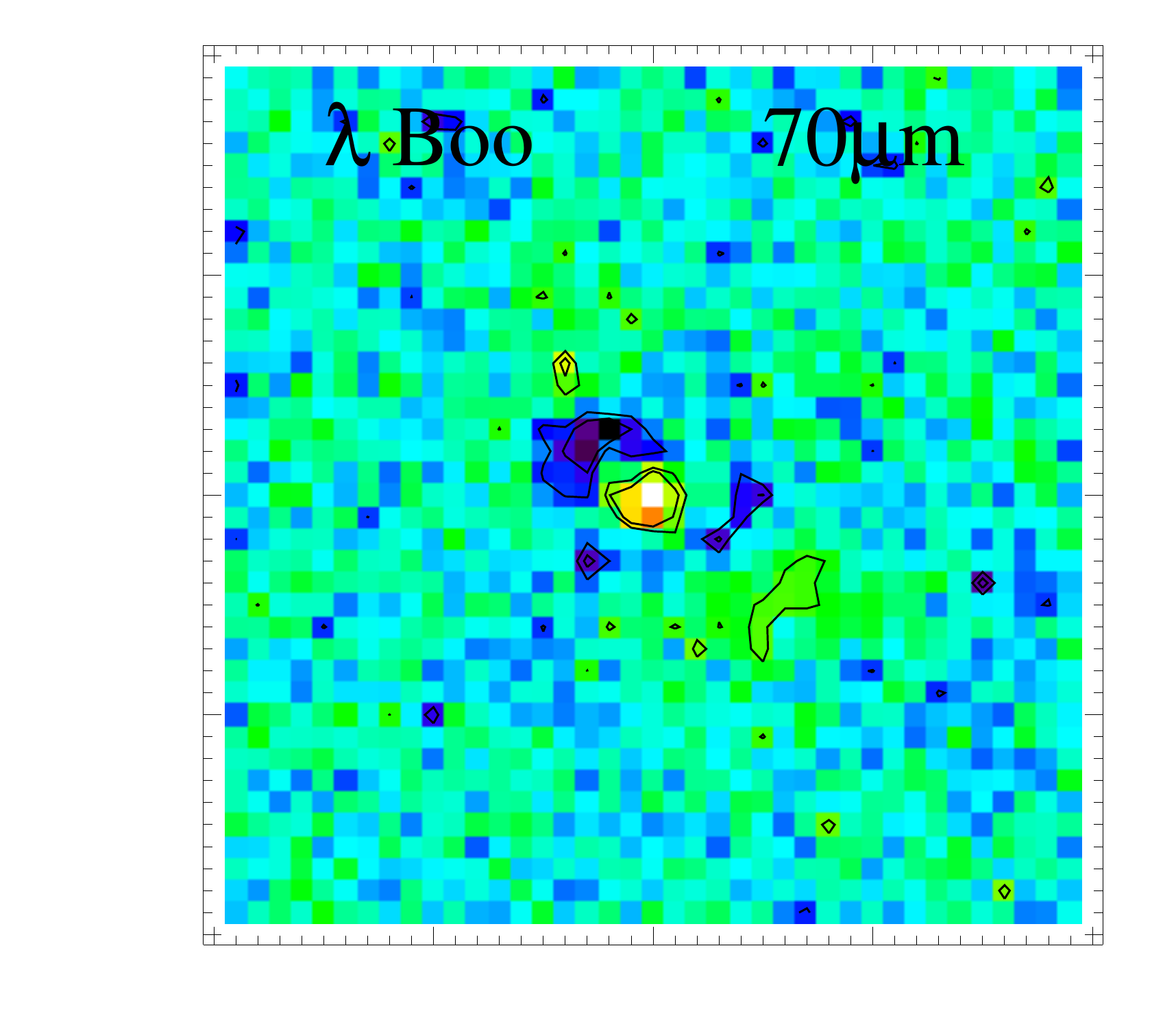} &
      \hspace{-0.5in} \includegraphics[width=0.31\textwidth]{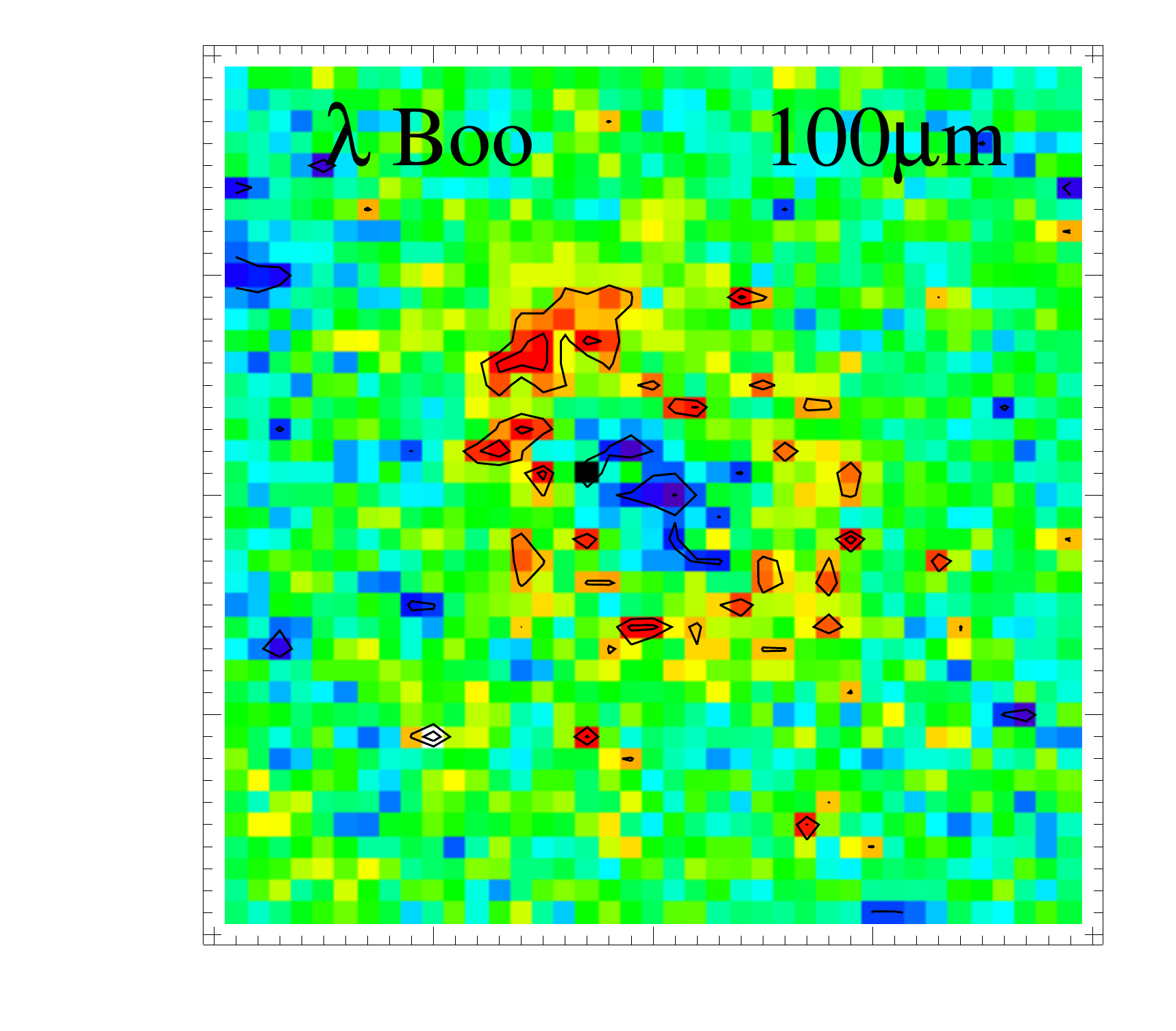} &
      \hspace{-0.5in} \includegraphics[width=0.31\textwidth]{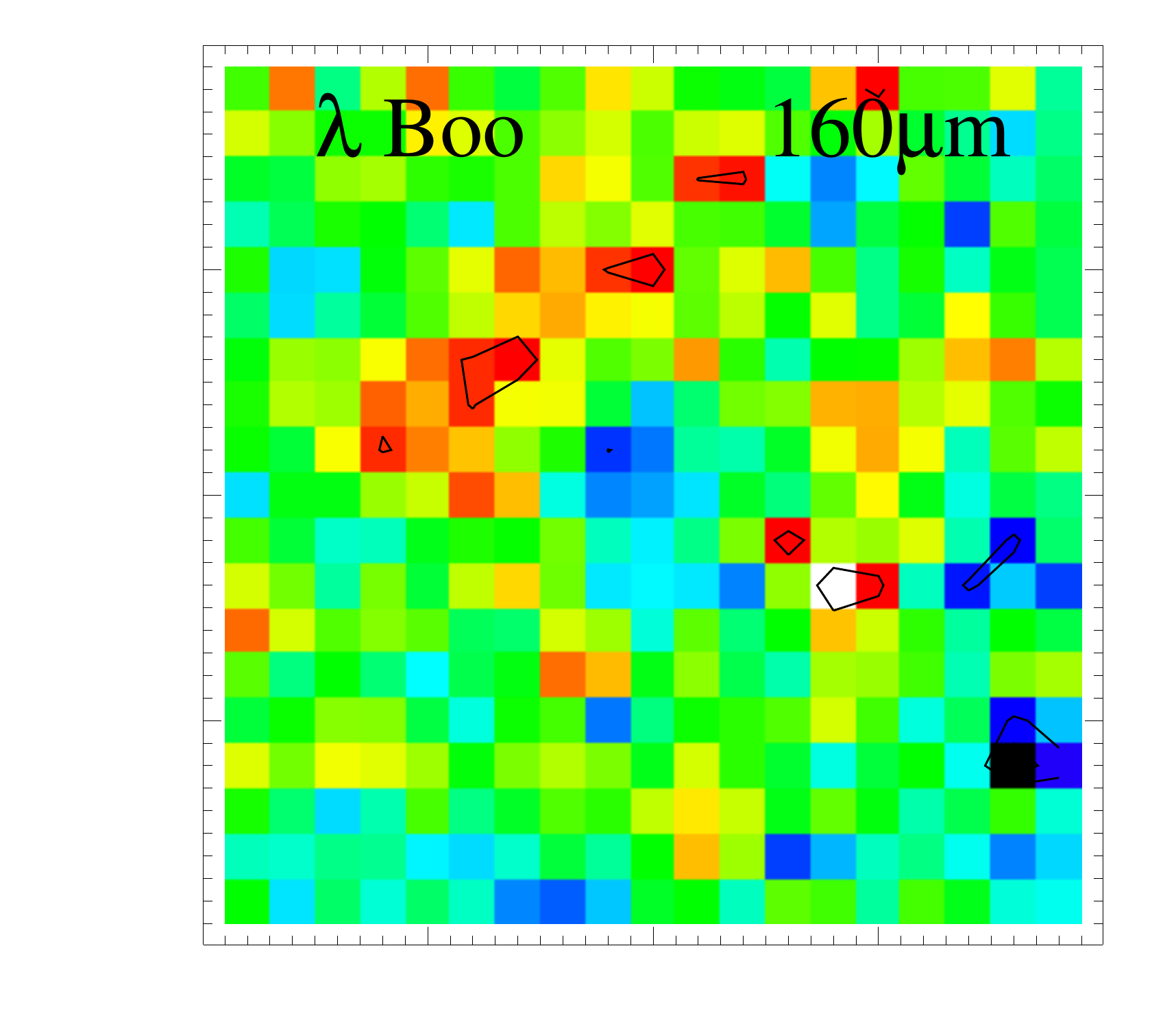} \\
      
 			\vspace{-0.2in}
      \hspace{-0.5in} \includegraphics[width=0.31\textwidth]{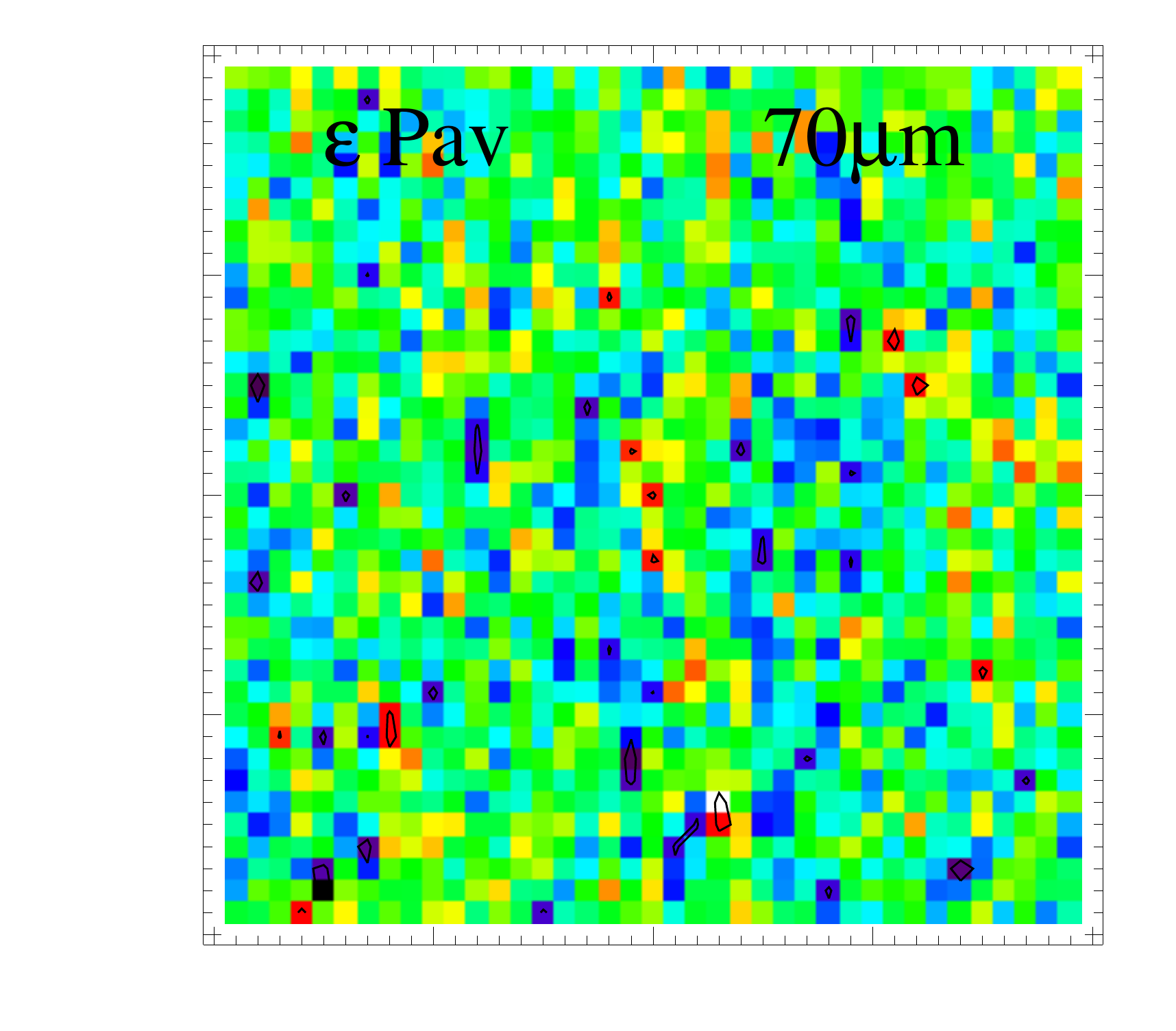} &
      \hspace{-0.5in} \includegraphics[width=0.31\textwidth]{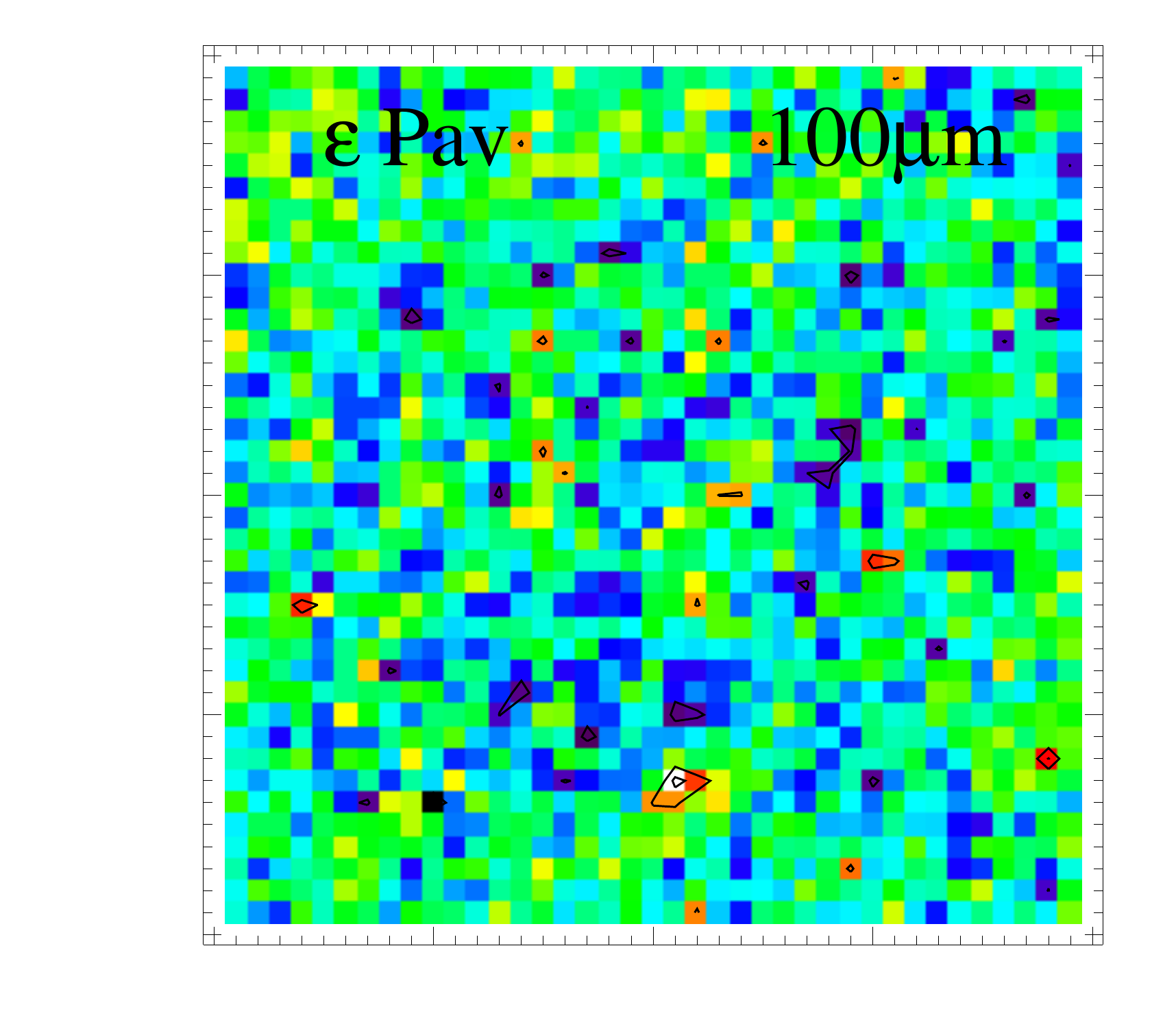} &
      \hspace{-0.5in} \includegraphics[width=0.31\textwidth]{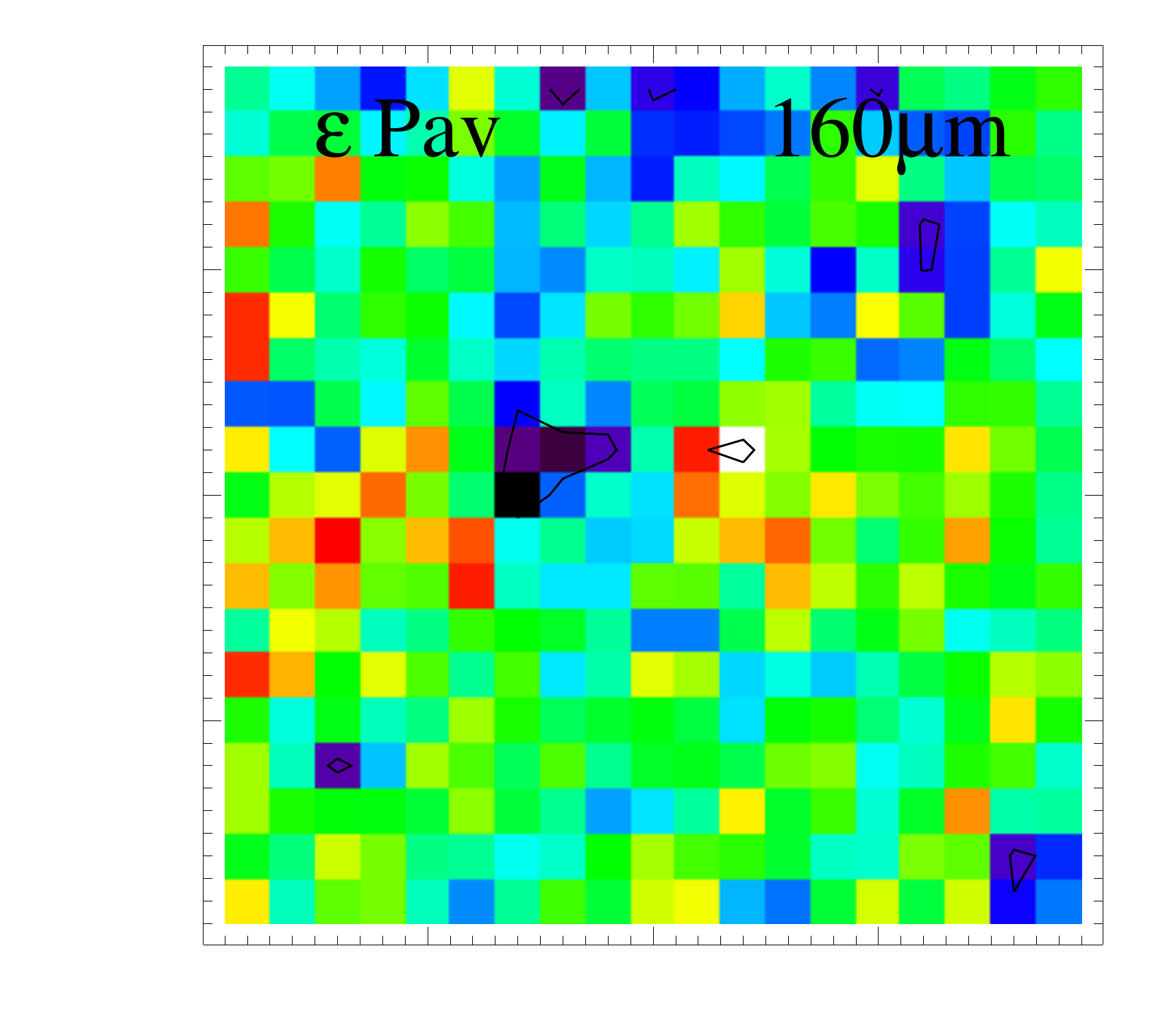} \\
      
			\vspace{-0.2in}
      \hspace{-0.5in} \includegraphics[width=0.31\textwidth]{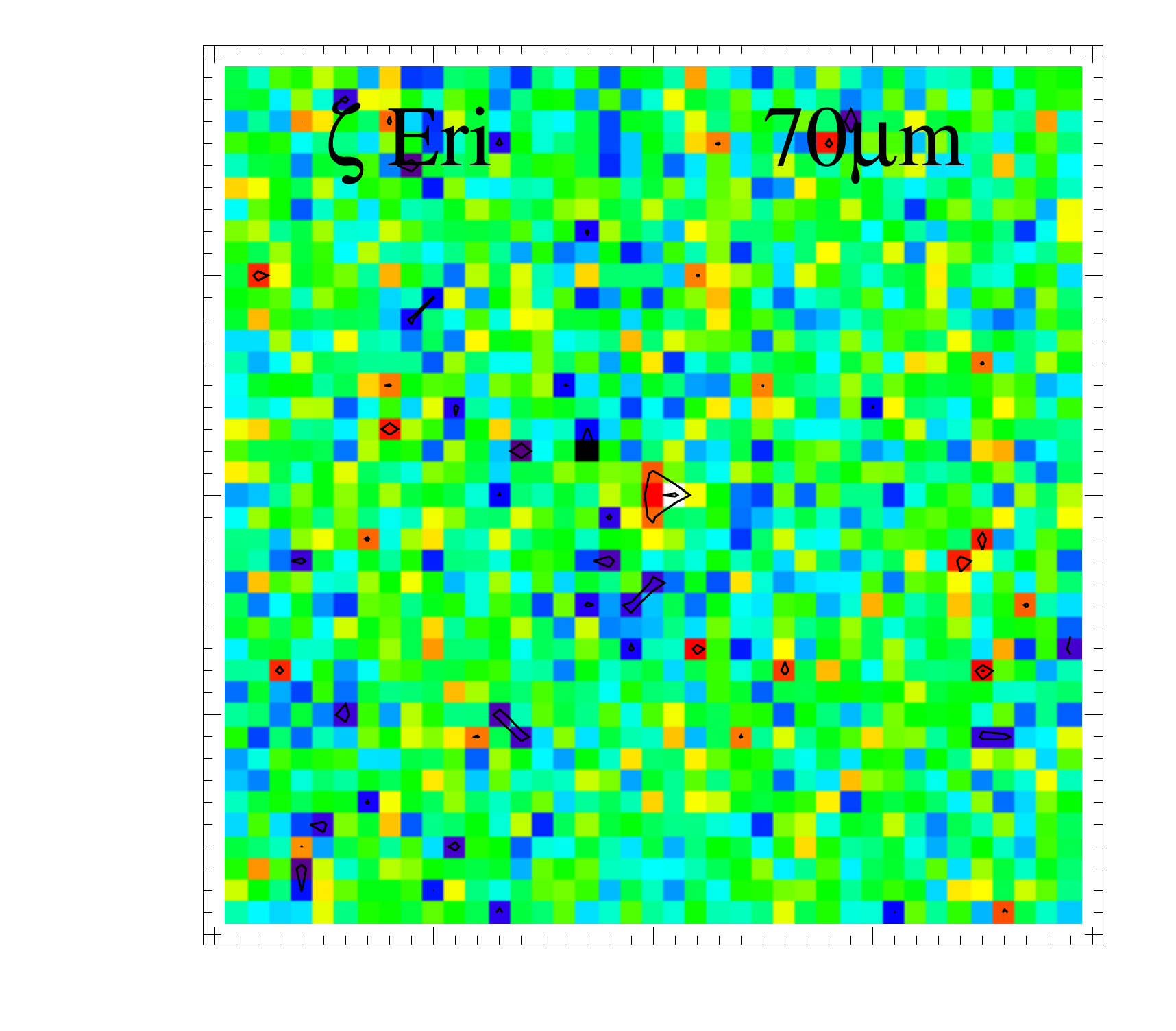} &
      \hspace{-0.5in} \includegraphics[width=0.31\textwidth]{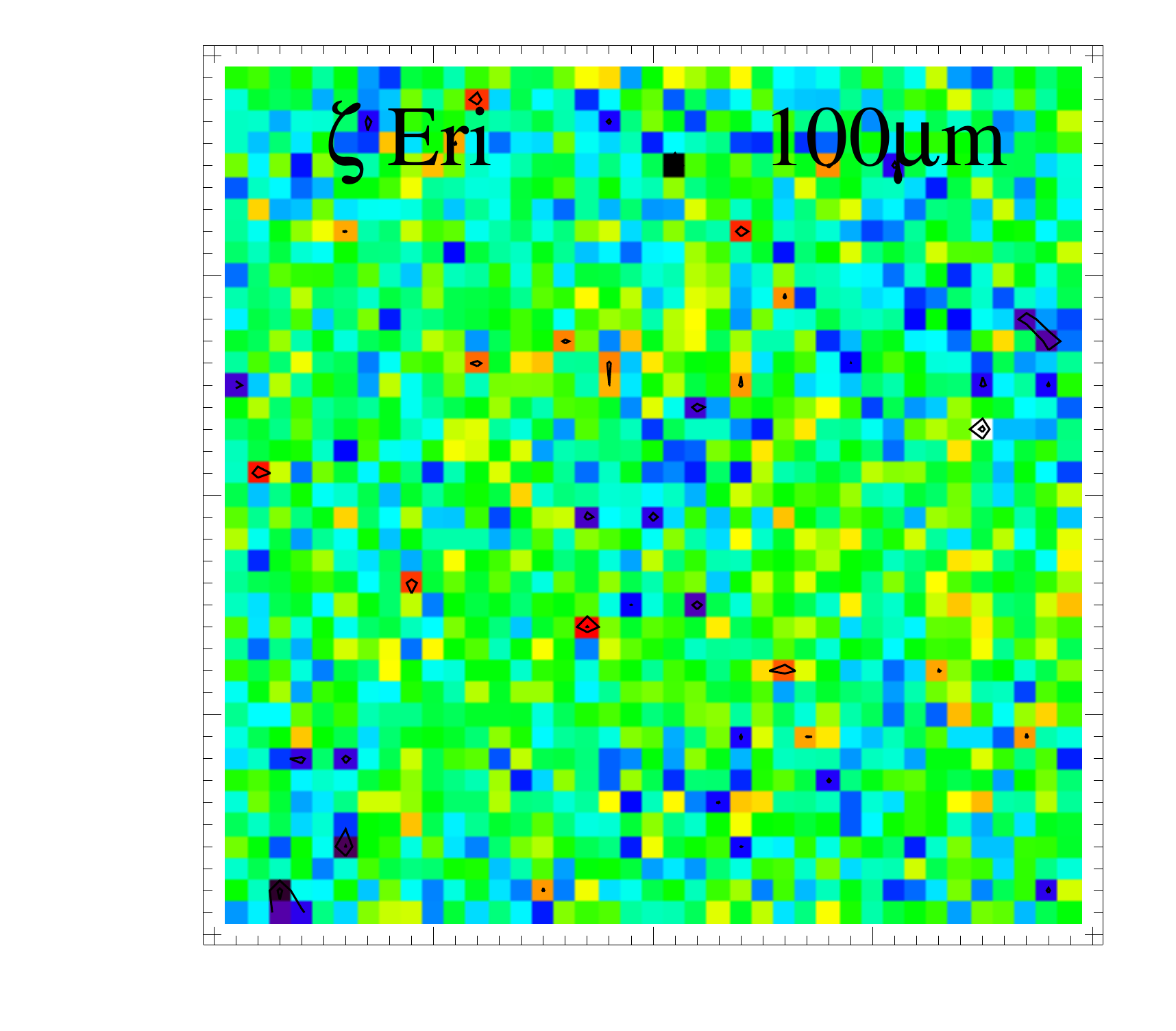} &
      \hspace{-0.5in} \includegraphics[width=0.31\textwidth]{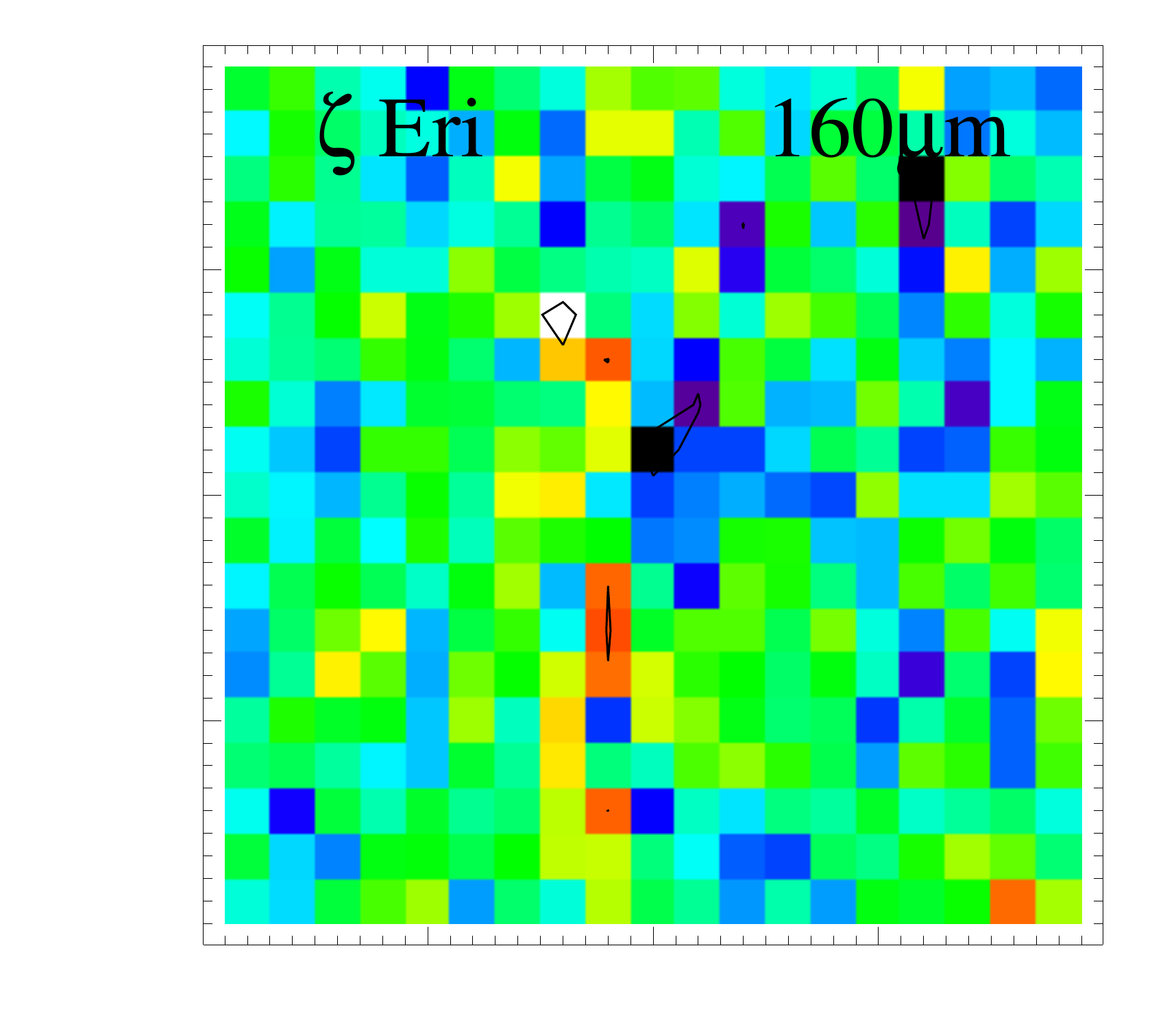} \\
          \end{tabular}

    \caption{Residuals after the model that best-fits all wavelengths has been subtracted from the image. 70$\mu$m images are on the left, 100$\mu$m images in the middle and 160$\mu$m on the right. The contours represent $\pm$2 and 3-$\sigma$ significance.}

   \label{fres100}

  \end{center}

\end{figure*}
\begin{figure*}

  \begin{center}


    \begin{tabular}{ccc}

			\vspace{-0.2in}
      \hspace{-0.5in} \includegraphics[width=0.31\textwidth]{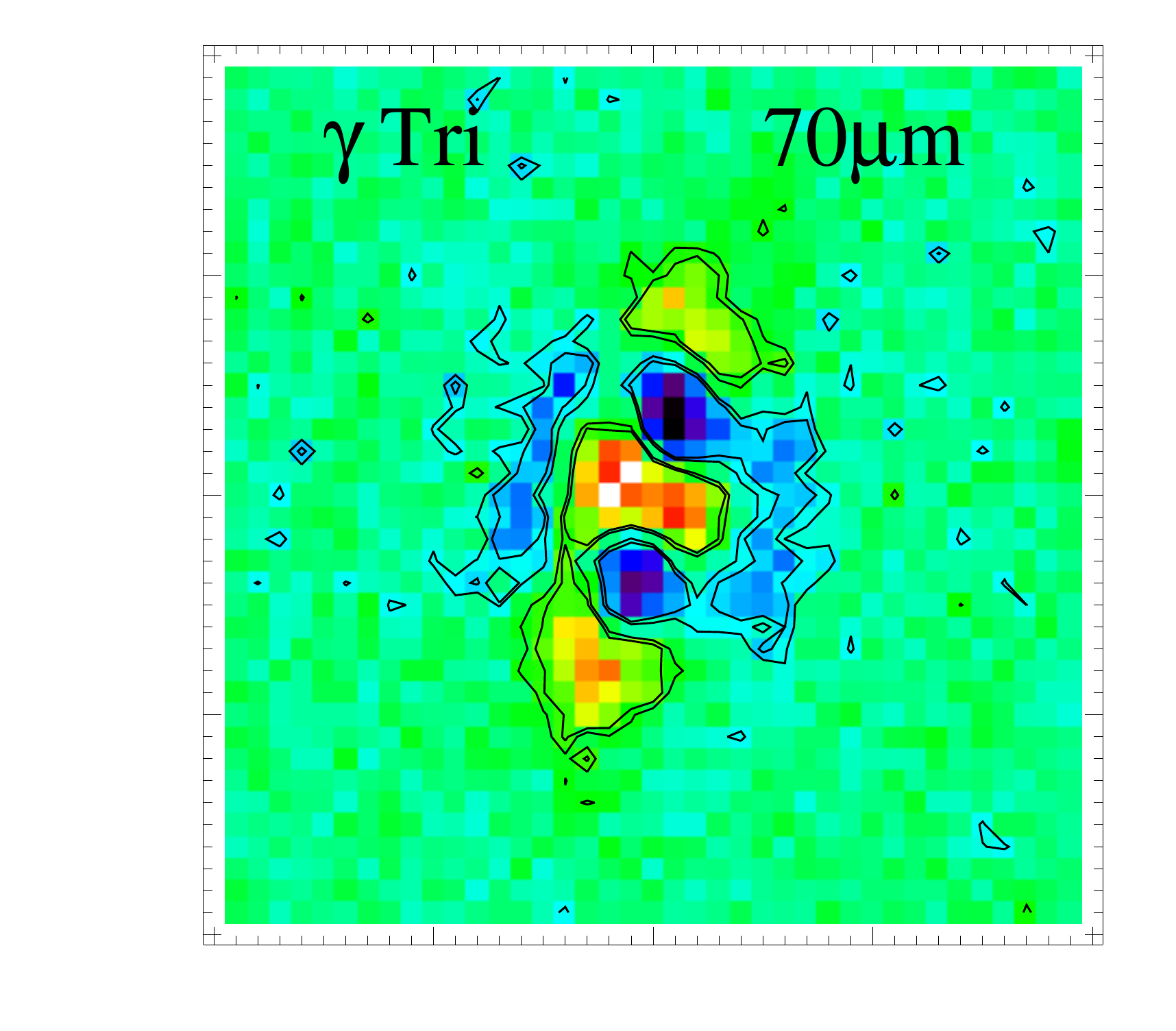} &
      \hspace{-0.5in} \includegraphics[width=0.31\textwidth]{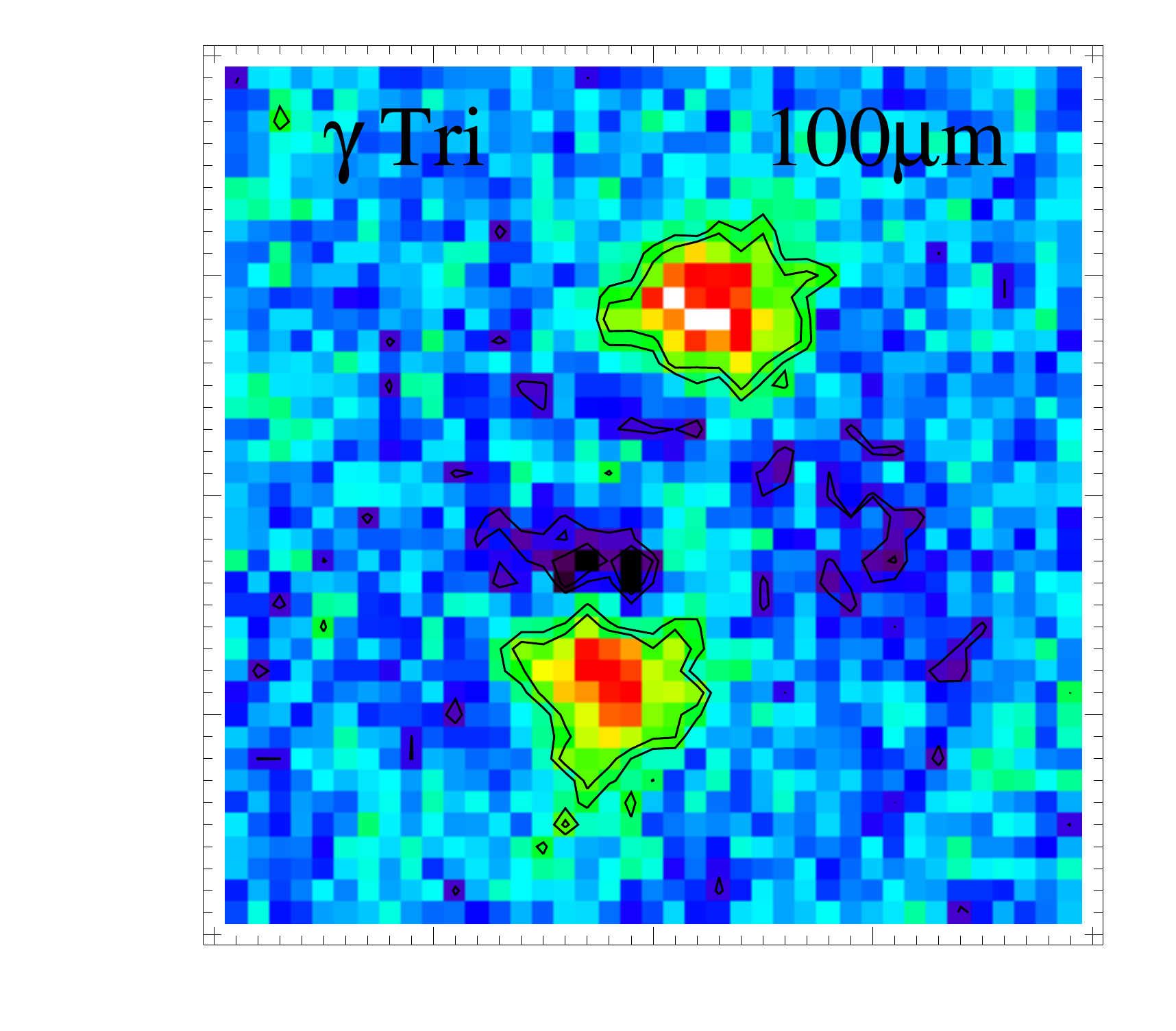} &
      \hspace{-0.5in} \includegraphics[width=0.31\textwidth]{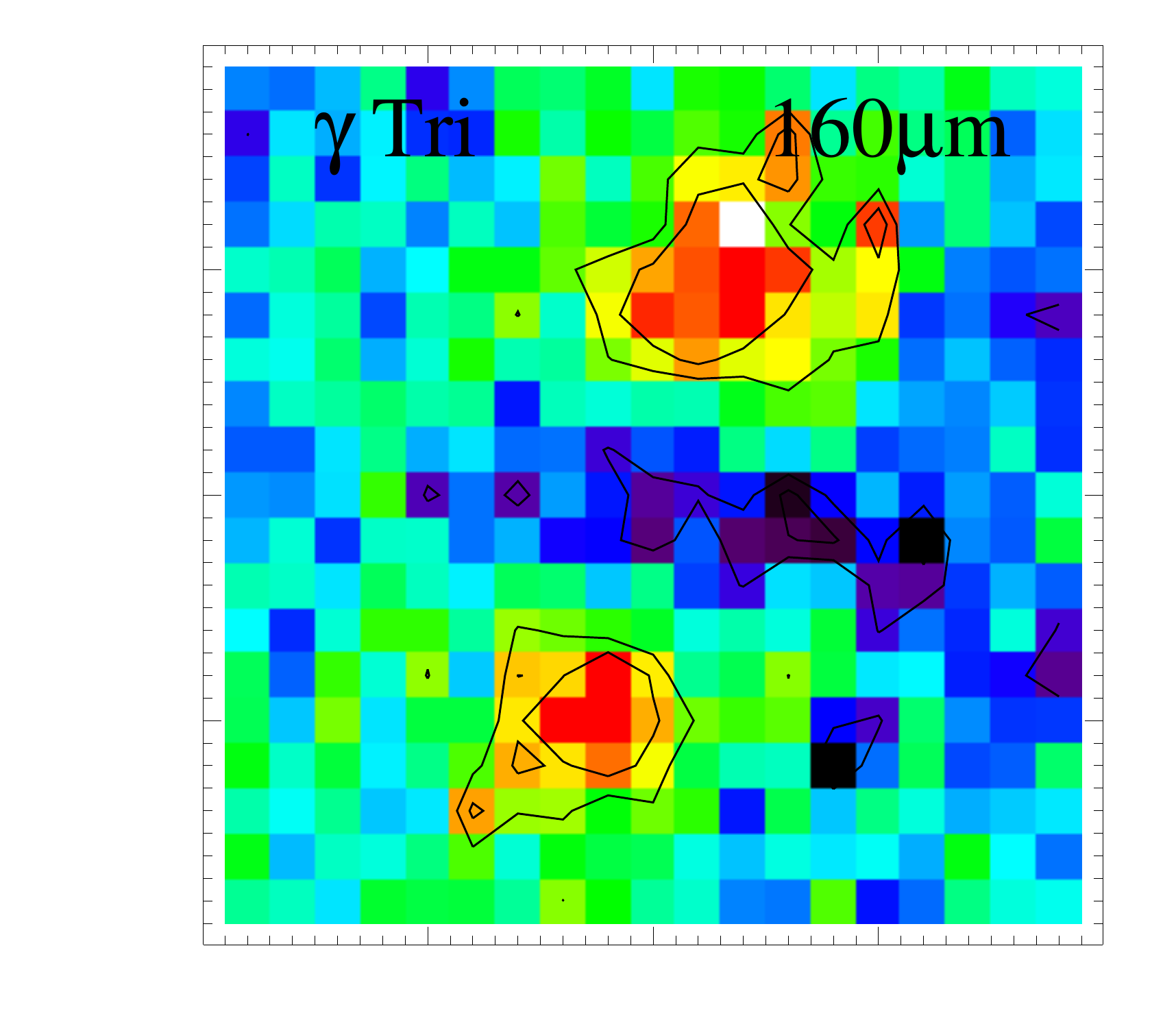} \\
      
			\vspace{-0.2in}
      \hspace{-0.5in} \includegraphics[width=0.31\textwidth]{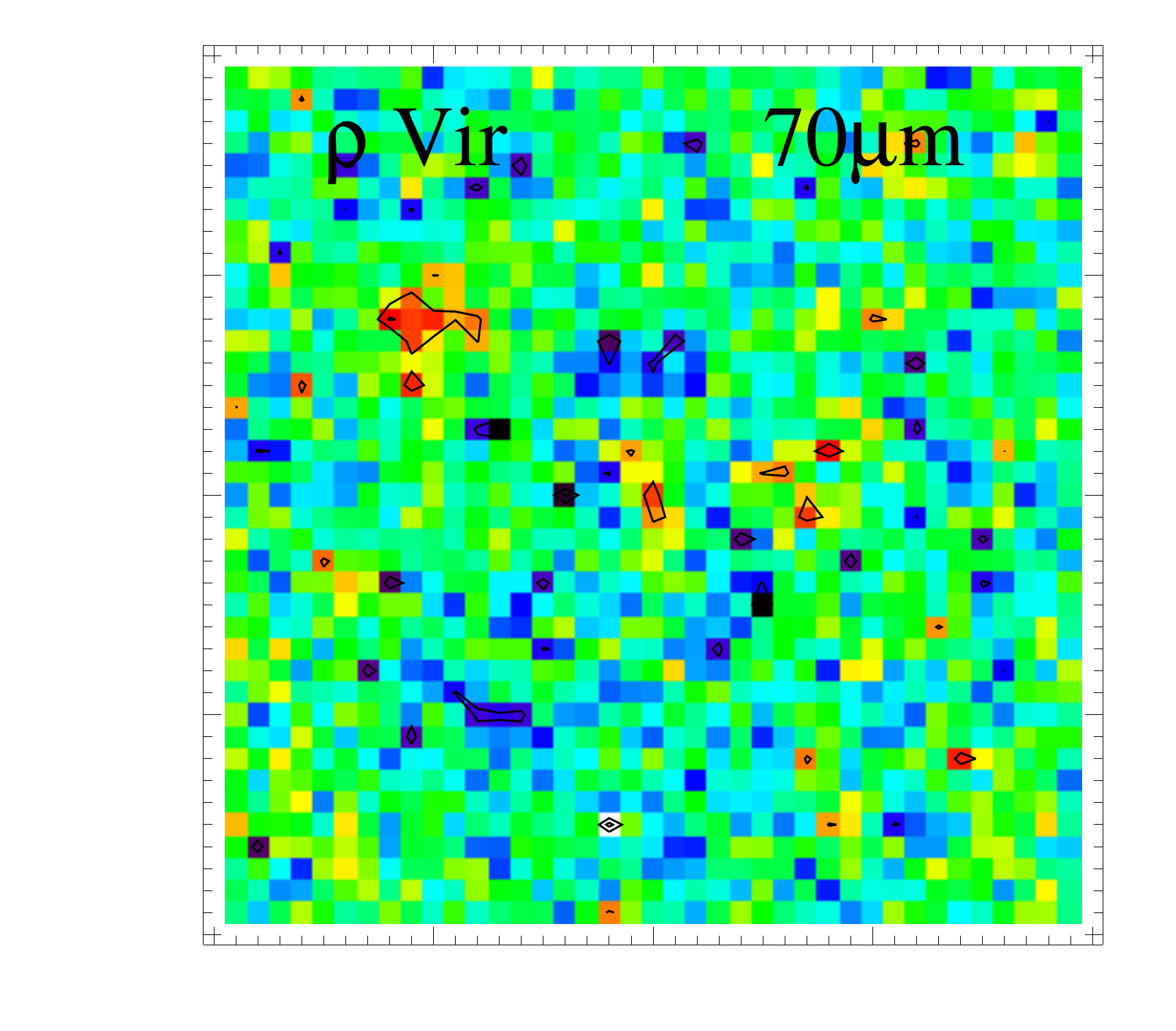} &
      \hspace{-0.5in} \includegraphics[width=0.31\textwidth]{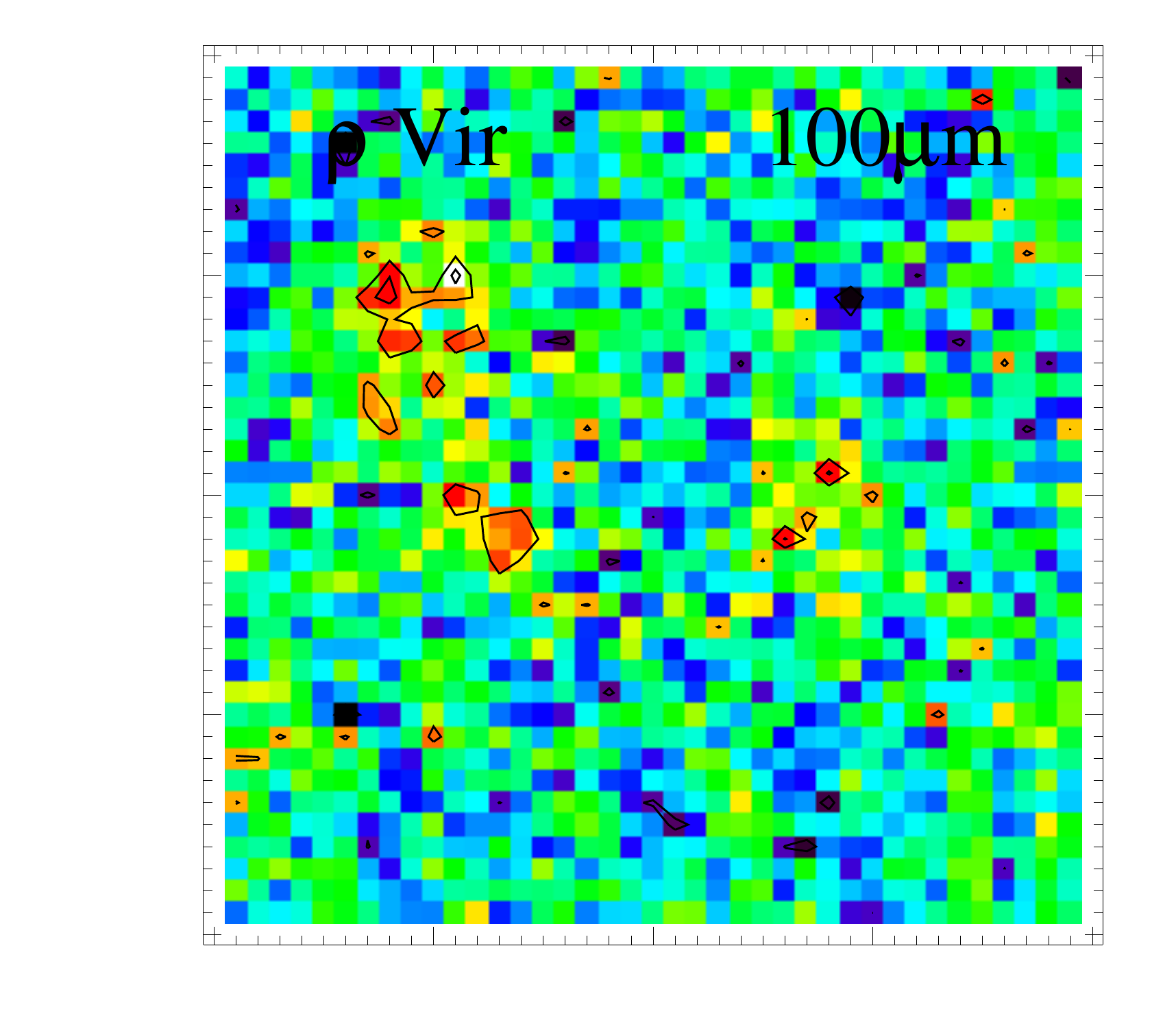} &
      \hspace{-0.5in} \includegraphics[width=0.31\textwidth]{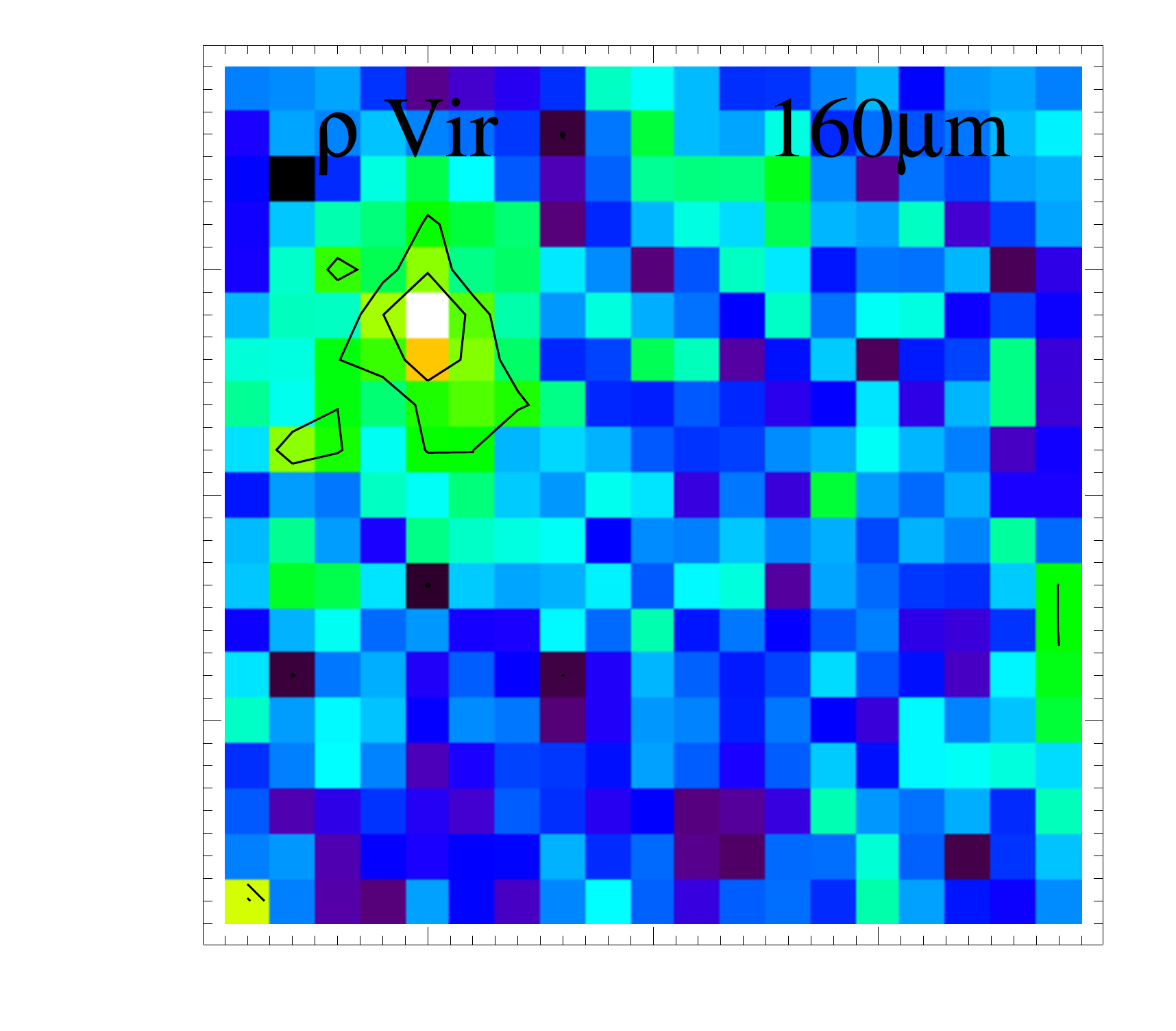} \\
      
			\vspace{-0.2in}
      \hspace{-0.5in} \includegraphics[width=0.31\textwidth]{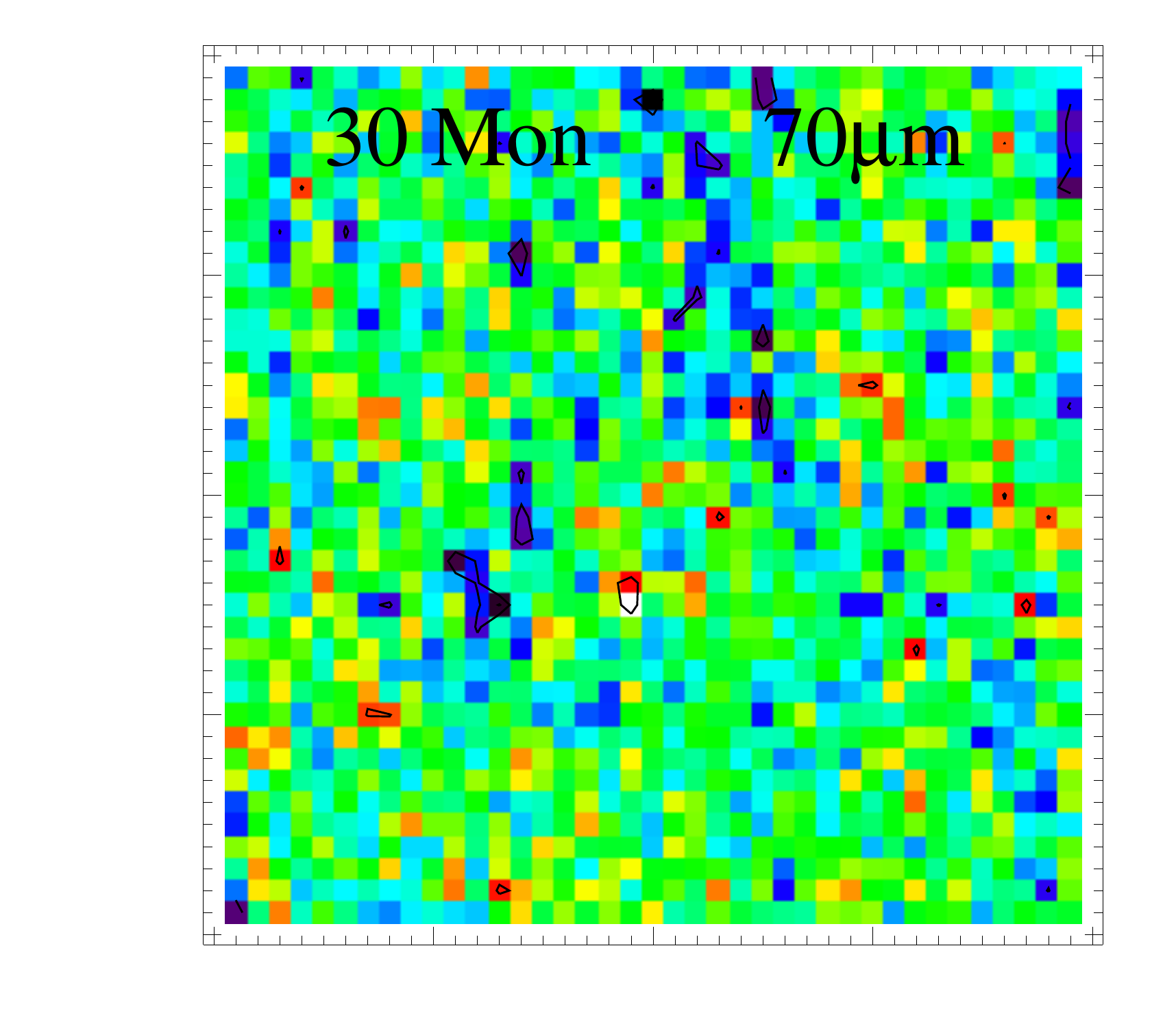} &
      \hspace{-0.5in} \includegraphics[width=0.31\textwidth]{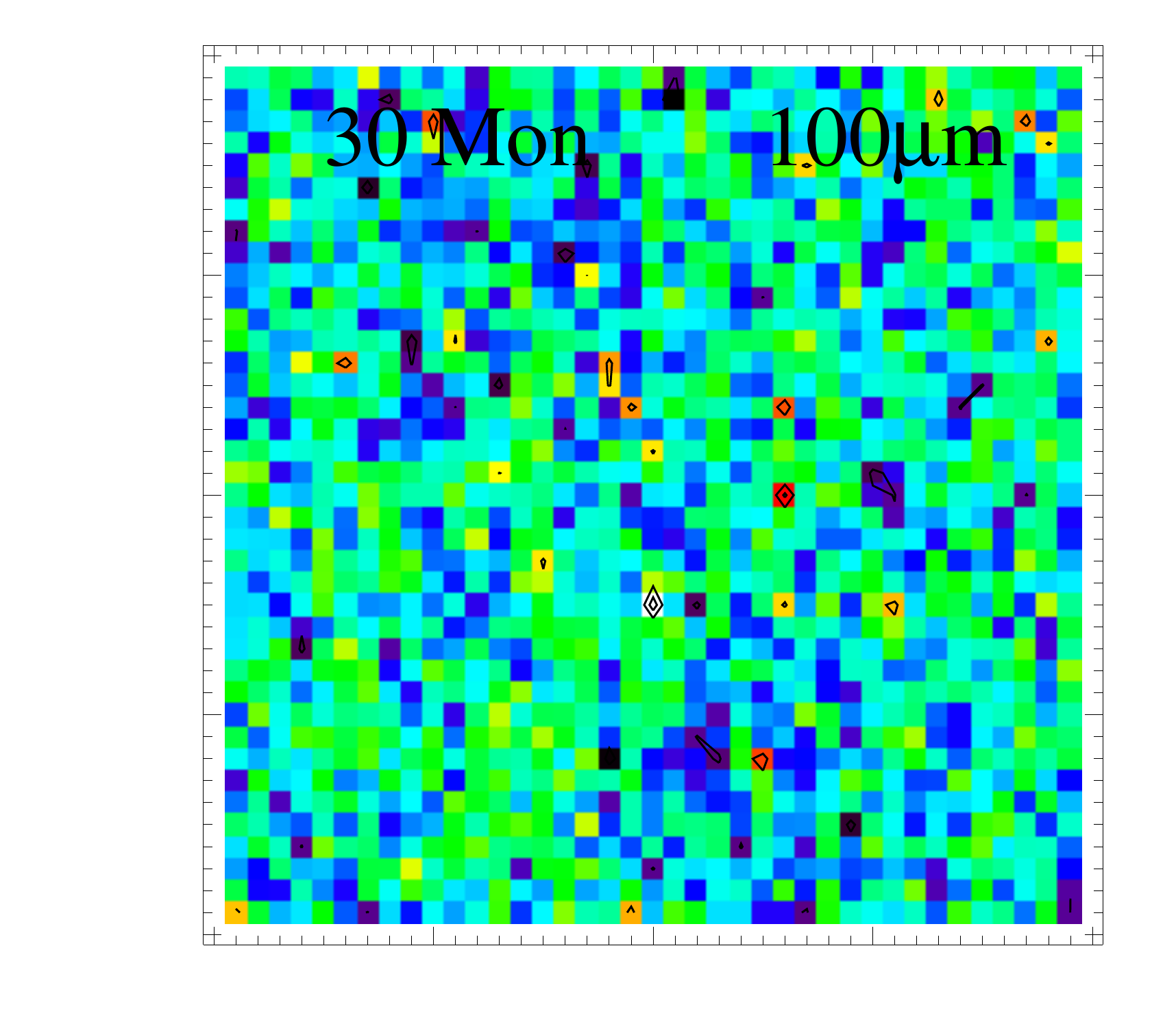} &
      \hspace{-0.5in} \includegraphics[width=0.31\textwidth]{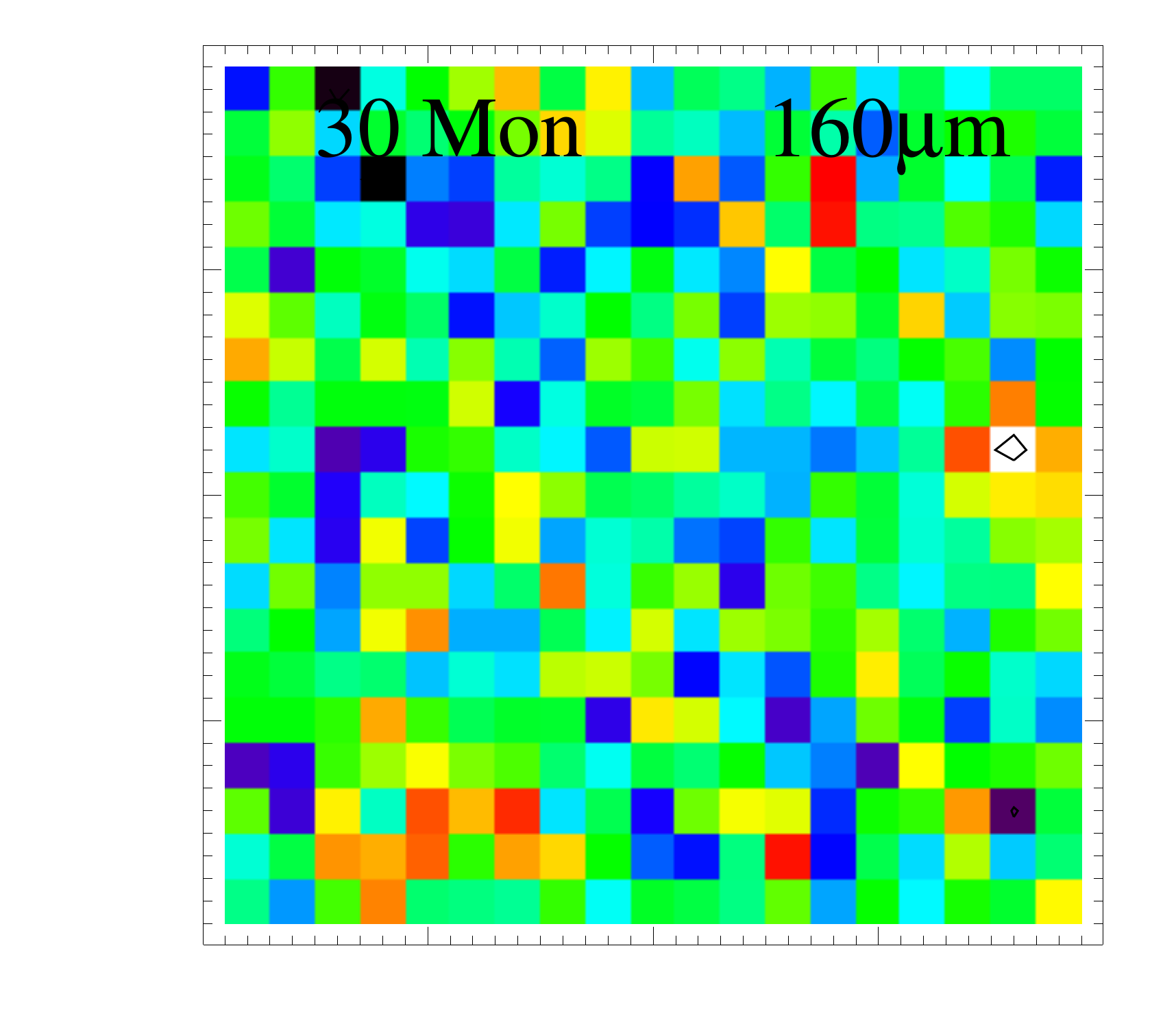} \\
      
			\vspace{-0.2in}
      \hspace{-0.5in} \includegraphics[width=0.31\textwidth]{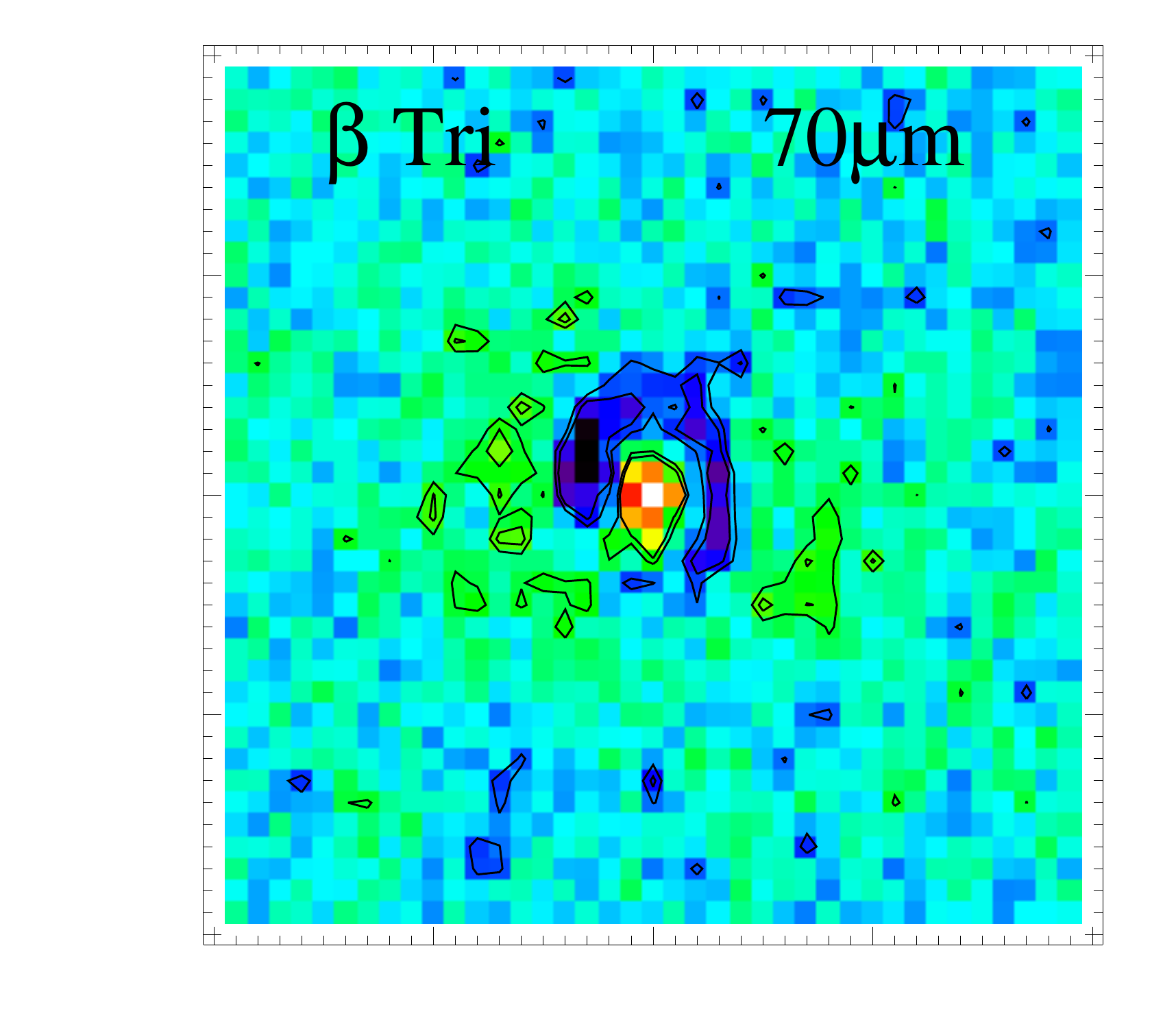} &
      \hspace{-0.5in} \includegraphics[width=0.31\textwidth]{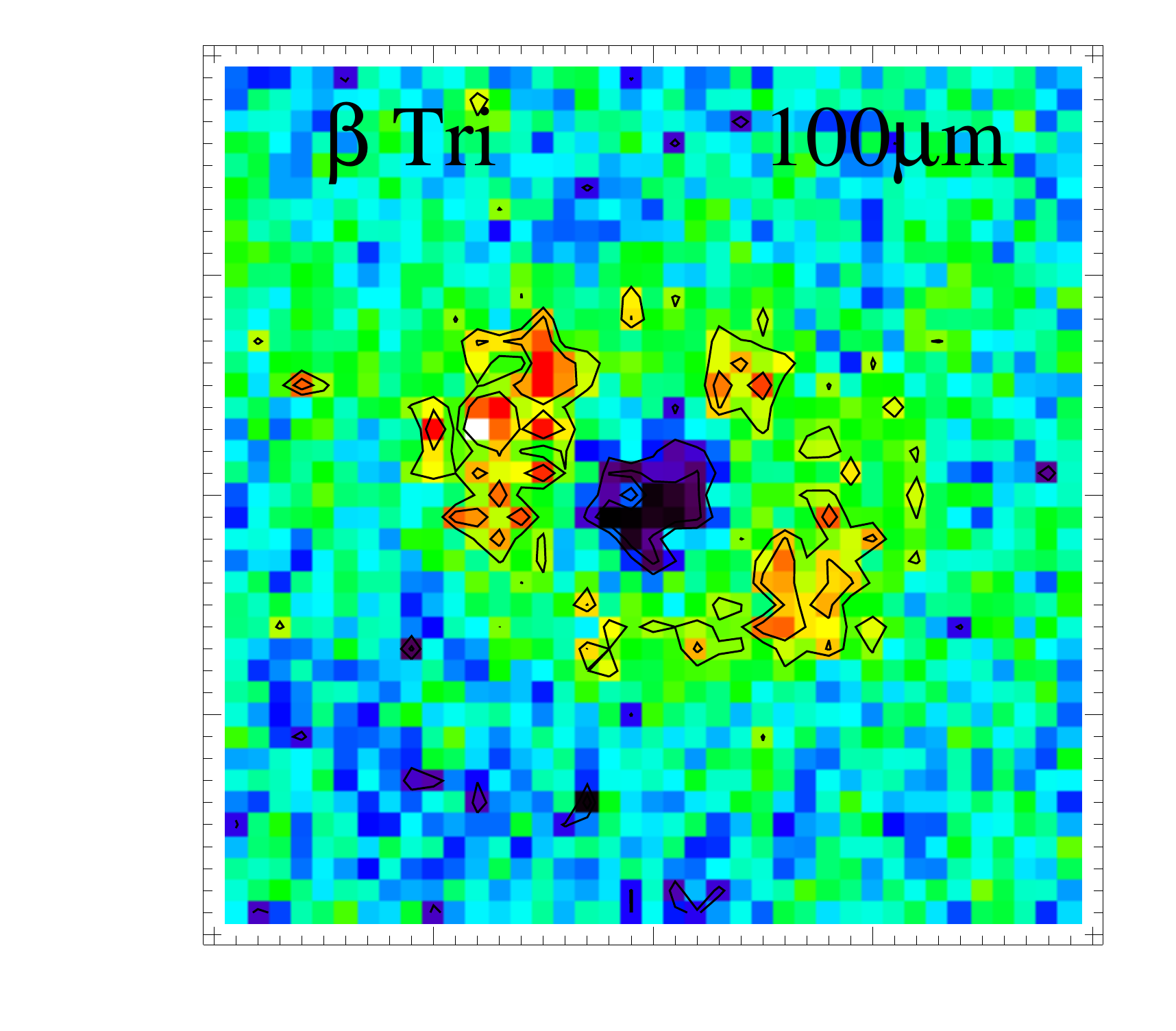} &
      \hspace{-0.5in} \includegraphics[width=0.31\textwidth]{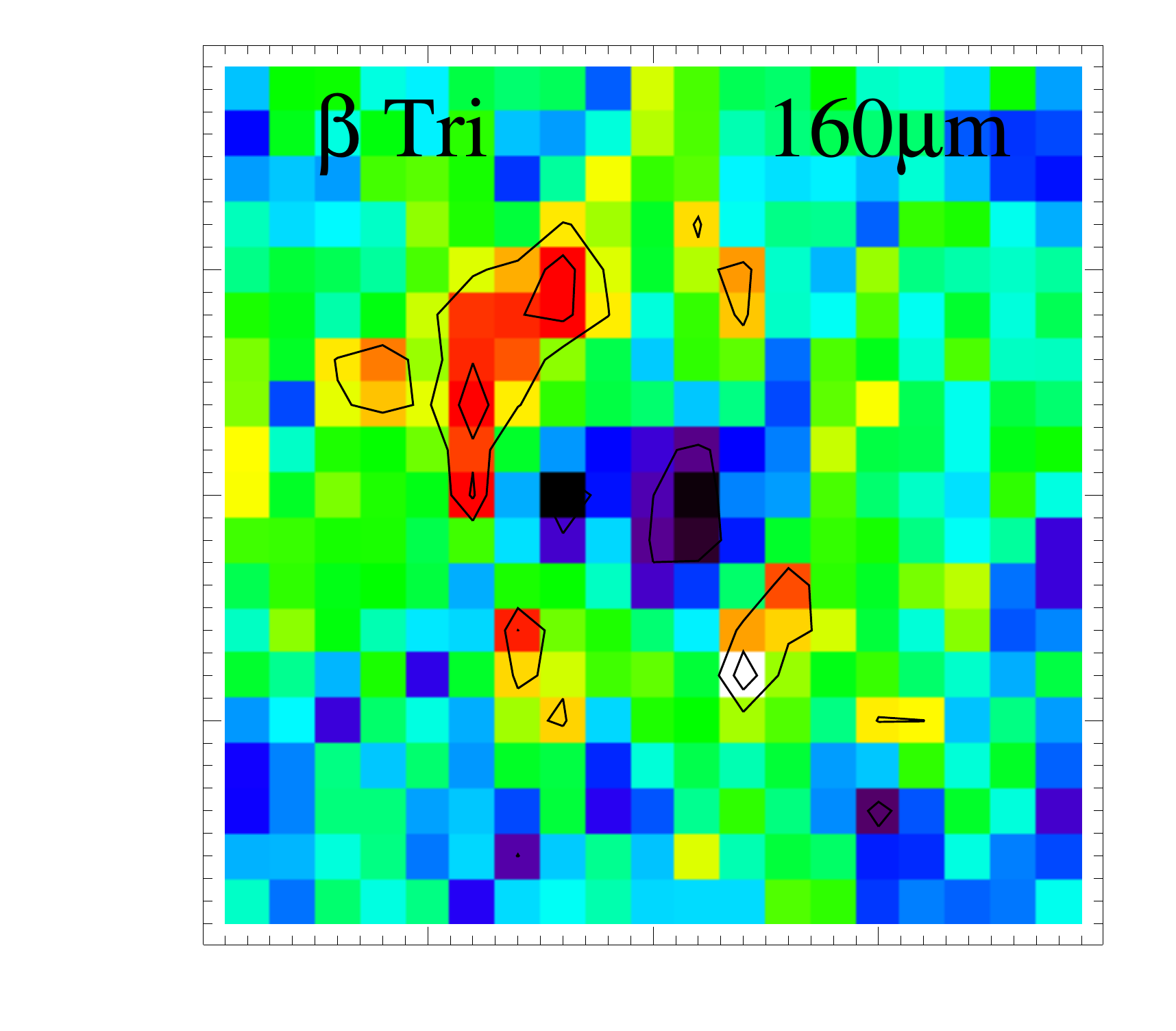} 
      
    \end{tabular}
    \\ \vspace{0.1in}
		\textbf{Figure \ref{fres100}.} (continued)

  \end{center}

\end{figure*}

\subsection{Model creation and fitting}
Since debris disc SEDs are generally well fit by single temperature blackbodies, we begin by assuming that the debris discs in each system can be approximated by a narrow belt at a distance, $R$, from the star with a width equal to 0.1$R$. This width is narrower than for some discs generally considered to be narrow, such as the asteroid and Kuiper belts, but even using a value as high as 0.4$R$ gives roughly similar results. Assuming all the dust is concentrated to a very narrow belt also has the benefit of allowing us to ignore the degeneracies between the distribution of dust with radius and the fall-off of temperature with radius and allows us to keep the number of free parameters low. 

Model fitting is done using a grid-based approach. Using the parameters listed above, a torus is set up and filled with particles. The model disc is then given an inclination to the line-of-sight, $I$, and a rotation, $PA_{ext}$, to set it up as it would be seen from the telescope. An example is shown in figure \ref{fa076ring} (left). A line-of-sight integrator method is then used to create the image \citep{wyatt99}. For each disc size and orientation the best fitting flux is then found for the model using the least squares fitting code MPFIT \citep{markwardt09}\footnote{Flux was not included as a grid parameter as it is strongly constrained by the images so would require a very fine grid to accurately determine its value and it is also much less likely that using a Levenberg-Marquardt technique to find the minimum $\chi^2$ might encounter a local minima, which is one of the main reasons for using a grid method to find the other parameters.}. The expected photospheric flux of the star is added in the central pixel and the resulting image is then convolved with the PSF. An example is shown in figure \ref{fa076ring} (right). Note the expected photospheric flux has to be scaled down to account for the filtering of the large scale beam (see section \ref{sobs}). 

For the majority of the stars $\alpha$ Boo was used as the PSF reference star. However, there is some variation at 70~$\mu$m between the PSF stars as noted in section \ref{sobs}. This variation can lead to differences in the quality of a fit, so we use the PSF that provides the best fit for a particular star. It was found that using $\gamma$ Dra as the 70~$\mu$m PSF reference star for $\alpha$ CrB and $\beta$ Uma resulted in an improved fit \citep[see][]{kennedy12a}.

In the cases of $\gamma$ Tri, 30 Mon and $\rho$ Vir, the SEDs indicate the presence of warm dust (see section \ref{ssedfit}). For the SED fitting we assumed this warm dust to indicate the presence of a narrow belt close to the star analogous to the asteroid belt. We could include this as a fixed flux component in the image modelling, however, since doing so involves extrapolating from an already uncertain mid-IR flux to PACS wavelengths, we chose to ignore this component. Doing so does not have a significant effect on our results as the inner components' fluxes are relatively low at these wavelengths.

This is all then rerun for different values of $R$, $I$ and $PA_{ext}$. A grid of around 50,000 models was used for each system, the exact number depending on how fine a grid was needed to accurately determine the errors on the parameters. 

For each grid model at each wavelength the reduced $\chi^2$ is found
\begin{eqnarray}
\chi_{\rm{red}}^2=\frac{\sum^{n_{\rm{p}}}_{i=0}((F_{\rm{im}}(i)-F_{\rm{sim}}(i))/\sigma)^2}{n_{\rm{p}}-n_{\rm{param}}-1}
\end{eqnarray}
where $\sigma$ is the pixel to pixel RMS, $F_{\rm{im}}(i)$ and $F_{\rm{sim}}(i)$ are the fluxes in pixel $i$ of the image and simulation respectively, $n_{\rm{p}}$ is the number of pixels used for the fitting process and $n_{\rm{param}}$ is the number of free parameters. The best fitting model is then the model with the minimum $\chi_{\rm{red}}^2$. The pixels used for this calculation are those within 1.5$\times$FWHM of the resolved disc (where this is calculated from a Gaussian fit) that also have a flux density greater than the 1-$\sigma$ pixel to pixel error. To determine whether or not this best fit model is in fact a good fit to the data, i.e. whether or not a single narrow ring is a good fit, we define a cut-off value for the $\chi_{\rm{red}}^2$ as the value for which there is only a 5\% probability that such a model is correct. Any model with a $\chi_{\rm{red}}^2$ greater than this is considered to be incorrect.

As mentioned in section \ref{sobs}, $\rho$ Vir has a background source very close to the star. If not taken into account, this can easily influence the fitted parameters. To stop this background source from influencing the model fitting, this region is masked for the purposes of calculating the $\chi^2$.

1-$\sigma$ errors on the parameters are then calculated by finding the maximum and minimum parameters for models with a $\chi^2$ within $\Delta(\chi^2)$ of the minimum $\chi^2$, where $\Delta(\chi^2)=3.53$ when there are 3 parameters \citep[see chapter 15 of][]{press92}. These parameter uncertainties work on the assumption that the best-fit model is the correct description of the disc. Therefore, they do not take into account disc geometries different to a narrow ring or the real optical properties of the grains and so the quoted uncertainties are likely to be somewhat underestimated. For this reason, in the cases where a disc has a best-fit model with a $\chi^2$ too high for the model to be considered believable, we do not provide parameter errors as they would be meaningless.

The parameters for the best-fit models are shown in table \ref{tbestfit}. This shows the best fits to each of the wavelengths individually and the grid model that best fits all of the wavelengths is shown in the rows marked `all'. The residuals left after the `all' best-fit model has been subtracted from the observations are shown in figure \ref{fres100}. Discussion of individual systems is left for section \ref{sindiv}.

\begin{table*}
\begin{minipage}{166mm}
	\caption{Best-fit parameters from modelling of the resolved images. The errors on the parameters are 1-$\sigma$ errors given by a $\Delta(\chi^2)$ cut. In cases where there are no upper or lower bounds, a narrow ring is not a good fit for that system. (See the text for further details.) For each model we show the reduced chi-squared, $\chi_{\rm{red}}^2$ and degrees of freedom and use these to determine whether or not this narrow ring model is a good fit to the observations.}
\begin{tabular}{llp{0.9cm}lp{1.8cm}p{1.5cm}lllp{0.7cm}}
	\hline
  Star & $\lambda$ ($\mu$m)& Radius ({\arcsec}) & Radius (AU) & Inclination ($^\circ$ from face on) & PA$_{ext}$ ($^\circ$ E of N) & F$_\nu$ (mJy) & $\chi_{\rm{red}}^2$ & d.o.f. & Good fit? \\
	\hline
$\alpha$ CrB & 70 & 2.0 & 45.6 $^{}_{}$ & 79.7 $^{}_{}$ & 170.3$^{}_{}$ & 418.2 & 2.213 & 279 & N \\
 & 100 & 2.3 & 52.6 $^{+2.4}_{-1.7}$ & 90.0 $^{+0.0}_{-10.3}$ & 170.3 $^{+2.1}_{-2.1}$ & 197.0 & 1.085 & 329 & Y \\
 & 160 & 3.4 & 77.2 $^{+17.8}_{-24.0}$ & 90.0 $^{+0.0}_{-25.0}$ & 170.3 $^{+14.7}_{-5.3}$ & 53.4 & 1.010 & 113 & Y \\
 & all & 2.0 & 46.1 $^{}_{}$ & 79.7 $^{}_{}$ & 170.3 $^{}_{}$ & - & 1.566 & 729 & N \\ \\
$\beta$ Uma & 70 & 1.9 & 46.4 $^{}_{}$ & 82.4 $^{}_{}$ & 113.8$^{}_{}$ & 312.3 & 1.233 & 270 & N \\
 & 100 & 2.1 & 52.0 $^{+3.0}_{-3.3}$ & 89.1 $^{+0.9}_{-15.5}$ & 108.5 $^{+2.7}_{-5.4}$ & 163.3 & 0.870 & 287 & Y \\
 & 160 & 2.0 & 48.3 $^{+6.7}_{-8.3}$ & 89.1 $^{+0.9}_{-30.9}$ & 38.5 $^{+32.3}_{-13.5}$ & 56.6 & 0.790 & 84 & Y \\
 & all & 1.9 & 46.8 $^{+1.3}_{-0.9}$ & 84.3 $^{+5.7}_{-4.4}$ & 113.8 $^{+0.0}_{-2.7}$ & - & 1.041 & 649 & Y \\ \\
$\lambda$ Boo & 70 & 2.6 & 80.1 $^{}_{}$ & 51.3 $^{}_{}$ & 42.4$^{}_{}$ & 318.9 & 1.776 & 375 & N \\
 & 100 & 3.3 & 99.7 $^{}_{}$ & 41.8 $^{}_{}$ & 42.4 $^{}_{}$ & 257.4 & 1.296 & 469 & N \\
 & 160 & 4.1 & 124.2 $^{+8.4}_{-8.5}$ & 41.8 $^{+8.8}_{-11.8}$ & 55.3 $^{+12.9}_{-14.7}$ & 144.6 & 0.694 & 177 & Y \\
 & all & 2.8 & 84.5 $^{}_{}$ & 48.4 $^{}_{}$ & 44.2 $^{}_{}$ & - & 1.891 & 1029 & N \\ \\
$\epsilon$ Pav & 70 & 3.2 & 104.4 $^{+8.5}_{-7.9}$ & 49.0 $^{+5.9}_{-8.8}$ & -4.1$^{+10.3}_{-10.3}$ & 51.2 & 0.904 & 214 & Y \\
 & 100 & 4.0 & 127.9 $^{+10.4}_{-13.3}$ & 37.2 $^{+8.8}_{-20.5}$ & 57.9 $^{+25.9}_{-25.9}$ & 50.3 & 1.072 & 205 & Y \\
 & 160 & 6.9 & 221.4 $^{+45.8}_{-49.1}$ & 63.6 $^{+26.4}_{-23.4}$ & 89.0 $^{+20.7}_{-15.5}$ & 31.1 & 0.912 & 80 & Y \\
 & all & 3.3 & 106.0 $^{+8.6}_{-4.9}$ & 34.3 $^{+8.8}_{-8.8}$ & 11.4 $^{+15.5}_{-15.5}$ & - & 1.073 & 507 & Y \\ \\
$\zeta$ Eri & 70 & 2.3 & 77.9 $^{+6.7}_{-7.3}$ & 42.4 $^{+7.9}_{-15.9}$ & 25.0$^{+15.0}_{-15.0}$ & 77.3 & 1.061 & 216 & Y \\
 & 100 & 3.8 & 127.7 $^{}_{}$ & 39.7 $^{}_{}$ & 25.0 $^{}_{}$ & 81.1 & 1.194 & 319 & N \\
 & 160 & 5.3 & 177.6 $^{+25.0}_{-27.0}$ & 45.0 $^{+15.9}_{-37.1}$ & 25.0 $^{+30.0}_{-35.0}$ & 50.6 & 0.930 & 106 & Y \\
 & all & 2.9 & 96.5 $^{}_{}$ & 45.0 $^{}_{}$ & 25.0 $^{}_{}$ & - & 1.290 & 649 & N \\ \\
$\gamma$ Tri & 70 & 3.7 & 126.5 $^{}_{}$ & 80.0 $^{}_{}$ & 160.5$^{}_{}$ & 770.8 & 10.207 & 525 & N \\
 & 100 & 4.3 & 148.3 $^{}_{}$ & 89.5 $^{}_{}$ & 163.7 $^{}_{}$ & 717.4 & 3.772 & 641 & N \\
 & 160 & 5.0 & 171.7 $^{}_{}$ & 89.5 $^{}_{}$ & 162.9 $^{}_{}$ & 450.6 & 1.229 & 251 & N \\
 & all & 4.0 & 137.4 $^{}_{}$ & 83.2 $^{}_{}$ & 162.1 $^{}_{}$ & - & 6.935 & 1425 & N \\ \\
$\rho$ Vir & 70 & 2.4 & 87.2 $^{+2.2}_{-1.6}$ & 70.9 $^{+2.6}_{-3.8}$ & 98.5$^{+0.0}_{-1.9}$ & 236.7 & 1.064 & 287 & Y \\
 & 100 & 3.0 & 109.4 $^{+3.5}_{-4.7}$ & 67.1 $^{+5.1}_{-3.8}$ & 90.8 $^{+3.8}_{-1.9}$ & 166.4 & 1.126 & 284 & Y \\
 & 160 & 2.4 & 87.2 $^{+27.8}_{-17.2}$ & 55.5 $^{+29.5}_{-20.5}$ & 83.1 $^{+51.9}_{-23.1}$ & 76.4 & 0.901 & 72 & Y \\
 & all & 2.5 & 92.3 $^{}_{}$ & 69.6 $^{}_{}$ & 96.5 $^{}_{}$ & - & 1.174 & 651 & N \\ \\
30 Mon & 70 & 1.8 & 67.9 $^{+1.7}_{-2.5}$ & 59.2 $^{+2.6}_{-5.1}$ & 164.1$^{+3.6}_{0.0}$ & 197.7 & 1.024 & 244 & Y \\
 & 100 & 2.2 & 83.7 $^{+8.7}_{-8.8}$ & 43.8 $^{+12.8}_{-15.4}$ & 167.7 $^{+17.9}_{-17.9}$ & 88.0 & 0.894 & 220 & Y \\
 & 160 & 5.8 & 216.3 $^{+13.7}_{-79.2}$ & 46.4 $^{+43.6}_{-56.4}$ & 153.3 $^{+46.7}_{-93.3}$ & 32.6 & 0.810 & 85 & Y \\
 & all & 1.8 & 68.7 $^{+1.7}_{-1.7}$ & 56.7 $^{+2.6}_{-5.1}$ & 167.7 $^{+0.0}_{-3.6}$ & - & 0.978 & 557 & Y \\ \\
$\beta$ Tri & 70 & 2.6 & 101.8 $^{}_{}$ & 40.3 $^{}_{}$ & 66.7$^{}_{}$ & 557.8 & 3.343 & 447 & N \\
 & 100 & 3.3 & 129.1 $^{}_{}$ & 41.3 $^{}_{}$ & 63.1 $^{}_{}$ & 442.2 & 1.568 & 543 & N \\
 & 160 & 4.1 & 161.2 $^{+3.8}_{-7.3}$ & 44.5 $^{+4.2}_{-4.2}$ & 54.7 $^{+6.0}_{-6.0}$ & 246.1 & 0.869 & 226 & Y \\
 & all & 2.8 & 109.0 $^{}_{}$ & 41.3 $^{}_{}$ & 64.3 $^{}_{}$ & - & 3.212 & 1224 & N \\ \hline
\end{tabular}
	\label{tbestfit}
\end{minipage}
\end{table*}

\subsection{Problems with a single narrow ring fit}
\label{sprob}
In this paper we have been focussing on fitting all the systems with a single, narrow ring to keep the model simple and avoid any need to make assumptions about the grain properties. This is seen to work well for spectral energy distributions. The same can be said for the case of $\beta$ Uma where the fits to different wavelengths result in similar best fitting radii. However, in most cases the individual images show an increase in best fit radius for increasing wavelengths. Since dust further from a star will be colder it will contribute more at longer wavelengths. A wide disc seen at low signal-to-noise will appear as rings with increasing radius for increasing wavelength. To test whether this is the case for any of our systems we have also calculated which model best fits all images simultaneously (table \ref{tbestfit}). The residuals and $\chi_{\rm{red}}^2$ for this best fit show a narrow ring to still be consistent with the data for $\epsilon$ Pav and 30 Mon. This multi-wavelength best fit is dominated by the 70~$\mu$m image due to its higher signal-to-noise ratio and better resolution. For $\zeta$ Eri and $\rho$ Vir the $\chi_{\rm{red}}^2$ is above the cut-off value but the residuals are still small and we consider these borderline cases that might be affected by an incomplete understanding of the noise.

Other systems show more obvious deviations from a single narrow ring. Judging by the $\chi_{\rm{red}}^2$ in table \ref{tbestfit} and the residuals in figure \ref{fres100} we can see that $\alpha$ CrB, $\lambda$ Boo, and $\beta$ Tri are poorly fit at 70~$\mu$m and $\gamma$ Tri is poorly fit at all wavelengths. This implies that these systems have wide belts and/or multiple narrow rings.

 \begin{figure}
 	\centering
 	\includegraphics[width=0.47\textwidth]{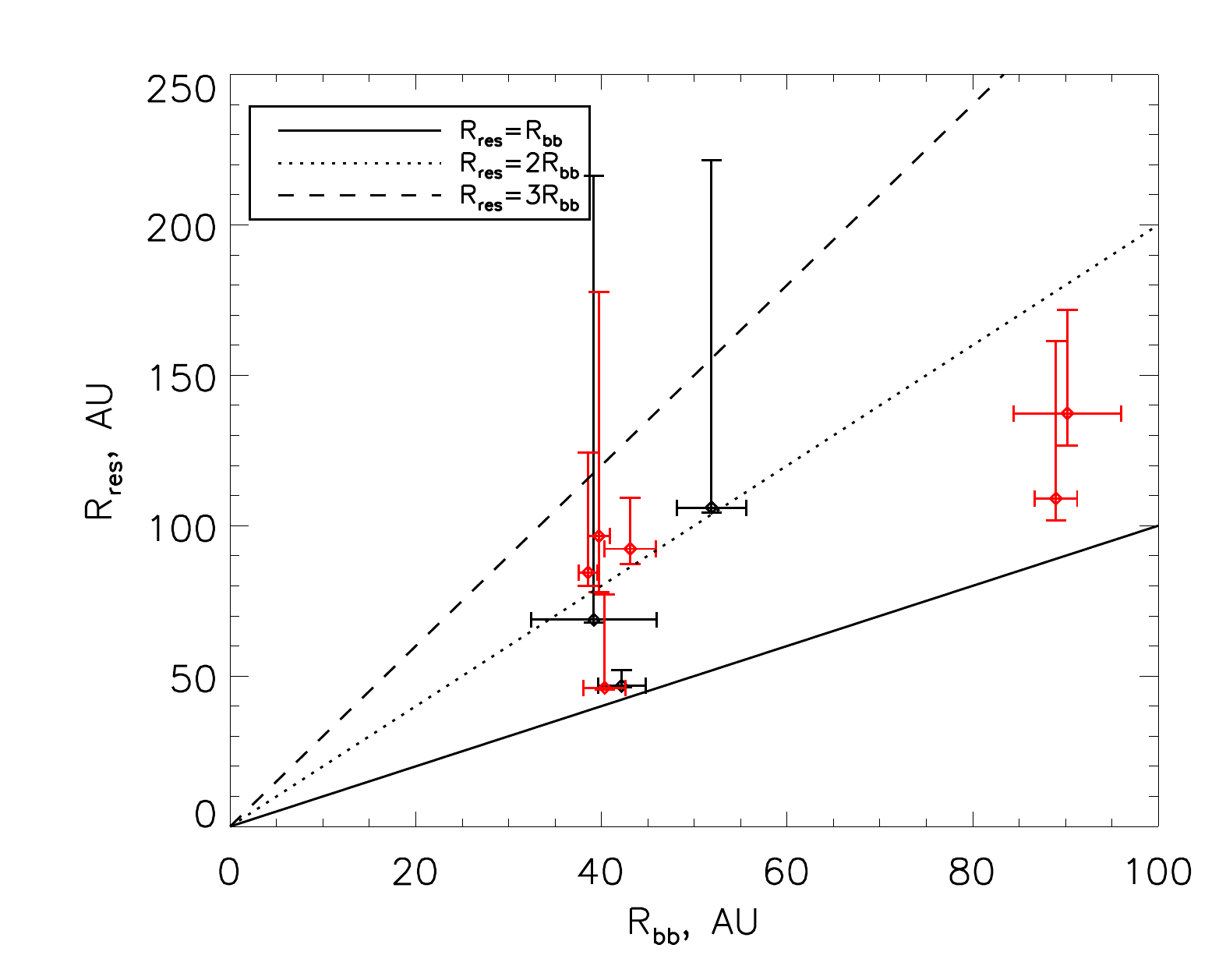}
 	\caption{Resolved radii versus blackbody radii. The discs in red are those for which a narrow ring seems to be a poor fit to the disc (see section \protect\ref{sprob}) and so the radii simply represents the peak of the emission rather than the location of all of the dust. The lines represent what we would expect for different values of $\Gamma$. }
 	\label{fradrat}
 \end{figure}

\begin{figure}
	\centering
	\includegraphics[width=0.47\textwidth]{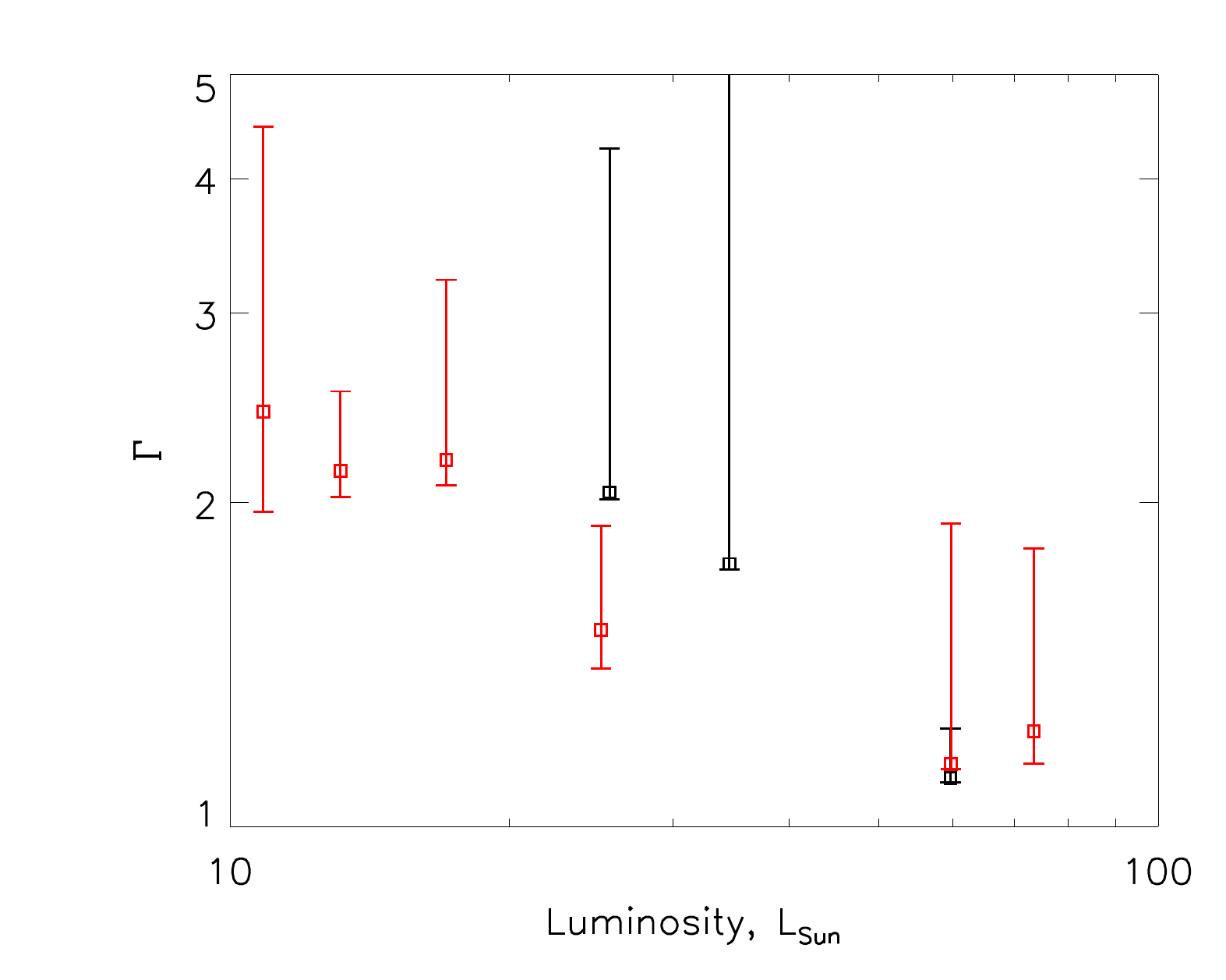}
	\caption{Ratio of resolved radii to blackbody radii against stellar luminosity.}
	\label{fradratlum}
\end{figure}

\begin{figure}
	\centering
	\includegraphics[width=0.47\textwidth]{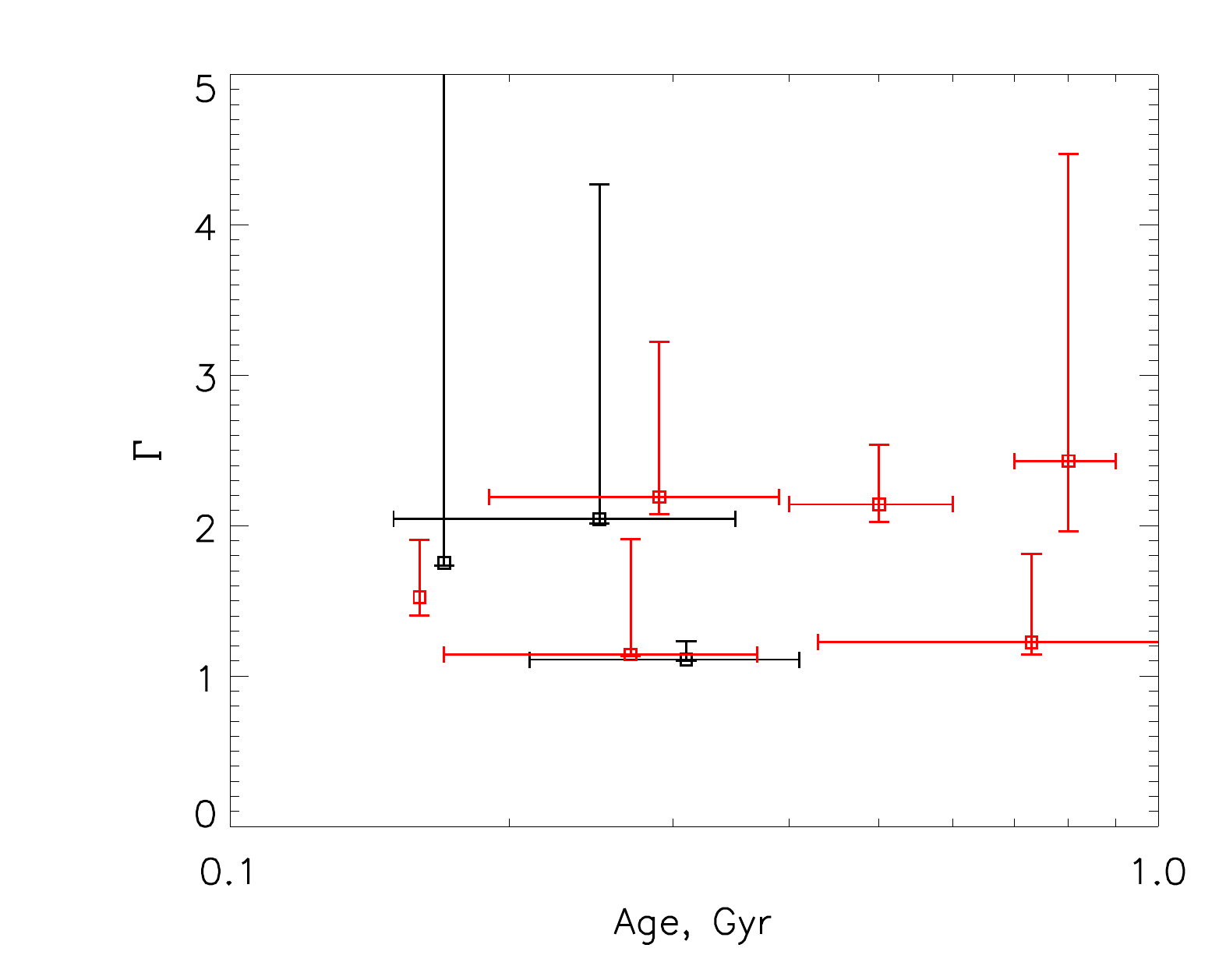}
	\caption{Ratio of resolved radii to blackbody radii against age.}
	\label{fradratage}
\end{figure}

\section{Debris disc radii}
\label{sddr}
From the previous sections we now have radii calculated from fitting a modified blackbody to the SED and radii calculated from fitting a narrow ring to the images. Figure \ref{fradrat} shows the resolved radii plotted against blackbody radii. Here we define a factor $\Gamma$ that is the ratio of the resolved radius to the blackbody radius. Those systems demonstrated to be poorly fit by a narrow ring in section \ref{sprob} are shown in red here and in figures \ref{fradratlum} and \ref{fradratage} to emphasize that these are not narrow ring systems. We still include them as the narrow ring fit will give the location that contributes most to the flux in the image and therefore the location of the grains that contribute most to the photometry at these wavelengths. As many of the systems are not well fit by a narrow ring the error bars in these plots represent the range of fitted values for the fits at individual wavelengths.

We can see that $\Gamma$ has some variation, with values between 1 and 2.5, which compares well with the literature \citep[e.g.][]{rodriguez12}. \citet{trilling07} study debris discs around binary systems and find that, if they assume blackbody grains, the dust for some of the systems seems to be in unstable circumbinary regions. If instead small grains are assumed, many of the unstable systems can be considered stable but a similar number of systems that were formerly considered stable now have dust in unstable regions. Our finding that there is a range of $\Gamma$ values with some systems surrounded by dust that is close to blackbody and others with dust much hotter than blackbody makes it possible to explain the findings of \citet{trilling07} without requiring the dust to be in unstable regions.

The distribution of $\Gamma$ does show an inverse correlation with luminosity, as can be seen from figure \ref{fradratlum}. This luminosity dependence is due to the the dependence of minimum grain size on luminosity as discussed further in the next section.

We might also expect some of this variation to be due to the different ages of the systems. It is not known for sure how debris discs become stirred and produce a collisional cascade. If this stirring is delayed \citep[e.g.][]{kenyon04a,wyatt08} then this could change the quantity of small dust in the system with time, which may in turn affect the radius ratio. Plotting $\Gamma$ ratio against age (figure \ref{fradratage}) shows no significant change with age and so the effects, if any, of delayed stirring would have to occur within the first 100~Myr.

\subsection{Theoretical explanations for the differences in radii}
\begin{figure}
	\centering
	\includegraphics[width=0.47\textwidth]{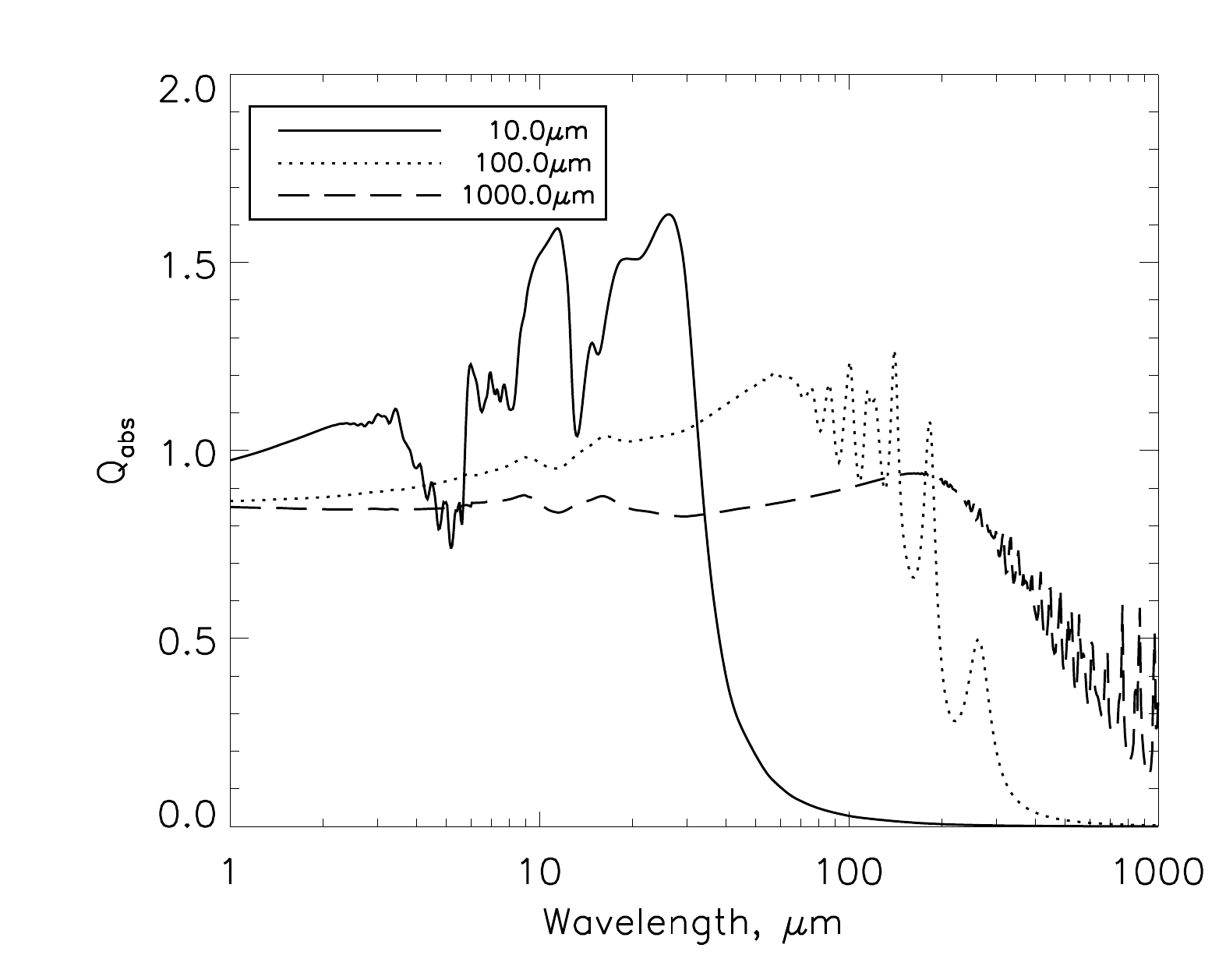}
	\caption{Absorption efficiency for amorphous silicate grains based on the core-mantle model of \protect\citet{li97}.}
	\label{fqabs}
\end{figure}

The reason for the differences between radii inferred by a blackbody fit and the actual location of the dust grains is the optical properties of the grains. The absorption efficiency, $Q_{\rm{abs}}$, approaches 0 for $\lambda\gg$ the size of the grain (figure \ref{fqabs}) and the same is true of the emission efficiencies. This means that the grain cannot emit a blackbody at all wavelengths and so its temperature increases in order to reach equilibrium temperature.

The grain size distributions of debris discs are assumed to follow a power law size distribution down to the smallest size grains in the disc, which is usually considered to be the size at which grains are blown out of the system due to radiation pressure ($D_{\rm{bl}}$). Therefore the smallest grains dominate the cross-sectional area of the system and thereby its temperature. The blowout size depends on the luminosity of the star and the different luminosities and temperatures of stars of different spectral types also affect the temperature of a disc for a given location. To demonstrate how a star's spectral type affects the $\Gamma$ value at different radii we create model discs and fit blackbodies to their SEDs, sampled at 8 wavelengths between 9 and 160$\mu$m, in a similar fashion to the SED fitting in section \ref{ssedfit} \citep[see also][]{bonsor10}.

To do this we use the compositional model of \citet{wyatt02}, which calculates the temperatures, $T$, and absorption efficiencies, $Q_{\rm{abs}}$, using core-mantle grains \citep{li97}. The temperature of the grains are found using an iterative process such that
\begin{align}
T(D,R)&=(\left<Q_{\rm{abs}}\right>_{T_\star}/\left<Q_{\rm{abs}}\right>_{T(D,R)})^{1/4}T_{\rm{bb}}
\label{ettrrg}
\end{align}
where $T_{\rm{bb}}$ is the black-body temperature and \citet{schmidt82} relations are used to give the temperature and luminosity of the star based on an assumed spectral type. The absorption efficiency is found using Mie theory, Rayleigh-Gans theory or geometric optics in the appropriate limits \citep{laor93} using optical properties from the compositional model.

The core-mantle model of \citet{li97} assumes the particles to be formed of a silicate core surrounded by an organic refractory mantle where the fraction of the total volume that is silicate is given by the parameter $q_{\rm{sil}}$. The silicate material may be either amorphous or crystalline. The particles will have a given porosity ($p$) defining how much of the particles' volume is empty space, and a given fraction of water ($q_{\rm{H_2O}}$) defining how much of the empty space is filled by ice. The size distribution of grains in the disc is assumed to follow a power law as $n(D)$d$D\propto D^{-\alpha}$d$D$ down to a minimum size, $D_{\rm{min}}$. For the following we assume, unless otherwise stated, the silicates are amorphous, $q_{\rm{sil}}=1/3$, $p=0$, $q_{\rm{H_2O}}=0$, $R$=100~AU, $D_{\rm{min}}=D_{\rm{bl}}$ and $\alpha=3.5$.

\begin{figure}
	\centering
	\includegraphics[width=0.47\textwidth]{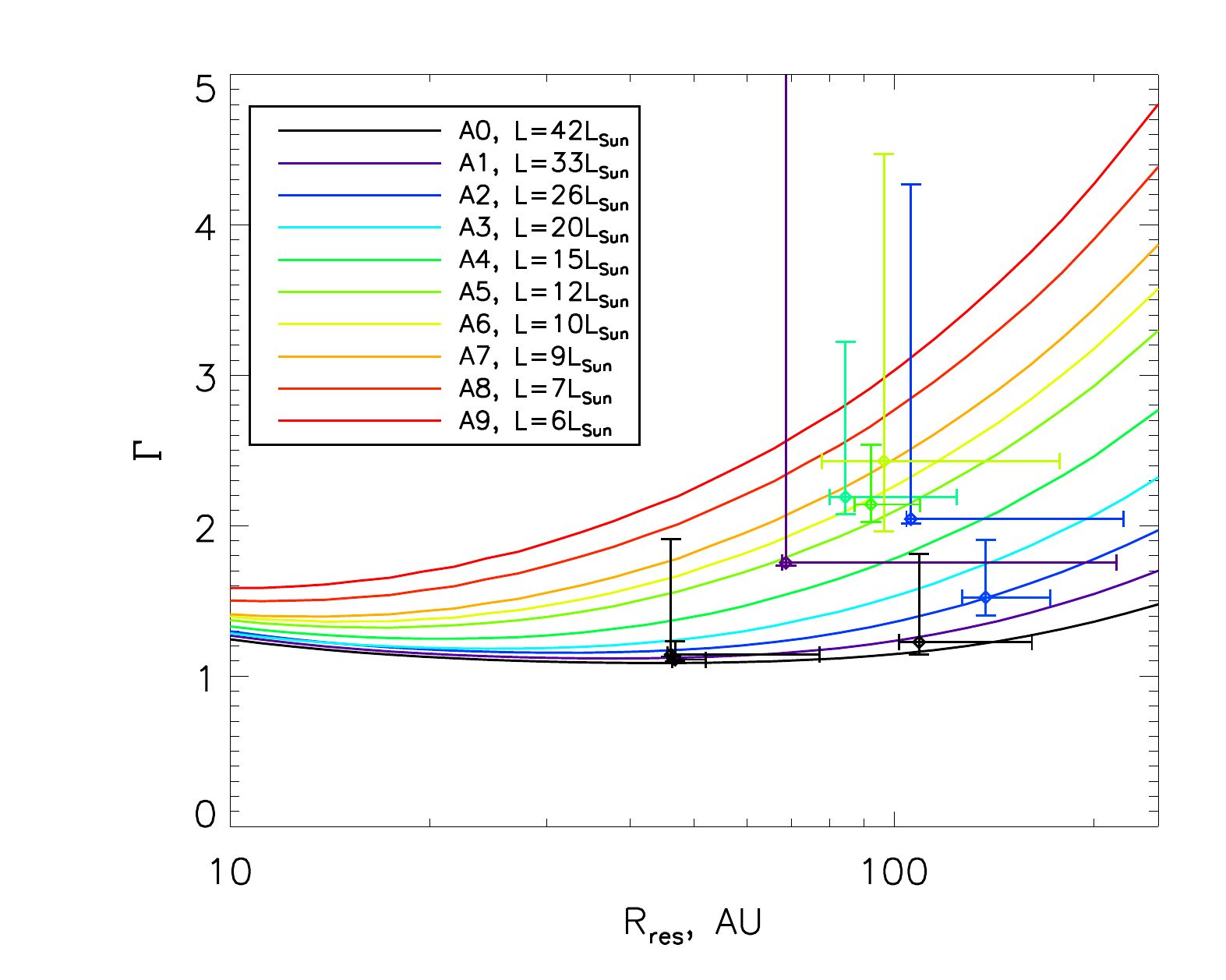}
	\caption{Ratio of resolved radii to blackbody radii against actual radii for simulated discs alongside the discs presented in this paper. The colour of the symbols and lines represents the luminosity of the star. This shows that although we do see a correlation between $\Gamma$ and spectral type, some differences still remain requiring further explanation.}
	\label{fsimratspr}
\end{figure}

Figure \ref{fsimratspr} shows how the radius ratio varies with radius for stars of different spectral type. This plot also shows the results for the resolved discs presented in this paper. The model stars use temperatures and luminosities that are the average for their spectral type, but significant variation is seen in these values for real stars. The symbols and lines are, therefore, colour coded by luminosity to ease comparison. From this plot we can see that the stars with $\Gamma\approx1$ are expected to have these values due to being early type stars and so have a large blowout size of $\approx$15~$\mu$m, thus all the grains in the system have temperatures close to blackbody. However, not all the systems fit so nicely with the expectations. Further differences are likely to be caused by differences in the size distribution. Although finding accurate fits to every system is beyond the scope of this paper, we can demonstrate how we would expect different parameters to vary the temperature of the disc \citep[see also][]{gaspar12}.

 \begin{figure}
 	\centering
 	\includegraphics[width=0.47\textwidth]{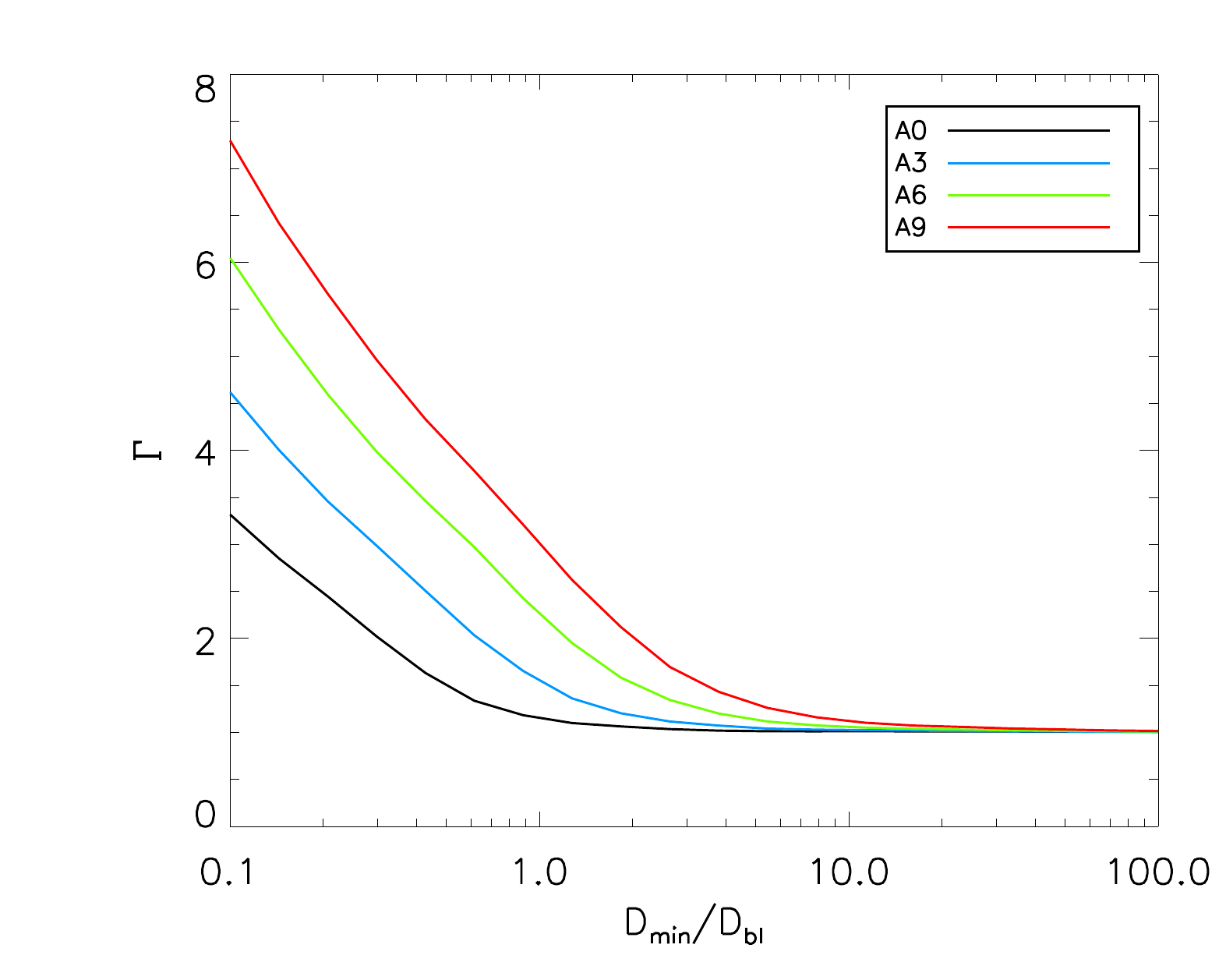}
 	\caption{Ratio of resolved radii to blackbody radii against actual ratio of minimum grain size in size distribution to the blowout size for simulated discs.}
 	\label{fsimratdmin}
 \end{figure}
 
 \begin{figure}
 	\centering
 	\includegraphics[width=0.47\textwidth]{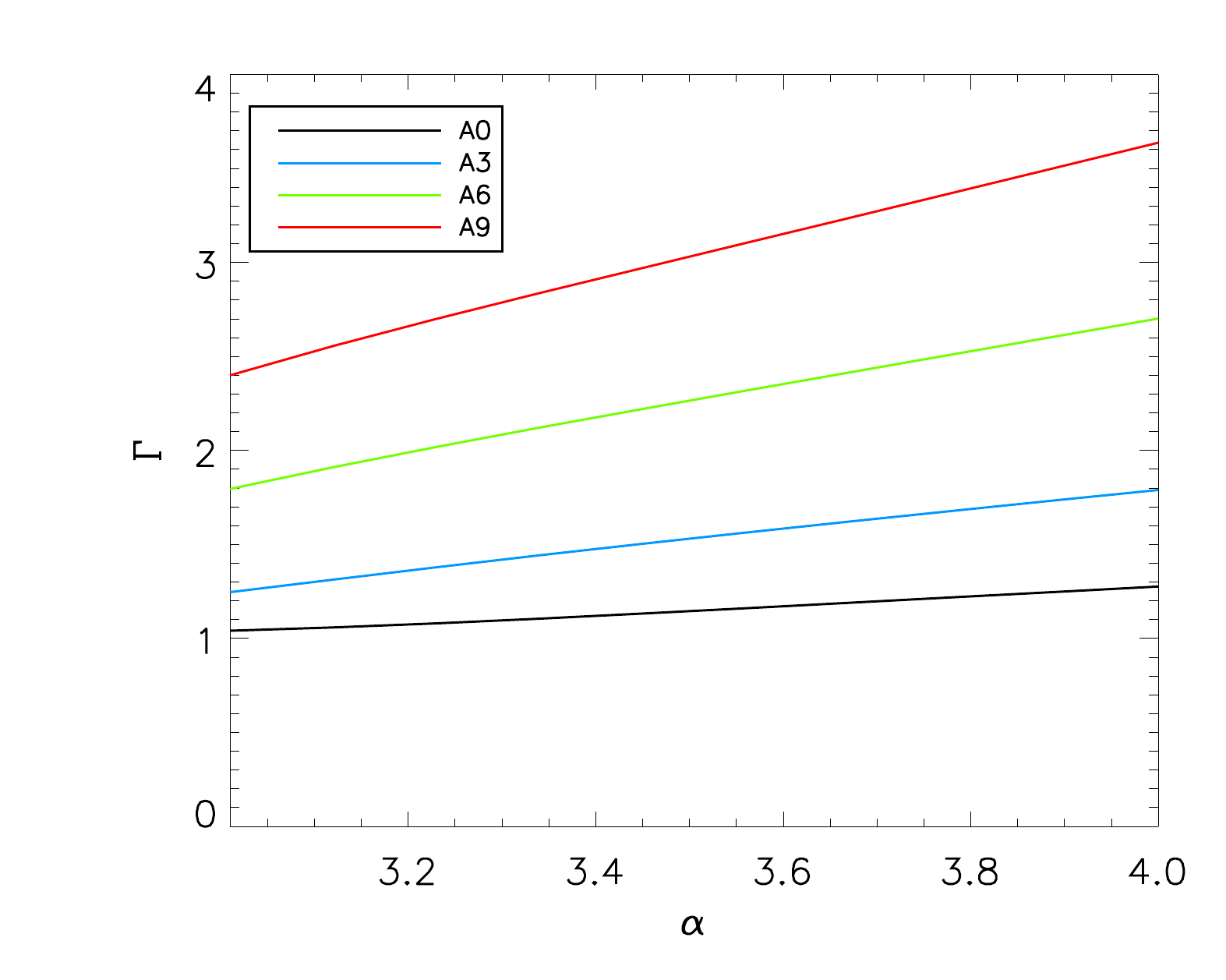}
 	\caption{Ratio of resolved radii to blackbody radii against the power law slope of the size distribution.}
 	\label{fsimratq}
 \end{figure}

\begin{figure}
	\centering
	\includegraphics[width=0.47\textwidth]{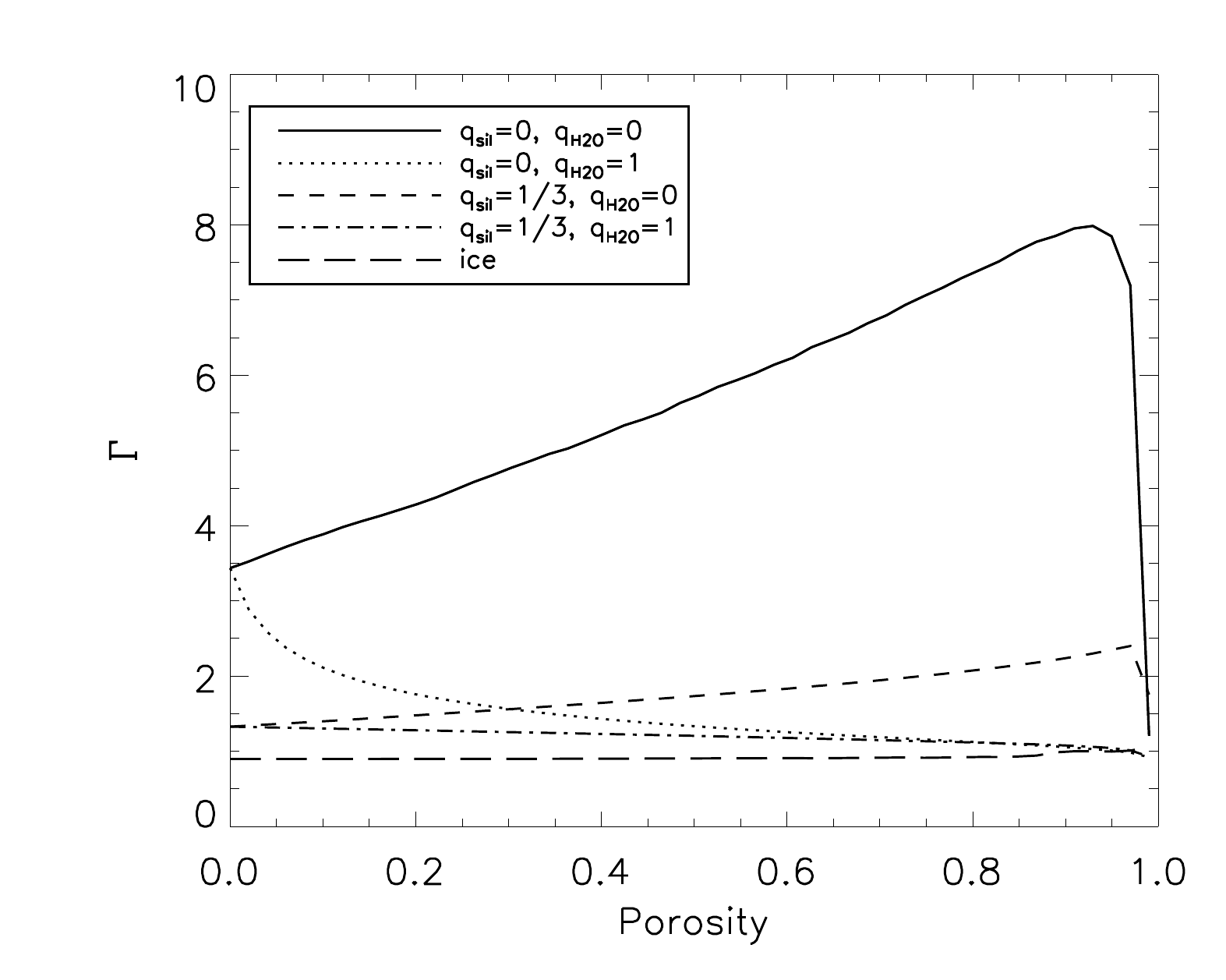}
	\caption{Ratio of resolved radii to blackbody radii against porosity for grains of different composition in a ring at a distance of 100~AU from a star of spectral type A2. The first four models are based on the \citet{li97} core-mantle model and the last shows an ice only grain. Note that for the two models with $q_{\rm{H_2O}}=1$, all the `porous space' is actually filled with ice. High values of $\Gamma$ are possible for organic grains with no silicates and no ice. Moderate changes in $\Gamma$ are also expected for grains with silicates and high porosity. Icy grains can result in values of $\Gamma$ less than 1.}
	\label{fsimratcomp}
\end{figure}

When modelling debris discs it is normally assumed that the discs can be represented by a power law slope of $\alpha=3.5$ \citep{dohnanyi69} going down to a minimum size given by the blowout radius. This may not necessarily be the case. Drag forces can cause a turnover in the size distribution \citep[e.g.][]{wyatt11} as can interactions between blowout grains and the low end of the size distribution \citep{krivov00,krivov05}, thus changing the minimum size grain in the distribution. Figure \ref{fsimratdmin} shows how the radius ratio would vary if the minimum size grain in the size distribution was different from the blowout size. Since smaller grains are hotter than blackbody, increasing the minimum grain size reduces the number of small grains and so reduces the radius ratio. The effect is stronger for stars of later spectral type since these stars have smaller blowout grain sizes than the early type stars. We also see that it is possible for $\Gamma$ to exceed unity for an A0 star if it has a minimum grain size smaller than the blowout size. 

If a more realistic tensile strength curve is used, the power law slope of the grain size distribution is steeper than the Dohnanyi relation of $n(D)$d$D\propto D^{-\alpha}$d$D$ \citep{durda97,obrien03}. A number of authors have shown that the distribution in debris discs has an index of 3.6 or more \citep{thebault03,krivov05,thebault07,lohne08,muller10,wyatt11}. \citet{gaspar12} show that this is a robust result for the outer disc zones probed in this paper, with a typical index of 3.65$\pm$0.05. For our simulations, figure \ref{fsimratq} shows the radius ratio increasing as the power law slope steepens, since a steeper slope means more small grains in the disc. 

Figure \ref{fsimratcomp} shows how the composition of the particles affects $\Gamma$. Large values of $\Gamma$ are feasible if the grains have no silicates and no ice, just organics. This results in a peak $\Gamma$ value of 8 for a porosity of 0.93 and a decreasing $\Gamma$ value for porosities beyond this as the grain becomes mostly empty space. However, it is expected that this kind of organic refractory material is formed by UV photoprocessing of ice mantles on silicate cores \citep[e.g.][]{greenberg82} and so it is unlikely that organic only grains would be present in these systems. Nonetheless, we present such models here to demonstrate the range of values that are possible. For silicate with organic grains it is still possible to moderately increase $\Gamma$ by increasing the porosity so long as no ice is present. For grains which only consist of ice it is possible to get values of $\Gamma$ less than 1 as ice is more efficient at emitting heat than absorbing starlight. 

Comparison of figures \ref{fsimratdmin}, \ref{fsimratq} and \ref{fsimratcomp} shows the minimum grain size to be the dominant disc factor in determining $\Gamma$, although unusual grain compositions can potentially have a large effect.
 
\section{Summary of Results for Individual Systems}
\label{sindiv}
\subsection{$\alpha$ CrB}
$\alpha$ CrB is resolved at 70 and 100~$\mu$m and is seen close to edge on with a best fit radius of 46~AU giving it $\Gamma=1.1$. A single ring fits the SED perfectly. However, residuals are clearly evident in the 70~$\mu$m fit, suggesting that the disc may be wider than modelled here. \citet{moerchen10} observed this star at mid-IR wavelengths. Although they do not find the excess to be statistically significant, they do find the 11~$\mu$m image to be resolved suggesting a radius of around 2~AU. Combined with the PACS images this also adds to the likelihood that this system is radially broad and the 11~$\mu$m observation is seeing the inner part of this wide disc. 

This system is an eclipsing binary and the inclination of 80$^\circ$ and the position angle of 170$^\circ$ show that the disc is consistent with being aligned with the plane of the binary stars \citep{tomkin86}. More detailed modelling of this system is undertaken by \citep[see][]{kennedy12a}.

\subsection{$\beta$ Uma}
$\beta$ Uma is resolved at 70 and 100~$\mu$m. Of all the systems in this paper, $\beta$ Uma has the most consistent fitted radii across all wavelengths showing the system to be well fit by a roughly edge on ring with a radius of 47~AU. This is in agreement with the earlier analysis of this data by \citet{matthews10}. This gives it $\Gamma=1.1$ as is expected for a star of spectral type A1.

\subsection{$\lambda$ Boo}
$\lambda$ Boo has a large extent at all wavelengths. A narrow ring is a very poor fit to the 70~$\mu$m image. When a simultaneous best fit is attempted, 2-$\sigma$ residuals are also seen in the 100 and 160~$\mu$m images. This shows that a much wider belt is necessary to explain these images, probably extending between at least 80 and 125~AU. The constraints on the inclination show it to be between 40 and 50$^\circ$. 

This star is the prototype $\lambda$ Bo{\"o}tis star i.e. it has a low abundance in refractory elements. It was proposed by \citep{martinez09} that infrared excesses around $\lambda$ Bo{\"o}tis stars might be explained by interactions with a diffuse interstellar cloud, although they noted that a debris disc model is more likely for stars like $\lambda$ Boo that are within the local bubble. The symmetry of the Herschel images supports the presence of a debris disc rather than heated ISM.

\subsection{$\epsilon$ Pav}
$\epsilon$ Pav is resolved at 70 and 100~$\mu$m. Emission at 160~$\mu$m appears to be offset to both 70 and 100~$\mu$m images, however this is only significant at the 2-$\sigma$ level and so we cannot draw any firm conclusions from it. The results shown in table \ref{tbestfit} show the fitted radius increasing with increasing wavelength from 105 to 220~AU and the inclinations and position angles vary between wavelengths. The increase in fitted radii could be indicative of a wide disc, however, we find that a narrow ring still gives a good simultaneous fit to all the data with a radius of 107~AU and inclination of 34$^\circ$. This results in a radii ratio of $\Gamma=2.1$, somewhat high given its luminosity, perhaps suggesting a highly porous or non-silicate composition or that some process has led to an increase in grains smaller than the blowout size that have yet to be blown out by radiation pressure.

\subsection{$\zeta$ Eri}
$\zeta$ Eri is resolved at 70 and 100~$\mu$m and marginally at 160~$\mu$m. The fitted radius is shown to increase from 80 to 175~AU. A combined fit at 96~AU with an inclination of 46$^\circ$ results in only minimal residuals but the high $\chi^2_{\rm{red}}$ suggests that some extension or a second ring is required to explain this disc. The SED is also suggestive of more dust close to the star. This star has the highest radius ratio of all the stars presented here with $\Gamma=2.4$, although this is unsurprising as it is also the least luminous star in this sample.

\subsection{$\gamma$ Tri}
$\gamma$ Tri has a large extent at all wavelengths and a narrow ring gives a poor fit to all images. Multiple temperatures are also required to fit to the SED. The fits suggest that the disc extends from 125~AU to at least 170~AU and probably much further with the disc being close to edge on.

\subsection{$\rho$ Vir}
$\rho$ Vir is resolved at 70 and 100~$\mu$m with a background source present to the NE consistent with a galaxy at redshift $z\approx1$. The fitted radii are consistent at 70 and 160~$\mu$m, but the 100~$\mu$m image requires a larger radius to explain the observations and some small but significant residuals are present when a combined fit is attempted. The combined best fit has a radius of 92~AU and inclination of 20$^\circ$. This gives the disc $\Gamma=2.2$ and if the dust in the inner disc is similar then this should have $\Gamma\approx1.5$ and an actual radius around 4~AU. 

\subsection{30 Mon}
30 Mon is resolved at 70~$\mu$m and marginally at 100~$\mu$m. The weak constraints at 100 and 160~$\mu$m mean that this system is easily fit by a narrow ring with a radius of 68~AU and inclination of 56$^\circ$. This system also has a high mid-IR excess with an inner ring expected at around 1~AU \citep{smith10,moerchen10}. The radius ratio of $\Gamma=1.8$ is higher than expected for a star of this spectral type, perhaps indicating a highly porous or non-silicate composition or a lower limit to the size distribution than would be expected if it were defined by the size of blowout grains.

\subsection{$\beta$ Tri}
$\beta$ Tri has a large extent at all wavelengths. A narrow ring is a very poor fit to the 70~$\mu$m image and some residuals are seen at 100~$\mu$m. When a simultaneous best fit is attempted, significant residuals are seen at all wavelengths. Like $\lambda$ Boo and $\gamma$ Tri, this disc must be much wider than we have modelled here. This system is a spectroscopic binary and the inclination of 41$^\circ$ and the position angle of 64$^\circ$ show that the disc is consistent with the plane of the binary stars \citep{pourbaix00}. A more thorough treatment of this debris disc and the effect of the binary system on the disc is given by \citet{kennedy12a}.

\section{Radial extension of debris discs}
\label{sext}
As discussed in section \ref{sprob}, the images for at least four of the systems presented in this paper are indicative of wide belts. \citet{kalas06} noted a similar distinction between narrow and wide belts with some systems having distinct outer edges and others with sensitivity-limited outer edges and disc widths $>50$~AU.

In our own Solar System the asteroid belt and Kuiper belt are both mostly constrained to narrow rings. The main asteroid belt is between 2.1 and 3.3~AU and is constrained on either side by the presence of Mars and Jupiter. The inner edge of the Kuiper belt is similarly constrained by the presence of Neptune. The outer edge is less well defined with high eccentricity objects in the hot classical belt and scattered disc. However the majority of objects reside between $\approx$37 and 50~AU \citep[based on the CFEPS L7 data,][]{petit11}. 

The distribution of dust is likely to be wider than this due to Poynting-Robertson drag causing dust to drift in towards the Sun and radiation pressure increasing the eccentricity of the dust \citep[e.g.][]{vitense10}. PR-drag is unlikely to be important for the systems presented in this paper as they are much more dense than the Kuiper belt and so the collision timescale is shorter than the PR-drag timescale \citep{wyatt05}. Radiation pressure may be more important and has two main effects on the distribution. Dust that is being blown out of the system forms a blowout grain halo and dust that is above the blowout limit is still weakly bound, forming a broad disc of eccentric grains. These effects are observed at these wavelengths around some stars, such as Vega \citep{su05,sibthorpe10}, but their contribution to the flux is small at the longest wavelengths. Blowout grain halos cannot be used to explain all wide discs since there are cases of wide discs being observed even at millimetre wavelengths \citep{hughes11,wilner12}, which probe grains much larger than the blowout size.   

There are a number of plausible reasons for distinct outer edges in debris discs. For the Solar System, \citet{lykawka08} have suggested that a planet outside the Kuiper belt could be responsible for the outer edge cut-off. A similar explanation has been suggested for the ring of Fomalhaut \citep{boley12}. An alternative dynamical explanation could be that objects forming outside this region were disrupted by interactions with other stars either in the birth cluster \citep{adams10} or since then \citep{ida00}.

Alternatively the edge of a debris disc may be a relic of the properties of the original protoplanetary disc. Stellar encounters during the early stages of disc formation could have stripped material from beyond this distance \citep{bate03} or radiation from a nearby massive star could have truncated the disc as is seen to occur in Orion \citep{johnstone98,mann10}. 

Wide discs may simply be discs that formed in the same way as narrow rings but were then subjected to dynamical interactions with the planetary system. They may be the result of dynamical scattering of objects onto eccentric orbits similar to the scattered disc component of the Kuiper belt but much more massive. However, such events are thought to be rare in mature systems \citep{booth09}. Even if such an event has occurred, we would have to be seeing these discs during or soon after the scattering event as such an event rapidly clears material out of a planetesimal disc \citep{booth09}. 

Combining this evidence suggests that these wide discs are most likely to be the result of stars that formed with large protoplanetary discs that were never truncated. This would then require the formation of planetesimals at distances up to $\approx200$~AU from the star, further complicating planetesimal formation theories that have trouble even explaining the formation of Kuiper belt objects. For instance, \citet{weidenschilling03} suggests that even if the Solar System's protoplanetary disc was $\approx$100~AU in radius, planetesimals are still unlikely to form beyond 50~AU. Nonetheless, recent studies of the formation of planetesimals through gravitational collapse could find ways that allow such objects to form at large distances \citep{johansen07,johansen12}.

\section{Conclusions}
\label{sconc}
The increased resolution of Herschel compared to previous infrared spacecraft has proven to be a great success in resolving debris discs. Observations with Herschel also allow us to fill in the spectrum between the mid-IR and sub-millimetre i.e. the region where the spectral energy distributions of cold discs peak. In this paper we have presented observations of nine debris discs observed as part of the DEBRIS survey, the cold discs of which have never previously been resolved. We fit the images with a narrow ring model to gain an insight into the parameters of the discs. The discs demonstrate the varied nature of planetary systems, ranging from around 50~AU to 150~AU in size.

Comparing resolved radii with the radii predicted by blackbody fits to the SED shows the resolved radii to be larger than the blackbody radii by factors between 1 and 2.5. Some of this variation can be attributed to differences in spectral types of the stars and location of the dust, with the ratio of resolved radii and blackbody radii being inversely proportional to the luminosity. Remaining differences from the expected ratio are most likely to be due to highly porous or non-silicate grains or differences to the expected minimum size grain in the size distribution, although further work is needed to narrow down the exact properties of the size distribution.

We also find that, although a single temperature blackbody is often a good fit to the SED, a narrow ring is sometimes not such a good fit to the images. $\beta$ Uma, $\epsilon$ Pav and 30 Mon all appear to be well fit by a narrow ring. $\zeta$ Eri and $\rho$ Vir show indications that they are probably broader than modelled here. $\alpha$ CrB, $\gamma$ Tri, $\lambda$ Boo and $\beta$ Tri are clear examples of discs that cannot be fit by a narrow ring model, with very significant residuals left after attempting to fit a narrow ring. They are likely to require a broad and/or multi-component model similar to $\gamma$ Dor \citep{broekhoven12} and $\beta$ Leo \citep{churcher11}.

The Herschel Space Observatory has provided us with an excellent opportunity to produce a consistent data set of resolved discs. Application of the analysis presented in this paper to the rest of Herschel's resolved discs will allow further exploration of debris disc properties not previously accessible with only photometric data and give us further insight into how disc properties vary around stars of different spectral types. 

\section*{Acknowledgements}
MB is funded through a Space Science Enhancement Program grant from the Canadian Space Agency. We are grateful to David Wilner for extensive comments on a draft of this paper. We thank the rest of the DEBRIS team for many and varied fruitful discussions especially Laura Churcher and Jean-Fran\c{c}ois Lestrade. This research has made use of the SIMBAD database, operated at CDS, Strasbourg, France. We thank the referee for constructive comments.

\bibliographystyle{mn2efix}
\bibliography{thesis}{}

\begin{thebibliography}{91}
\expandafter\ifx\csname natexlab\endcsname\relax\def\natexlab#1{#1}\fi

\bibitem[{{Adams}(2010)}]{adams10}
{Adams} F.~C., 2010, \araa, 48, 47

\bibitem[{{Aumann} {et~al}\mbox{.}(1984){Aumann}, {Beichman}, {Gillett}, {de
  Jong}, {Houck}, {Low}, {Neugebauer}, {Walker}, \& {Wesselius}}]{aumann84}
{Aumann} H.~H. {et~al.}, 1984, \apjl, 278, L23

\bibitem[{{Backman} \& {Paresce}(1993)}]{backman93}
{Backman} D.~E., {Paresce} F., 1993, in Protostars and Planets III, {E.~H.~Levy
  \& J.~I.~Lunine}, ed., Univ. Arizona Press, Tucson, pp. 1253--1304

\bibitem[{{Bate} {et~al}\mbox{.}(2003){Bate}, {Bonnell}, \& {Bromm}}]{bate03}
{Bate} M.~R., {Bonnell} I.~A., {Bromm} V., 2003, \mnras, 339, 577

\bibitem[{{Berta} {et~al}\mbox{.}(2011){Berta}, {Magnelli}, {Nordon}, {Lutz},
  {Wuyts}, {Altieri}, {Andreani}, {Aussel}, {Casta{\~n}eda}, {Cepa}, {Cimatti},
  {Daddi}, {Elbaz}, {F{\"o}rster Schreiber}, {Genzel}, {Le Floc'h}, {Maiolino},
  {P{\'e}rez-Fournon}, {Poglitsch}, {Popesso}, {Pozzi}, {Riguccini},
  {Rodighiero}, {Sanchez-Portal}, {Sturm}, {Tacconi}, \&
  {Valtchanov}}]{berta11}
{Berta} S. {et~al.}, 2011, \aap, 532, A49

\bibitem[{{Blain} {et~al}\mbox{.}(2002){Blain}, {Smail}, {Ivison}, {Kneib}, \&
  {Frayer}}]{blain02}
{Blain} A.~W., {Smail} I., {Ivison} R.~J., {Kneib} J.-P., {Frayer} D.~T., 2002,
  \physrep, 369, 111

\bibitem[{{Boley} {et~al}\mbox{.}(2012){Boley}, {Payne}, {Corder}, {Dent},
  {Ford}, \& {Shabram}}]{boley12}
{Boley} A.~C., {Payne} M.~J., {Corder} S., {Dent} W.~R.~F., {Ford} E.~B.,
  {Shabram} M., 2012, \apjl, 750, L21

\bibitem[{{Bonsor} \& {Wyatt}(2010)}]{bonsor10}
{Bonsor} A., {Wyatt} M., 2010, \mnras, 409, 1631

\bibitem[{{Booth} {et~al}\mbox{.}(2009){Booth}, {Wyatt}, {Morbidelli},
  {Moro-Mart{\'{\i}}n}, \& {Levison}}]{booth09}
{Booth} M., {Wyatt} M.~C., {Morbidelli} A., {Moro-Mart{\'{\i}}n} A., {Levison}
  H.~F., 2009, \mnras, 399, 385

\bibitem[{{Broekhoven-Fiene} {et~al}\mbox{.}(2012){Broekhoven-Fiene},
  {Matthews}, {Kennedy}, {Booth}, {Sibthorpe}, {Lawler}, {Kavelaars}, {Wyatt},
  {Qi}, {Koning}, {Su}, {Rieke}, {Wilner}, \& {Greaves}}]{broekhoven12}
{Broekhoven-Fiene} H. {et~al.}, 2012, \apj, submitted

\bibitem[{{Brott} \& {Hauschildt}(2005)}]{brott05}
{Brott} I., {Hauschildt} P.~H., 2005, in ESA Special Publication, Vol. 576, The
  Three-Dimensional Universe with Gaia, {C.~Turon, K.~S.~O'Flaherty, \&
  M.~A.~C.~Perryman}, ed., p. 565

\bibitem[{{Chen} {et~al}\mbox{.}(2006){Chen}, {Sargent}, {Bohac}, {Kim},
  {Leibensperger}, {Jura}, {Najita}, {Forrest}, {Watson}, {Sloan}, \&
  {Keller}}]{chen06}
{Chen} C.~H. {et~al.}, 2006, \apjs, 166, 351

\bibitem[{{Churcher} {et~al}\mbox{.}(2011){Churcher}, {Wyatt}, {Duch{\^e}ne},
  {Sibthorpe}, {Kennedy}, {Matthews}, {Kalas}, {Greaves}, {Su}, \&
  {Rieke}}]{churcher11}
{Churcher} L.~J. {et~al.}, 2011, \mnras, 417, 1715

\bibitem[{{Cutri} {et~al}\mbox{.}(2003){Cutri}, {Skrutskie}, {van Dyk},
  {Beichman}, {Carpenter}, {Chester}, {Cambresy}, {Evans}, {Fowler}, {Gizis},
  {Howard}, {Huchra}, {Jarrett}, {Kopan}, {Kirkpatrick}, {Light}, {Marsh},
  {McCallon}, {Schneider}, {Stiening}, {Sykes}, {Weinberg}, {Wheaton},
  {Wheelock}, \& {Zacarias}}]{cutri03}
{Cutri} R.~M. {et~al.}, 2003, {2MASS All Sky Catalog of point sources.}
  NASA/IPAC Infrared Science Archive

\bibitem[{{Dohnanyi}(1969)}]{dohnanyi69}
{Dohnanyi} J.~S., 1969, \jgr, 74, 2531

\bibitem[{{Durda} \& {Dermott}(1997)}]{durda97}
{Durda} D.~D., {Dermott} S.~F., 1997, Icarus, 130, 140

\bibitem[{{Eiroa} {et~al}\mbox{.}(2010){Eiroa}, {Fedele}, {Maldonado},
  {Gonz{\'a}lez-Garc{\'{\i}}a}, {Rodmann}, {Heras}, {Pilbratt}, {Augereau},
  {Mora}, {Montesinos}, {Ardila}, {Bryden}, {Liseau}, {Stapelfeldt},
  {Launhardt}, {Solano}, {Bayo}, {Absil}, {Ar{\'e}valo}, {Barrado},
  {Beichmann}, {Danchi}, {Del Burgo}, {Ertel}, {Fridlund}, {Fukagawa},
  {Guti{\'e}rrez}, {Gr{\"u}n}, {Kamp}, {Krivov}, {Lebreton}, {L{\"o}hne},
  {Lorente}, {Marshall}, {Mart{\'{\i}}nez-Arn{\'a}iz}, {Meeus}, {Montes},
  {Morbidelli}, {M{\"u}ller}, {Mutschke}, {Nakagawa}, {Olofsson}, {Ribas},
  {Roberge}, {Sanz-Forcada}, {Th{\'e}bault}, {Walker}, {White}, \&
  {Wolf}}]{eiroa10}
{Eiroa} C. {et~al.}, 2010, \aap, 518, L131+

\bibitem[{{Fruchter} \& {Hook}(2002)}]{fruchter02}
{Fruchter} A.~S., {Hook} R.~N., 2002, \pasp, 114, 144

\bibitem[{{G{\'a}sp{\'a}r} {et~al}\mbox{.}(2012){G{\'a}sp{\'a}r}, {Psaltis},
  {{\"O}zel}, {Rieke}, \& {Cooney}}]{gaspar12}
{G{\'a}sp{\'a}r} A., {Psaltis} D., {{\"O}zel} F., {Rieke} G.~H., {Cooney} A.,
  2012, \apj, 749, 14

\bibitem[{{Greaves} {et~al}\mbox{.}(2005){Greaves}, {Holland}, {Wyatt}, {Dent},
  {Robson}, {Coulson}, {Jenness}, {Moriarty-Schieven}, {Davis}, {Butner},
  {Gear}, {Dominik}, \& {Walker}}]{greaves05}
{Greaves} J.~S. {et~al.}, 2005, \apjl, 619, L187

\bibitem[{{Greenberg}(1982)}]{greenberg82}
{Greenberg} J.~M., 1982, {Dust in dense clouds - One stage in a cycle},
  {Beckman} J.~E., {Phillips} J.~P., eds., Cambridge University Press, pp.
  261--306

\bibitem[{{Griffin} {et~al}\mbox{.}(2010){Griffin}, {Abergel}, {Abreu}, {Ade},
  {Andr{\'e}}, {Augueres}, {Babbedge}, {Bae}, {Baillie}, {Baluteau}, {Barlow},
  {Bendo}, {Benielli}, {Bock}, {Bonhomme}, {Brisbin}, {Brockley-Blatt},
  {Caldwell}, {Cara}, {Castro-Rodriguez}, {Cerulli}, {Chanial}, {Chen},
  {Clark}, {Clements}, {Clerc}, {Coker}, {Communal}, {Conversi}, {Cox},
  {Crumb}, {Cunningham}, {Daly}, {Davis}, {de Antoni}, {Delderfield}, {Devin},
  {di Giorgio}, {Didschuns}, {Dohlen}, {Donati}, {Dowell}, {Dowell}, {Duband},
  {Dumaye}, {Emery}, {Ferlet}, {Ferrand}, {Fontignie}, {Fox}, {Franceschini},
  {Frerking}, {Fulton}, {Garcia}, {Gastaud}, {Gear}, {Glenn}, {Goizel},
  {Griffin}, {Grundy}, {Guest}, {Guillemet}, {Hargrave}, {Harwit}, {Hastings},
  {Hatziminaoglou}, {Herman}, {Hinde}, {Hristov}, {Huang}, {Imhof}, {Isaak},
  {Israelsson}, {Ivison}, {Jennings}, {Kiernan}, {King}, {Lange}, {Latter},
  {Laurent}, {Laurent}, {Leeks}, {Lellouch}, {Levenson}, {Li}, {Li},
  {Lilienthal}, {Lim}, {Liu}, {Lu}, {Madden}, {Mainetti}, {Marliani}, {McKay},
  {Mercier}, {Molinari}, {Morris}, {Moseley}, {Mulder}, {Mur}, {Naylor},
  {Nguyen}, {O'Halloran}, {Oliver}, {Olofsson}, {Olofsson}, {Orfei}, {Page},
  {Pain}, {Panuzzo}, {Papageorgiou}, {Parks}, {Parr-Burman}, {Pearce},
  {Pearson}, {P{\'e}rez-Fournon}, {Pinsard}, {Pisano}, {Podosek}, {Pohlen},
  {Polehampton}, {Pouliquen}, {Rigopoulou}, {Rizzo}, {Roseboom}, {Roussel},
  {Rowan-Robinson}, {Rownd}, {Saraceno}, {Sauvage}, {Savage}, {Savini},
  {Sawyer}, {Scharmberg}, {Schmitt}, {Schneider}, {Schulz}, {Schwartz},
  {Shafer}, {Shupe}, {Sibthorpe}, {Sidher}, {Smith}, {Smith}, {Smith},
  {Spencer}, {Stobie}, {Sudiwala}, {Sukhatme}, {Surace}, {Stevens}, {Swinyard},
  {Trichas}, {Tourette}, {Triou}, {Tseng}, {Tucker}, {Turner}, {Vaccari},
  {Valtchanov}, {Vigroux}, {Virique}, {Voellmer}, {Walker}, {Ward}, {Waskett},
  {Weilert}, {Wesson}, {White}, {Whitehouse}, {Wilson}, {Winter}, {Woodcraft},
  {Wright}, {Xu}, {Zavagno}, {Zemcov}, {Zhang}, \& {Zonca}}]{griffin10}
{Griffin} M.~J. {et~al.}, 2010, \aap, 518, L3

\bibitem[{{Habing} {et~al}\mbox{.}(2001){Habing}, {Dominik}, {Jourdain de
  Muizon}, {Laureijs}, {Kessler}, {Leech}, {Metcalfe}, {Salama},
  {Siebenmorgen}, {Trams}, \& {Bouchet}}]{habing01}
{Habing} H.~J. {et~al.}, 2001, \aap, 365, 545

\bibitem[{{Hauck} \& {Mermilliod}(1998)}]{hauck98}
{Hauck} B., {Mermilliod} M., 1998, \aaps, 129, 431

\bibitem[{{H{\o}g} {et~al}\mbox{.}(2000){H{\o}g}, {Fabricius}, {Makarov},
  {Urban}, {Corbin}, {Wycoff}, {Bastian}, {Schwekendiek}, \& {Wicenec}}]{hog00}
{H{\o}g} E. {et~al.}, 2000, \aap, 355, L27

\bibitem[{{Hughes} {et~al}\mbox{.}(2011){Hughes}, {Wilner}, {Andrews},
  {Williams}, {Su}, {Murray-Clay}, \& {Qi}}]{hughes11}
{Hughes} A.~M., {Wilner} D.~J., {Andrews} S.~M., {Williams} J.~P., {Su}
  K.~Y.~L., {Murray-Clay} R.~A., {Qi} C., 2011, \apj, 740, 38

\bibitem[{{Ida} {et~al}\mbox{.}(2000){Ida}, {Larwood}, \& {Burkert}}]{ida00}
{Ida} S., {Larwood} J., {Burkert} A., 2000, \apj, 528, 351

\bibitem[{{Ishihara} {et~al}\mbox{.}(2010){Ishihara}, {Onaka}, {Kataza},
  {Salama}, {Alfageme}, {Cassatella}, {Cox}, {Garc{\'{\i}}a-Lario},
  {Stephenson}, {Cohen}, {Fujishiro}, {Fujiwara}, {Hasegawa}, {Ita}, {Kim},
  {Matsuhara}, {Murakami}, {M{\"u}ller}, {Nakagawa}, {Ohyama}, {Oyabu}, {Pyo},
  {Sakon}, {Shibai}, {Takita}, {Tanab{\'e}}, {Uemizu}, {Ueno}, {Usui}, {Wada},
  {Watarai}, {Yamamura}, \& {Yamauchi}}]{ishihara10}
{Ishihara} D. {et~al.}, 2010, \aap, 514, A1

\bibitem[{{Johansen} {et~al}\mbox{.}(2007){Johansen}, {Oishi}, {Low}, {Klahr},
  {Henning}, \& {Youdin}}]{johansen07}
{Johansen} A., {Oishi} J.~S., {Low} M., {Klahr} H., {Henning} T., {Youdin} A.,
  2007, \nat, 448, 1022

\bibitem[{{Johansen} {et~al}\mbox{.}(2012){Johansen}, {Youdin}, \&
  {Lithwick}}]{johansen12}
{Johansen} A., {Youdin} A.~N., {Lithwick} Y., 2012, \aap, 537, A125

\bibitem[{{Johnstone} {et~al}\mbox{.}(1998){Johnstone}, {Hollenbach}, \&
  {Bally}}]{johnstone98}
{Johnstone} D., {Hollenbach} D., {Bally} J., 1998, \apj, 499, 758

\bibitem[{{Kalas} {et~al}\mbox{.}(2005){Kalas}, {Graham}, \&
  {Clampin}}]{kalas05}
{Kalas} P., {Graham} J.~R., {Clampin} M., 2005, \nat, 435, 1067

\bibitem[{{Kalas} {et~al}\mbox{.}(2006){Kalas}, {Graham}, {Clampin}, \&
  {Fitzgerald}}]{kalas06}
{Kalas} P., {Graham} J.~R., {Clampin} M.~C., {Fitzgerald} M.~P., 2006, \apjl,
  637, L57

\bibitem[{{Kalas} \& {Jewitt}(1995)}]{kalas95}
{Kalas} P., {Jewitt} D., 1995, \aj, 110, 794

\bibitem[{{Kennedy} {et~al}\mbox{.}(2012{\natexlab{a}}){Kennedy}, {Wyatt},
  {Sibthorpe}, {Duch{\^e}ne}, {Kalas}, {Matthews}, {Greaves}, {Su}, \&
  {Fitzgerald}}]{kennedy12}
{Kennedy} G.~M. {et~al.}, 2012{\natexlab{a}}, \mnras, 421, 2264

\bibitem[{{Kennedy} {et~al}\mbox{.}(2012{\natexlab{b}}){Kennedy}, {Wyatt},
  {Sibthorpe}, {Phillips}, {Matthews}, \& {Greaves}}]{kennedy12a}
{Kennedy} G.~M., {Wyatt} M.~C., {Sibthorpe} B., {Phillips} N.~M., {Matthews}
  B., {Greaves} J.~S., 2012{\natexlab{b}}, \mnras, in press

\bibitem[{{Kenyon} \& {Bromley}(2004)}]{kenyon04a}
{Kenyon} S.~J., {Bromley} B.~C., 2004, \aj, 127, 513

\bibitem[{{Krivov} {et~al}\mbox{.}(2000){Krivov}, {Mann}, \&
  {Krivova}}]{krivov00}
{Krivov} A.~V., {Mann} I., {Krivova} N.~A., 2000, \aap, 362, 1127

\bibitem[{{Krivov} {et~al}\mbox{.}(2005){Krivov}, {Srem{\v c}evi{\'c}}, \&
  {Spahn}}]{krivov05}
{Krivov} A.~V., {Srem{\v c}evi{\'c}} M., {Spahn} F., 2005, Icarus, 174, 105

\bibitem[{{Lagrange} {et~al}\mbox{.}(2010){Lagrange}, {Bonnefoy}, {Chauvin},
  {Apai}, {Ehrenreich}, {Boccaletti}, {Gratadour}, {Rouan}, {Mouillet},
  {Lacour}, \& {Kasper}}]{lagrange10}
{Lagrange} A.-M. {et~al.}, 2010, Science, 329, 57

\bibitem[{{Laor} \& {Draine}(1993)}]{laor93}
{Laor} A., {Draine} B.~T., 1993, \apj, 402, 441

\bibitem[{{Lebouteiller} {et~al}\mbox{.}(2011){Lebouteiller}, {Barry}, {Spoon},
  {Bernard-Salas}, {Sloan}, {Houck}, \& {Weedman}}]{lebouteiller11}
{Lebouteiller} V., {Barry} D.~J., {Spoon} H.~W.~W., {Bernard-Salas} J., {Sloan}
  G.~C., {Houck} J.~R., {Weedman} D.~W., 2011, \apjs, 196, 8

\bibitem[{{Li} \& {Greenberg}(1997)}]{li97}
{Li} A., {Greenberg} J.~M., 1997, \aap, 323, 566

\bibitem[{{L{\"o}hne} {et~al}\mbox{.}(2008){L{\"o}hne}, {Krivov}, \&
  {Rodmann}}]{lohne08}
{L{\"o}hne} T., {Krivov} A.~V., {Rodmann} J., 2008, \apj, 673, 1123

\bibitem[{{Lykawka} \& {Mukai}(2008)}]{lykawka08}
{Lykawka} P.~S., {Mukai} T., 2008, \aj, 135, 1161

\bibitem[{{Mann} \& {Williams}(2010)}]{mann10}
{Mann} R.~K., {Williams} J.~P., 2010, \apj, 725, 430

\bibitem[{{Markwardt}(2009)}]{markwardt09}
{Markwardt} C.~B., 2009, in Astronomical Society of the Pacific Conference
  Series, {D.~A.~Bohlender, D.~Durand, \& P.~Dowler}, ed., Vol. 411, pp. 251--+

\bibitem[{{Mart{\'{\i}}nez-Galarza}
  {et~al}\mbox{.}(2009){Mart{\'{\i}}nez-Galarza}, {Kamp}, {Su},
  {G{\'a}sp{\'a}r}, {Rieke}, \& {Mamajek}}]{martinez09}
{Mart{\'{\i}}nez-Galarza} J.~R., {Kamp} I., {Su} K.~Y.~L., {G{\'a}sp{\'a}r} A.,
  {Rieke} G., {Mamajek} E.~E., 2009, \apj, 694, 165

\bibitem[{{Matthews} {et~al}\mbox{.}(2010){Matthews}, {Sibthorpe}, {Kennedy},
  {Phillips}, {Churcher}, {Duch{\^e}ne}, {Greaves}, {Lestrade},
  {Moro-Mart{\'i}n}, {Wyatt}, {Bastien}, {Biggs}, {Bouvier}, {Butner}, {Dent},
  {di Francesco}, {Eisl{\"o}ffel}, {Graham}, {Harvey}, {Hauschildt}, {Holland},
  {Horner}, {Ibar}, {Ivison}, {Johnstone}, {Kalas}, {Kavelaars}, {Rodriguez},
  {Udry}, {van der Werf}, {Wilner}, \& {Zuckerman}}]{matthews10}
{Matthews} B.~C. {et~al.}, 2010, \aap, 518, L135+

\bibitem[{{Mermilliod}(2006)}]{mermilliod06}
{Mermilliod} J.~C., 2006, VizieR Online Data Catalog, 2168, 0

\bibitem[{{Moerchen} {et~al}\mbox{.}(2010){Moerchen}, {Telesco}, \&
  {Packham}}]{moerchen10}
{Moerchen} M.~M., {Telesco} C.~M., {Packham} C., 2010, \apj, 723, 1418

\bibitem[{{Morales} {et~al}\mbox{.}(2011){Morales}, {Rieke}, {Werner},
  {Bryden}, {Stapelfeldt}, \& {Su}}]{morales11}
{Morales} F.~Y., {Rieke} G.~H., {Werner} M.~W., {Bryden} G., {Stapelfeldt}
  K.~R., {Su} K.~Y.~L., 2011, \apjl, 730, L29+

\bibitem[{{Moshir} {et~al}\mbox{.}(1990){Moshir}, Copan, Conrow, McCallon,
  Hacking, Gregorich, Rohrbach, Melnyk, Rice, Fullmer, \& Chester}]{moshir90}
{Moshir} M. {et~al.}, 1990, in IRAS Faint Source Catalogue, version 2.0 (1990),
  p.~0

\bibitem[{{Mouillet} {et~al}\mbox{.}(1997){Mouillet}, {Larwood}, {Papaloizou},
  \& {Lagrange}}]{mouillet97}
{Mouillet} D., {Larwood} J.~D., {Papaloizou} J.~C.~B., {Lagrange} A.~M., 1997,
  \mnras, 292, 896

\bibitem[{{M{\"u}ller} {et~al}\mbox{.}(2010){M{\"u}ller}, {L{\"o}hne}, \&
  {Krivov}}]{muller10}
{M{\"u}ller} S., {L{\"o}hne} T., {Krivov} A.~V., 2010, \apj, 708, 1728

\bibitem[{{O'Brien} \& {Greenberg}(2003)}]{obrien03}
{O'Brien} D.~P., {Greenberg} R., 2003, Icarus, 164, 334

\bibitem[{{Ott}(2010)}]{ott10}
{Ott} S., 2010, in Astronomical Society of the Pacific Conference Series, Vol.
  434, Astronomical Data Analysis Software and Systems XIX, {Y.~Mizumoto,
  K.-I.~Morita, \& M.~Ohishi}, ed., p. 139

\bibitem[{{Perryman} \& {ESA}(1997)}]{perryman97a}
{Perryman} M.~A.~C., {ESA}, eds., 1997, ESA Special Publication, Vol. 1200,
  {The HIPPARCOS and TYCHO catalogues. Astrometric and photometric star
  catalogues derived from the ESA HIPPARCOS Space Astrometry Mission}

\bibitem[{{Petit} {et~al}\mbox{.}(2011){Petit}, {Kavelaars}, {Gladman},
  {Jones}, {Parker}, {Van Laerhoven}, {Nicholson}, {Mars}, {Rousselot},
  {Mousis}, {Marsden}, {Bieryla}, {Taylor}, {Ashby}, {Benavidez}, {Campo
  Bagatin}, \& {Bernabeu}}]{petit11}
{Petit} J.-M. {et~al.}, 2011, \aj, 142, 131

\bibitem[{{Phillips} {et~al}\mbox{.}(2010){Phillips}, {Greaves}, {Dent},
  {Matthews}, {Holland}, {Wyatt}, \& {Sibthorpe}}]{phillips10}
{Phillips} N.~M., {Greaves} J.~S., {Dent} W.~R.~F., {Matthews} B.~C., {Holland}
  W.~S., {Wyatt} M.~C., {Sibthorpe} B., 2010, \mnras, 403, 1089

\bibitem[{{Pilbratt} {et~al}\mbox{.}(2010){Pilbratt}, {Riedinger}, {Passvogel},
  {Crone}, {Doyle}, {Gageur}, {Heras}, {Jewell}, {Metcalfe}, {Ott}, \&
  {Schmidt}}]{pilbratt10}
{Pilbratt} G.~L. {et~al.}, 2010, \aap, 518, L1

\bibitem[{{Poglitsch} {et~al}\mbox{.}(2010){Poglitsch}, {Waelkens}, {Geis},
  {Feuchtgruber}, {Vandenbussche}, {Rodriguez}, {Krause}, {Renotte}, {van
  Hoof}, {Saraceno}, {Cepa}, {Kerschbaum}, {Agn{\`e}se}, {Ali}, {Altieri},
  {Andreani}, {Augueres}, {Balog}, {Barl}, {Bauer}, {Belbachir}, {Benedettini},
  {Billot}, {Boulade}, {Bischof}, {Blommaert}, {Callut}, {Cara}, {Cerulli},
  {Cesarsky}, {Contursi}, {Creten}, {De Meester}, {Doublier}, {Doumayrou},
  {Duband}, {Exter}, {Genzel}, {Gillis}, {Gr{\"o}zinger}, {Henning},
  {Herreros}, {Huygen}, {Inguscio}, {Jakob}, {Jamar}, {Jean}, {de Jong},
  {Katterloher}, {Kiss}, {Klaas}, {Lemke}, {Lutz}, {Madden}, {Marquet},
  {Martignac}, {Mazy}, {Merken}, {Montfort}, {Morbidelli}, {M{\"u}ller},
  {Nielbock}, {Okumura}, {Orfei}, {Ottensamer}, {Pezzuto}, {Popesso},
  {Putzeys}, {Regibo}, {Reveret}, {Royer}, {Sauvage}, {Schreiber}, {Stegmaier},
  {Schmitt}, {Schubert}, {Sturm}, {Thiel}, {Tofani}, {Vavrek}, {Wetzstein},
  {Wieprecht}, \& {Wiezorrek}}]{poglitsch10}
{Poglitsch} A. {et~al.}, 2010, \aap, 518, L2

\bibitem[{{Pourbaix}(2000)}]{pourbaix00}
{Pourbaix} D., 2000, \aaps, 145, 215

\bibitem[{Press {et~al}\mbox{.}(1992)Press, Teukolsky, Vetterling, \&
  Flannery}]{press92}
Press W.~H., Teukolsky S.~A., Vetterling W.~T., Flannery B.~P., 1992, Numerical
  recipes in C (2nd ed.): the art of scientific computing. Cambridge University
  Press, New York, NY, USA

\bibitem[{{Quillen}(2006)}]{quillen06}
{Quillen} A.~C., 2006, \mnras, 372, L14

\bibitem[{{Regibo} {et~al}\mbox{.}(2012){Regibo}, {Vandenbussche}, {Waelkens},
  {Acke}, {Sibthorpe}, {Nottebaere}, {Voet}, {Di Francesco}, {Fridlund},
  {Gear}, {Ivison}, \& {Olofsson}}]{regibo12}
{Regibo} S. {et~al.}, 2012, \aap, 541, A3

\bibitem[{{Rieke} {et~al}\mbox{.}(2004){Rieke}, {Young}, {Engelbracht},
  {Kelly}, {Low}, {Haller}, {Beeman}, {Gordon}, {Stansberry}, {Misselt},
  {Cadien}, {Morrison}, {Rivlis}, {Latter}, {Noriega-Crespo}, {Padgett},
  {Stapelfeldt}, {Hines}, {Egami}, {Muzerolle}, {Alonso-Herrero}, {Blaylock},
  {Dole}, {Hinz}, {Le Floc'h}, {Papovich}, {P{\'e}rez-Gonz{\'a}lez}, {Smith},
  {Su}, {Bennett}, {Frayer}, {Henderson}, {Lu}, {Masci}, {Pesenson}, {Rebull},
  {Rho}, {Keene}, {Stolovy}, {Wachter}, {Wheaton}, {Werner}, \&
  {Richards}}]{rieke04}
{Rieke} G.~H. {et~al.}, 2004, \apjs, 154, 25

\bibitem[{{Rodriguez} \& {Zuckerman}(2012)}]{rodriguez12}
{Rodriguez} D.~R., {Zuckerman} B., 2012, \apj, 745, 147

\bibitem[{{Schmidt-Kaler}(1982)}]{schmidt82}
{Schmidt-Kaler} T., 1982, {Landolt-B{\"o}rnstein: Numerical Data and Functional
  Relationships in Science and Technology, New Series, Group VI}, Vol.~2,
  Springer Verlag, p.~14

\bibitem[{{Sibthorpe} {et~al}\mbox{.}(2010){Sibthorpe}, {Vandenbussche},
  {Greaves}, {Pantin}, {Olofsson}, {Acke}, {Barlow}, {Blommaert}, {Bouwman},
  {Brandeker}, {Cohen}, {De Meester}, {Dent}, {di Francesco}, {Dominik},
  {Fridlund}, {Gear}, {Glauser}, {Gomez}, {Hargrave}, {Harvey}, {Henning},
  {Heras}, {Hogerheijde}, {Holland}, {Ivison}, {Leeks}, {Lim}, {Liseau},
  {Matthews}, {Naylor}, {Pilbratt}, {Polehampton}, {Regibo}, {Royer},
  {Sicilia-Aguilar}, {Swinyard}, {Waelkens}, {Walker}, \&
  {Wesson}}]{sibthorpe10}
{Sibthorpe} B. {et~al.}, 2010, \aap, 518, L130

\bibitem[{{Smith} \& {Wyatt}(2010)}]{smith10}
{Smith} R., {Wyatt} M.~C., 2010, \aap, 515, A95

\bibitem[{{Su} {et~al}\mbox{.}(2005){Su}, {Rieke}, {Misselt}, {Stansberry},
  {Moro-Martin}, {Stapelfeldt}, {Werner}, {Trilling}, {Bendo}, {Gordon},
  {Hines}, {Wyatt}, {Holland}, {Marengo}, {Megeath}, \& {Fazio}}]{su05}
{Su} K.~Y.~L. {et~al.}, 2005, \apj, 628, 487

\bibitem[{{Th{\'e}bault} \& {Augereau}(2007)}]{thebault07}
{Th{\'e}bault} P., {Augereau} J.-C., 2007, \aap, 472, 169

\bibitem[{{Th{\'e}bault} {et~al}\mbox{.}(2003){Th{\'e}bault}, {Augereau}, \&
  {Beust}}]{thebault03}
{Th{\'e}bault} P., {Augereau} J.~C., {Beust} H., 2003, \aap, 408, 775

\bibitem[{{Tomkin} \& {Popper}(1986)}]{tomkin86}
{Tomkin} J., {Popper} D.~M., 1986, \aj, 91, 1428

\bibitem[{{Trilling} {et~al}\mbox{.}(2007){Trilling}, {Stansberry},
  {Stapelfeldt}, {Rieke}, {Su}, {Gray}, {Corbally}, {Bryden}, {Chen}, {Boden},
  \& {Beichman}}]{trilling07}
{Trilling} D.~E. {et~al.}, 2007, \apj, 658, 1289

\bibitem[{{van Leeuwen}(2007)}]{leeuwen07}
{van Leeuwen} F., 2007, \aap, 474, 653

\bibitem[{{Vican}(2012)}]{vican12}
{Vican} L., 2012, \aj, 143, 135

\bibitem[{{Vitense} {et~al}\mbox{.}(2010){Vitense}, {Krivov}, \&
  {L{\"o}hne}}]{vitense10}
{Vitense} C., {Krivov} A.~V., {L{\"o}hne} T., 2010, \aap, 520, A32

\bibitem[{{Weidenschilling}(2003)}]{weidenschilling03}
{Weidenschilling} S.~J., 2003, Icarus, 165, 438

\bibitem[{{Williams} \& {Andrews}(2006)}]{williams06}
{Williams} J.~P., {Andrews} S.~M., 2006, \apj, 653, 1480

\bibitem[{{Wilner} {et~al}\mbox{.}(2012){Wilner}, {Andrews}, {MacGregor}, \&
  {Hughes}}]{wilner12}
{Wilner} D.~J., {Andrews} S.~M., {MacGregor} M.~A., {Hughes} A.~M., 2012,
  \apjl, 749, L27

\bibitem[{{Wright} {et~al}\mbox{.}(2010){Wright}, {Eisenhardt}, {Mainzer},
  {Ressler}, {Cutri}, {Jarrett}, {Kirkpatrick}, {Padgett}, {McMillan},
  {Skrutskie}, {Stanford}, {Cohen}, {Walker}, {Mather}, {Leisawitz}, {Gautier},
  {McLean}, {Benford}, {Lonsdale}, {Blain}, {Mendez}, {Irace}, {Duval}, {Liu},
  {Royer}, {Heinrichsen}, {Howard}, {Shannon}, {Kendall}, {Walsh}, {Larsen},
  {Cardon}, {Schick}, {Schwalm}, {Abid}, {Fabinsky}, {Naes}, \&
  {Tsai}}]{wright10}
{Wright} E.~L. {et~al.}, 2010, \aj, 140, 1868

\bibitem[{{Wyatt}(2005)}]{wyatt05}
{Wyatt} M.~C., 2005, \aap, 433, 1007

\bibitem[{{Wyatt}(2008)}]{wyatt08}
{Wyatt} M.~C., 2008, \araa, 46, 339

\bibitem[{{Wyatt} {et~al}\mbox{.}(2011){Wyatt}, {Clarke}, \& {Booth}}]{wyatt11}
{Wyatt} M.~C., {Clarke} C.~J., {Booth} M., 2011, Celestial Mechanics and
  Dynamical Astronomy, 111, 1

\bibitem[{{Wyatt} \& {Dent}(2002)}]{wyatt02}
{Wyatt} M.~C., {Dent} W.~R.~F., 2002, \mnras, 334, 589

\bibitem[{{Wyatt} {et~al}\mbox{.}(1999){Wyatt}, {Dermott}, {Telesco}, {Fisher},
  {Grogan}, {Holmes}, \& {Pi{\~n}a}}]{wyatt99}
{Wyatt} M.~C., {Dermott} S.~F., {Telesco} C.~M., {Fisher} R.~S., {Grogan} K.,
  {Holmes} E.~K., {Pi{\~n}a} R.~K., 1999, \apj, 527, 918

\bibitem[{{Wyatt} {et~al}\mbox{.}(2012){Wyatt}, {Kennedy}, {Sibthorpe},
  {Moro-Mart{\'{\i}}n}, {Lestrade}, {Ivison}, {Matthews}, {Udry}, {Greaves},
  {Kalas}, {Lawler}, {Su}, {Rieke}, {Booth}, {Bryden}, {Horner}, {Kavelaars},
  \& {Wilner}}]{wyatt12}
{Wyatt} M.~C. {et~al.}, 2012, \mnras, 3237

\bibitem[{{Wyatt} {et~al}\mbox{.}(2007){Wyatt}, {Smith}, {Su}, {Rieke},
  {Greaves}, {Beichman}, \& {Bryden}}]{wyatt07a}
{Wyatt} M.~C., {Smith} R., {Su} K.~Y.~L., {Rieke} G.~H., {Greaves} J.~S.,
  {Beichman} C.~A., {Bryden} G., 2007, \apj, 663, 365

\bibitem[{{Zuckerman} \& {Song}(2004)}]{zuckerman04}
{Zuckerman} B., {Song} I., 2004, \apj, 603, 738

\end{thebibliography}

\bsp

\end{document}